\documentclass[12pt,english,preprint]{article}
\usepackage[T1]{fontenc}
\usepackage[latin9]{inputenc}
\usepackage{geometry}
\usepackage{amstext}
\usepackage{amssymb}
\usepackage{graphicx}
\usepackage{setspace}
\usepackage{esint}
\makeatletter
\newcommand{\lyxaddress}[1]{
\par {\raggedright #1
\vspace{1.4em}
\noindent\par}
}

\@ifundefined{date}{}{\date{}}
\makeatother

\usepackage{babel}
\begin{document}

\title{RG flow equations for the proper-vertices\\
of the background effective average action}

\author{Alessandro Codello%
\thanks{email: codello@sissa.it%
}}

\maketitle

\lyxaddress{\begin{center}
\emph{SISSA}\\
\emph{via Bonomea 265,} \emph{I-34136} \emph{Trieste, Italy}
\par\end{center}}
\begin{abstract}
We derive a system of coupled flow equations for the proper-vertices
of the background effective average action and we give an explicit
representation of these by means of diagrammatic and momentum space
techniques. This explicit representation can be used as a new computational
technique that enables the projection of the flow of a large new class
of truncations of the background effective average action. In particular,
these can be single-- or bi--field truncations of local or non--local
character. As an application we study non--abelian gauge theories.
We show how to use this new technique to calculate the beta function
of the gauge coupling (without employing the heat kernel expansion)
under various approximations. In particular, one of these approximations
leads to a derivation of beta functions similar to those proposed
as candidates for an ``all--orders'' beta function. Finally, we
discuss some possible phenomenology related to these flows.

\newpage

\tableofcontents{}

\newpage
\end{abstract}

\section{Introduction}

The effective average action (EAA) formalism is a promising approach
to QFT which recently has seen important developments and applications
\cite{Berges_Tetradis_Wetterich_2002,Gies_2006}. The EAA is a $k$
dependent functional that interpolates smoothly between the bare action
for $k=\Lambda$ and the standard effective action for $k=0$; from
this point of view it offers a new approach to quantization. When
applied to theories with local symmetries, as non-abelian gauge theories
\cite{Reuter_Wetterich_1994a,Reuter_1995}, non-linear sigma models
\cite{Codello_Percacci_2009}, gravity \cite{Reuter_1996}, or membranes
\cite{Codello_Zanusso_2011} it can be implemented using the background
field method \cite{Reuter_Wetterich_1994a}. This defines the background
EAA (bEAA).

In this formalisms one usually makes a truncation ansatz for the bEAA
and then performs calculations with the aid of heat kernel techniques
\cite{Codello_Percacci_Rahmede_2009}. Generally, the range of applicability
of heat kernel methods is confined to special cases, where the Hessian
of the bEAA is of generalized Laplacian type \cite{Benedetti_Groh_Machado_Saueressig_2010}.
This technology permits the study of truncations where the bEAA is
the sum of local operators; these are interesting from the point of
view of renormalizability, but are not enough to capture the full
form of the bEAA, which in general contains non-local operators. These
non-local terms are of fundamental importance, if one wants to construct
RG trajectories that reach the far infrared (IR) $k=0$. It has been
shown in a workable examples \cite{Codello_2010} that only by employing
these kind of truncations it is possible to construct an EAA that
correctly reproduces the effective action in the limit $k\rightarrow0$.
Another important and complementary issue that requires stronger computational
tools is the one related to ``bi--field'' truncations \cite{Reuter_Bi_Field}:
since (as we will review in section 2) the full bEAA is a functional
of both fluctuation and background fields, its RG flow takes place
in the enlarged theory space where this functional naturally lives.
Bi--field truncations generally give rise to operators that are difficult
to treat using heat kernel methods.

In this paper we present a novel formalisms based on the hierarchy
of coupled flow equations satisfied by the one-particle-irreducible
(1PI) proper-vertices of the bEAA. We give an explicit representation
of these equations by means of diagrammatic and momentum space techniques.
This explicit representation can be used as a new computational technique
that enables the projection of the flow of a large new class of truncations
of the bEAA. These include both the non-local and bi-field truncations
mentioned above.

The hierarchy of flow equations for the proper-vertices contains vertices
with both fluctuation and background legs. The non-trivial part of
the formalism is the explicit momentum space representation of vertices
with background legs, since these contain legs attached to the cutoff
action (which is function of the background field). These terms are
not present in the non-background formalism of the standard EAA and
constitute the necessary contributions to make the RG flow covariant.
A key element in the derivation of this representation is the perturbative
expansion of the (un-traced) heat kernel \cite{Codello_Zanusso_2012}.

As an application we study non-abelian gauge theories. First we show
how standard heat kernel computation of the one--loop beta function
of the gauge coupling is reproduced by our formalism, then we extend
the study under two approximations, a single-- and a bi--field one,
to obtain RG improved beta functions. One of these turns out to be
very similar to the ``all-orders'' beta function proposed by Ryttov
and Sannino \cite{Ryttov_Sannino_2008}. Finally, we discuss some
possible phenomenology related to these RG flows.

In the second section of the paper we review the basic properties
of the bEAA and we derive the hierarchy of coupled flow equations
for the proper-vertices and we expose the relative momentum space
representation by introducing the diagrammatic rules. The details
of the derivation of the non-trivial momentum space representation
of vertices with background legs are given in appendix A. In the third
section we apply our formalisms to non-abelian gauge theories to show
how it works in practice. 

\newpage

\section{Flow equations for proper-vertices of the bEAA}

In this section we first review the construction of the bEAA and the
derivation of the exact RG flow equation it satisfies; successively
we derive the hierarchy of flow equations for the proper-vertices
of bEAA and we expose their momentum space representation and relative
diagrammatic rules.

\subsection{Construction of bEAA}

The crucial point in the construction of the EAA for theories with
local gauge symmetries is, obviously, the preservation of gauge invariance
during the RG coarse-graining procedure. A possible way to cut--off
field modes covariantly is to define the cutoff using covariant differential
operators. But if we try to introduce a cutoff by simply taking it
as a function of a covariant differential operator, constructed with
the quantum fields, we will spoil the simple one-loop structure of
the exact flow that the EAA obeys. To obtain a one-loop like flow
equation, the cutoff action has to be quadratic in the quantum fields.
Still, the EAA will not be gauge invariant because of the non-covariant
coupling of the quantum fields to the source, a problem that affects
also the standard effective action. A way out of this is to employ
the background field method, as was first done in \cite{Reuter_Wetterich_1994a,Reuter_Wetterich_1993a},
to define what we will call the background effective average action
(bEAA).

The theories that we have in mind are non-abelian gauge theories,
gravity and non-linear sigma models. In this paper we will use the
language of non-abelian gauge theories to keep the notation as light
as possible, but it is clear that everything is readily translated
to the other cases. 

In the background field formalism \cite{Abbott_1981} the quantum
field $A_{\mu}$ is linearly split between the background field $\bar{A}_{\mu}$
and the fluctuation field $a_{\mu}$:
\begin{equation}
A_{\mu}=\bar{A}_{\mu}+a_{\mu}\,.\label{gauge_1}
\end{equation}
The cutoff action $\Delta S_{k}$ is taken to be quadratic in the
fluctuation field, while the cutoff kernel $R_{k}[\bar{A}]$ is constructed
employing the background field alone:
\begin{equation}
\Delta S_{k}[\varphi;\bar{A}]=\frac{1}{2}\int d^{d}x\,\varphi\, R_{k}[\bar{A}]\varphi\,,\label{gauge_2}
\end{equation}
where $\varphi=(a_{\mu},\bar{c},c)$ is the fluctuation multiplet,
combining the fluctuation field $a_{\mu}$ and the ghost fields $\bar{c}$
and $c$. To construct the bEAA we introduce the cutoff action (\ref{gauge_2})
into the integro-differential definition of the standard background
effective action%
\footnote{Functional derivatives of a functionals depending on many arguments
are indicated by the notation $\Gamma^{(n_{1},n_{2},...)}[....]$.%
}:
\begin{equation}
e^{-\Gamma_{k}[\varphi;\bar{A}]}=\int D\chi\exp\left(-S[\chi+\varphi;\bar{A}]-\Delta S_{k}[\chi;\bar{A}]+\int d^{d}x\,\Gamma_{k}^{(1;0)}[\varphi;\bar{A}]\chi\right)\,,\label{gauge_3}
\end{equation}
where the fluctuation field multiplet $\chi$ has vanishing vacuum
expectation value $\left\langle \chi\right\rangle =0$. The role of
the cutoff action is to suppress field modes with momenta smaller
than the RG scale $k$, the shape of the cutoff kernel must be constructed
in a way consistent with this property; more details on this can be
found in \cite{Berges_Tetradis_Wetterich_2002,Gies_2006}. With these
definitions the bEAA is a functional that interpolates smoothly between
the bare action (\ref{gauge_4}) for $k\rightarrow\infty$ (fact that
can be easily checked) and the full background effective action for
$k\rightarrow0$. This fact, together with the exact flow equation
bEAA satisfies (that we derive in the next subsection), offers the
starting point to develop a formalisms that can be used to define
and construct QFT characterized by local gauge symmetries. The computational
technique introduced in this paper has to be seen as a tool to pursue
this research route. 

In (\ref{gauge_3}) $S[\varphi;\bar{A}]$ is the bare action which
is the sum of an invariant action (which can be the classical non-abelian
gauge theory action), the background gauge-fixing action and the background
ghost action: 
\begin{equation}
S[\varphi;\bar{A}]=S[\bar{A}+a]+S_{gf}[a;\bar{A}]+S_{gh}[a,\bar{c},c;\bar{A}]\,.\label{gauge_4}
\end{equation}
The background gauge-fixing action is
\begin{equation}
S_{gf}[a;\bar{A}]=\frac{1}{2\alpha}\int d^{d}x\,\bar{D}_{\mu}a^{\mu}\bar{D}_{\nu}a^{\nu}\,,\label{gauge_5}
\end{equation}
where $\alpha$ is the gauge-fixing parameter, while the background
ghost action is
\begin{equation}
S_{gh}[a,\bar{c},c;\bar{A}]=\int d^{d}x\,\bar{D}_{\mu}\bar{c}\, D^{\mu}c=\int d^{d}x\,\bar{D}_{\mu}\bar{c}\left(\bar{D}^{\mu}+ga^{\mu}\right)c\,.\label{gauge_6}
\end{equation}
With these definitions, the bEAA (\ref{gauge_3}) is invariant under
combined physical%
\footnote{See appendix B for our conventions.%
},
\begin{equation}
\delta_{\theta}A_{\mu}=D_{\mu}\theta\qquad\delta_{\theta}\bar{c}=[\bar{c},\theta]\qquad\delta_{\theta}c=[c,\theta]\qquad\delta_{\theta}\bar{A}_{\mu}=0\,,\label{gauge_6A}
\end{equation}
and background,
\begin{equation}
\bar{\delta}_{\theta}A_{\mu}=\delta_{\theta}\bar{c}=\bar{\delta}_{\theta}c=0\qquad\qquad\bar{\delta}_{\theta}\bar{A}_{\mu}=\bar{D}_{\mu}\theta\,,\label{gauge_6B}
\end{equation}
gauge transformations:
\begin{equation}
(\delta_{\theta}+\bar{\delta}_{\theta})\Gamma_{k}[\varphi;\bar{A}]=0\,.\label{gauge_6.01}
\end{equation}
For a proof of (\ref{gauge_6.01}) and for more details about the
(background) gauge invariance of the bEAA we refer to the literature
\cite{Ellwanger_1994}. See also \cite{Lavrov_Shapiro_2013}.

It is possible to define a gauge invariant functional of the background
field, that we will call ``gauge invariant effective average action''
(gEAA), by setting to zero the fluctuation multiplet $\varphi=0$
in the bEAA:
\begin{equation}
\bar{\Gamma}_{k}[\bar{A}]\equiv\Gamma_{k}[0;\bar{A}]\,.\label{gauge_6.1}
\end{equation}
This definition is equivalent to a parametrization of the bEAA as
the sum of a functional of the full quantum field $A_{\mu}=\bar{A}_{\mu}+a_{\mu}$,
the gEAA (\ref{gauge_6.1}), and a ``remainder functional'' $\hat{\Gamma}_{k}[\varphi;\bar{A}]$
(rEAA) which remains a functional of both the fluctuation multiplet
and the background field: 
\begin{equation}
\Gamma_{k}[\varphi;\bar{A}]=\bar{\Gamma}_{k}[\bar{A}+a]+\hat{\Gamma}_{k}[\varphi;\bar{A}]\,.\label{gauge_6.2}
\end{equation}
To recover (\ref{gauge_6.1}) we must have $\hat{\Gamma}_{k}[0;\bar{A}]=0$.
The gEAA defined in this way is a functional invariant under physical
gauge transformations (\ref{gauge_6A}):
\begin{equation}
\delta_{\theta}\bar{\Gamma}_{k}[\bar{A}]=0\,,\label{gauge_6.3}
\end{equation}
while the rEAA remains a functional invariant under combined physical
and background gauge transformations, as the full bEAA, and is subject
to modified Ward-Takahashi identities. We refer to \cite{Reuter_Wetterich_1994a}
for more details on this point. What is important is that the gEAA
flows, in the IR $k\rightarrow0$, to the standard gauge invariant
effective action of the background field formalism, which can thus
be computed using the bEAA formalism.

\subsubsection{Exact flow equation for the bEAA}

We derive now the exact flow equation satisfied by the bEAA \cite{Reuter_Wetterich_1994a}.
Differentiating with respect to the ``RG time'' $t=\log k/k_{0}$
both sides of the integro-differential equation (\ref{gauge_3}) we
find:
\begin{eqnarray}
e^{-\Gamma_{k}[\varphi;\bar{A}]}\partial_{t}\Gamma_{k}[\varphi;\bar{A}] & = & \int D\chi\left(\partial_{t}\Delta S_{k}[\chi;\bar{A}]-\int d^{d}x\,\partial_{t}\Gamma_{k}^{(1;0)}[\varphi;\bar{A}]\chi\right)\times\nonumber \\
 &  & \times e^{-S[\varphi+\chi;\bar{A}]-\Delta S_{k}[\chi;\bar{A}]+\int\Gamma_{k}^{(1;0)}[\varphi;\bar{A}]\chi}\,.\label{gauge_7}
\end{eqnarray}
We can re-express the terms on the rhs of (\ref{gauge_7}) as expectation
values and use (\ref{gauge_2}) to rewrite (\ref{gauge_7}) as:
\begin{eqnarray}
\partial_{t}\Gamma_{k}[\varphi;\bar{A}] & = & \left\langle \partial_{t}\Delta S_{k}[\chi;\bar{A}]\right\rangle -\int d^{d}x\,\partial_{t}\Gamma_{k}^{(1;0)}[\varphi;\bar{A}]\left\langle \chi\right\rangle \nonumber \\
 & = & \frac{1}{2}\int d^{d}x\left\langle \chi\chi\right\rangle \partial_{t}R_{k}\,,\label{gauge_8}
\end{eqnarray}
where we used $\left\langle \chi\right\rangle =0$. The two-point
function of $\chi$ in (\ref{gauge_8}) can be written in terms of
the inverse Hessian of the bEAA plus the cutoff action (where functional
derivatives are taken with respect to $\varphi$):
\begin{equation}
\left\langle \chi\chi\right\rangle =\left(\Gamma_{k}^{(2;0)}[\varphi;\bar{A}]+\Delta S_{k}^{(2;0)}[\varphi;\bar{A}]\right)^{-1}=\left(\Gamma_{k}^{(2;0)}[\varphi;\bar{A}]+R_{k}[\bar{A}]\right)^{-1}\,,\label{gauge_9}
\end{equation}
where in the second step we used the fact that the Hessian of cutoff
action (\ref{gauge_2}) is the cutoff kernel $R_{k}[\bar{A}]$. Inserting
(\ref{gauge_9}) back into (\ref{gauge_8}), and writing a functional
trace in place of the integral, finally gives:
\begin{equation}
\partial_{t}\Gamma_{k}[\varphi;\bar{A}]=\frac{1}{2}\mathrm{Tr}\left(\Gamma_{k}^{(2;0)}[\varphi;\bar{A}]+R_{k}[\bar{A}]\right)^{-1}\,\partial_{t}R_{k}[\bar{A}]\,.\label{gauge_10}
\end{equation}
Equation (\ref{gauge_10}) is the exact RG flow equation for the bEAA
\cite{Reuter_Wetterich_1994a}. The flow equation (\ref{gauge_10})
has the same general properties as the flow equation for the standard
effective average action \cite{Berges_Tetradis_Wetterich_2002}. In
the following it will be useful to define the field dependent ``regularized
propagator'': 
\begin{equation}
G_{k}[\varphi;\bar{A}]=\left(\Gamma_{k}^{(2;0)}[\varphi;\bar{A}]+R_{k}[\bar{A}]\right)^{-1}\,,\label{gauge_12}
\end{equation}
to rewrite the flow equation (\ref{gauge_10}) in the following compact
form:
\begin{equation}
\partial_{t}\Gamma_{k}[\varphi;\bar{A}]=\frac{1}{2}\mathrm{Tr}\, G_{k}[\varphi;\bar{A}]\partial_{t}R_{k}[\bar{A}]\,.\label{gauge_13}
\end{equation}
The flow equation satisfied by the gEAA is readily obtained combining
(\ref{gauge_6.2}) and (\ref{gauge_13}):
\begin{equation}
\partial_{t}\bar{\Gamma}_{k}[\bar{A}]=\partial_{t}\Gamma_{k}[0;\bar{A}]=\frac{1}{2}\textrm{Tr}\, G_{k}[0;\bar{A}]\partial_{t}R_{k}[\bar{A}]\,.\label{gauge_14}
\end{equation}
By construction the flow equation for the gEAA respects gauge symmetry,
in the sense that the trace on the rhs of (\ref{gauge_14}), the ``beta
functional'', is a gauge invariant functional of $\bar{A}_{\mu}$.
Still, it is important to realize that equation (\ref{gauge_14})
is not a closed equation since it involves the Hessian of the bEAA
$\Gamma_{k}^{(2;0)}[0;\bar{A}]$ taken with respect to the fluctuation
field. This fact implies that, even if in the limit $k\rightarrow0$
the gEAA flows to the gauge invariant effective action $\bar{\Gamma}[\bar{A}]\equiv\Gamma[0;\bar{A}]$,
for $k\neq0$ it is necessary to consider the flow in the extended
theory space where the bEAA lives, composed of all functionals of
the fields $\varphi$ and $\bar{A}_{\mu}$ invariant under simultaneous
physical and background gauge transformations.

When we set $\varphi=0$ the Hessian $\Gamma_{k}^{(2;0)}[0,\bar{A}]$
in (\ref{gauge_14}) becomes ``super-diagonal'', since the ghost
action is bilinear, and we can immediately perform the multiplet trace
in the flow equation. Using the shorthands $\Gamma_{k,aa}=\Gamma_{k}^{(2,0,0;0)}[0,0,0;\bar{A}]$,
$\Gamma_{k,\bar{c}c}=\Gamma_{k}^{(0,1,1;0)}[0,0,0;\bar{A}]$, $\Delta S_{k,aa}=\Delta S_{k}^{(2,0,0;0)}[0,0,0;\bar{A}]$
and $\Delta S_{k,\bar{c}c}=\Delta S_{k}^{(0,1,1;0)}[0,0,0;\bar{A}]$
we can write the flow equation in the following matrix form:
\begin{eqnarray}
\partial_{t}\bar{\Gamma}_{k}[\bar{A}] & = & \frac{1}{2}\textrm{Tr}\left(\begin{array}{ccc}
\Gamma_{k,aa}+\Delta S_{k,aa} & 0 & 0\\
0 & 0 & \Gamma_{k,\bar{c}c}+\Delta S_{k,\bar{c}c}\\
0 & -\left(\Gamma_{k,\bar{c}c}+\Delta S_{k,\bar{c}c}\right) & 0
\end{array}\right)^{-1}\nonumber \\
 &  & \times\left(\begin{array}{ccc}
\partial_{t}\Delta S_{k,aa} & 0 & 0\\
0 & 0 & \partial_{t}\Delta S_{k,\bar{c}c}\\
0 & -\partial_{t}\Delta S_{k,\bar{c}c} & 0
\end{array}\right)\nonumber \\
 & = & \frac{1}{2}\textrm{Tr}\left(\begin{array}{ccc}
G_{k,aa} & 0 & 0\\
0 & 0 & -G_{k,\bar{c}c}\\
0 & G_{k,\bar{c}c} & 0
\end{array}\right)\left(\begin{array}{ccc}
\partial_{t}\Delta S_{k,aa} & 0 & 0\\
0 & 0 & \partial_{t}\Delta S_{k,\bar{c}c}\\
0 & -\partial_{t}\Delta S_{k,\bar{c}c} & 0
\end{array}\right)\nonumber \\
 & = & \frac{1}{2}\textrm{Tr}\left(\begin{array}{ccc}
G_{k,aa}\partial_{t}\Delta S_{k,aa} & 0 & 0\\
0 & G_{k,\bar{c}c}\partial_{t}\Delta S_{k,\bar{c}c} & 0\\
0 & 0 & G_{k,\bar{c}c}\partial_{t}\Delta S_{k,\bar{c}c}
\end{array}\right)\nonumber \\
 & = & \frac{1}{2}\textrm{Tr}\, G_{k,aa}\partial_{t}\Delta S_{k,aa}-\textrm{Tr}\, G_{k,\bar{c}c}\partial_{t}\Delta S_{k,\bar{c}c}\,,\label{gauge_14.1}
\end{eqnarray}
where we also defined $G_{k,aa}=(\Gamma_{k,aa}+\Delta S_{k,aa})^{-1}$
and $G_{k,\bar{c}c}=(\Gamma_{k,\bar{c}c}+\Delta S_{k,\bar{c}c})^{-1}$.
In (\ref{gauge_14.1}) we used the property that the trace over anti-commuting
fields carries an additional minus sign. The last line of equation
(\ref{gauge_14.1}) gives the flow equation for the gEAA in its commonly
used form \cite{Reuter_Wetterich_1994a}:
\begin{eqnarray}
\partial_{t}\bar{\Gamma}_{k}[\bar{A}] & = & \frac{1}{2}\textrm{Tr}\left(\Gamma_{k}^{(2,0,0;0)}[0;\bar{A}]+\Delta S_{k}^{(2,0,0;0)}[0,0,0;\bar{A}]\right)^{-1}\partial_{t}\Delta S_{k}^{(2,0,0;0)}[0,0,0;\bar{A}]\nonumber \\
 &  & -\textrm{Tr}\left(\Gamma_{k}^{(0,1,1;0)}[0;\bar{A}]+\Delta S_{k}^{(0,1,1;0)}[0,0,0;\bar{A}]\right)^{-1}\partial_{t}\Delta S_{k}^{(0,1,1;0)}[0,0,0;\bar{A}]\,.\label{gauge_15}
\end{eqnarray}
One can obtain equation (\ref{gauge_15}) as an RG improvement of
the one--loop gEAA obtained from (\ref{gauge_3}) by saddle point
expansion. One finds:
\begin{eqnarray}
\bar{\Gamma}_{k}[\bar{A}] & = & S[\bar{A}]+\frac{1}{2}\textrm{Tr}\log\left(S^{(2)}[\bar{A}]+S_{gf}^{(2;0)}[0;\bar{A}]+S_{gh}^{(2,0,0;0)}[0,0,0;\bar{A}]+\Delta S_{k}^{(2,0,0;0)}[0,0,0;\bar{A}]\right)\nonumber \\
 &  & -\textrm{Tr}\log\left(S_{gh}^{(0,1,1;0)}[0,0,0;\bar{A}]+\Delta S_{k}^{(0,1,1;0)}[0,0,0;\bar{A}]\right)\,;\label{gauge_15.1}
\end{eqnarray}
if we now substitute in the traces on the rhs the Hessian of bare
action $S[\varphi;\bar{A}]$ with the Hessians of the bEAA $\Gamma_{k}[\varphi;\bar{A}]$
and we differentiate with respect to $t=\log k/k_{0}$ on both sides,
we recover the exact RG flow equation (\ref{gauge_15}). We note that
one may just substitute $S[\bar{A}]$ with $\bar{\Gamma}_{k}[\bar{A}]$
to obtain directly a closed equation involving only the gEAA; this
seams to suggests that truncations where the rEAA is approximated
with the classical gauge-fixing and ghost actions may be a consistent
approximation when considering the flow of the gEAA.

The flow equation for rEAA can be deduced differentiating equation
(\ref{gauge_6.2}):
\begin{equation}
\partial_{t}\hat{\Gamma}_{k}[\varphi;\bar{A}]=\partial_{t}\bar{\Gamma}_{k}[\bar{A}+a]-\partial_{t}\Gamma_{k}[\varphi;\bar{A}]\,,\label{gauge_16}
\end{equation}
where we used (\ref{gauge_10}) and (\ref{gauge_15}).

\subsection{Flow equations for the proper-vertices of the bEAA}

In this section we derive the system of equations governing the RG
flow of the proper-vertices of both the bEAA and the gEAA. To obtain
these equations we take functional derivatives of the flow equation
(\ref{gauge_10}) satisfied by the bEAA with respect to the fields
$\varphi$ and $\bar{A}_{\mu}$. When we differentiate with respect
to the background field, we have to remember that the cutoff terms
present in the flow equation depend explicitly on it: this adds additional
terms to the flow equations for the proper-vertices that are not present
in the non-background formalism. We will see that these terms are
crucial to preserve gauge covariance of the gEAA along the flow.

\subsubsection{Derivation}

To start, we take a functional derivative of the flow equation (\ref{gauge_10})
with respect to the fluctuation field multiplet, or with respect to
the background field, to obtain the following flow equations for the
one-vertices of the bEAA:
\begin{eqnarray}
\partial_{t}\Gamma_{k}^{(1;0)}[\varphi;\bar{A}] & = & -\frac{1}{2}\textrm{Tr}\, G_{k}[\varphi;\bar{A}]\,\Gamma_{k}^{(3;0)}[\varphi;\bar{A}]G_{k}[\varphi;\bar{A}]\partial_{t}R_{k}[\bar{A}]\nonumber \\
\partial_{t}\Gamma_{k}^{(0;1)}[\varphi;\bar{A}] & = & -\frac{1}{2}\textrm{Tr}\, G_{k}[\varphi;\bar{A}]\left(\Gamma_{k}^{(2;1)}[\varphi;\bar{A}]+R_{k}^{(1)}[\bar{A}]\right)G_{k}[\varphi;\bar{A}]\partial_{t}R_{k}[\bar{A}]\nonumber \\
 &  & +\frac{1}{2}\textrm{Tr}\, G_{k}[\varphi;\bar{A}]\partial_{t}R_{k}^{(1)}[\bar{A}]\,.\label{gauge_27}
\end{eqnarray}
Note that in the second equation of (\ref{gauge_27}), where we differentiated
with respect to the background field, there are additional terms containing
functional derivatives of the cutoff kernel $R_{k}[\bar{A}]$. Taking
further derivatives of equation (\ref{gauge_27}), with respect to
both the fluctuation field multiplet and background field, gives the
following flow equations for the two-vertices%
\footnote{Here and in other equations of this section we omit, for clarity,
to write explicitly the arguments of the functionals when these are
understood.%
}:
\begin{eqnarray}
\partial_{t}\Gamma_{k}^{(2;0)} & = & \textrm{Tr}\, G_{k}\,\Gamma_{k}^{(3;0)}G_{k}\,\Gamma_{k}^{(3;0)}G_{k}\partial_{t}R_{k}-\frac{1}{2}\textrm{Tr}\, G_{k}\,\Gamma_{k}^{(4;0)}G_{k}\partial_{t}R_{k}\nonumber \\
\partial_{t}\Gamma_{k}^{(1;1)} & = & \textrm{Tr}\, G_{k}\left(\Gamma_{k}^{(2;1)}+R_{k}^{(1)}\right)G_{k}\,\Gamma_{k}^{(3;0)}G_{k}\partial_{t}R_{k}\nonumber \\
 &  & +\textrm{Tr}\, G_{k}\,\Gamma_{k}^{(3;0)}G_{k}\left(\Gamma_{k}^{(2;1)}+R_{k}^{(1)}\right)G_{k}\partial_{t}R_{k}\nonumber \\
 &  & -\frac{1}{2}\textrm{Tr}\, G_{k}\Gamma_{k}^{(3;1)}G_{k}\partial_{t}R_{k}-\frac{1}{2}\textrm{Tr}\, G_{k}\,\Gamma_{k}^{(3;0)}G_{k}\partial_{t}R_{k}^{(1)}\nonumber \\
\partial_{t}\Gamma_{k}^{(0;2)} & = & \textrm{Tr}\, G_{k}\left(\Gamma_{k}^{(2;1)}+R_{k}^{(1)}\right)G_{k}\left(\Gamma_{k}^{(2;1)}+R_{k}^{(1)}\right)G_{k}\partial_{t}R_{k}\nonumber \\
 &  & -\frac{1}{2}\textrm{Tr}\, G_{k}\left(\Gamma_{k}^{(2;2)}+R_{k}^{(2)}\right)G_{k}\partial_{t}R_{k}\nonumber \\
 &  & -\textrm{Tr}\, G_{k}\left(\Gamma_{k}^{(2;1)}+R_{k}^{(1)}\right)G_{k}\partial_{t}R_{k}^{(1)}+\frac{1}{2}\textrm{Tr}\, G_{k}\partial_{t}R_{k}^{(2)}\,.\label{gauge_28}
\end{eqnarray}
Proceeding in this way we generate a hierarchy of flow equations for
the proper-vertices $\Gamma_{k}^{(n;m)}[\varphi;\bar{A}]$ of the
bEAA. In general the flow equation for $\Gamma_{k}^{(n;m)}[\varphi;\bar{A}]$
involves proper-vertices up to $\Gamma_{k}^{(n+2;m)}[\varphi;\bar{A}]$
and functional derivatives of the cutoff kernel up to $R_{k}^{(m)}[\bar{A}]$.

Note that, as they stand in (\ref{gauge_27}) and (\ref{gauge_28}),
every equation of the hierarchy contains the same information as the
original flow equation (\ref{gauge_10}). To profit of the vertex
expansion we perform a Taylor expansion of the functional $\Gamma_{k}[\varphi;\bar{A}]$
around $\varphi=0$ and $\bar{A}_{\mu}=0$:
\begin{equation}
\Gamma_{k}[\varphi;\bar{A}]=\sum_{n,m=0}^{\infty}\frac{1}{n!m!}\int_{x_{1}...x_{n}y_{1}...y_{m}}\gamma_{k,x_{1}...x_{n}y_{1}...y_{m}}^{(n;m)}\varphi_{x_{1}}...\varphi_{x_{n}}\bar{A}_{y_{1}}...\bar{A}_{y_{m}}\,,\label{gauge_29}
\end{equation}
where we defined the zero-field proper-vertices as follows:
\begin{equation}
\gamma_{k,x_{1}...x_{n}y_{1}...y_{m}}^{(n;m)}\equiv\Gamma_{k}^{(n;m)}[0;0]_{x_{1}...x_{n}y_{1}...y_{m}}\,.\label{gauge_29.1}
\end{equation}
If we evaluate the hierarchy of flow equations at $\varphi=0$ and
$\bar{A}_{\mu}=0$ it becomes an infinite system of coupled integro-differential
equations for the zero-field proper-vertices $\gamma_{k}^{(n;m)}$.
From the Taylor expansion (\ref{gauge_29}) that defines these vertices,
we see that this system can be used to project the RG flow of all
terms of the bEAA which are analytic in the fields $\varphi$ and
$\bar{A}_{\mu}$. In particular these terms can be of non-local character.

Note that in the above considerations is not necessary to expand around
the zero-background configuration $\bar{A}_{\mu}=0$; one can choose
to expand around any background, preferably where one is able to perform
computations. An example is a constant magnetic field configuration
or, in the gravitational case, a sphere or an Einstein space.

As for the bEAA, we can derive a hierarchy of flow equations for the
proper-vertices of the gEAA. In this case the functional depends only
on the background field. Taking a functional derivative of (\ref{gauge_14})
with respect to this field gives the following flow equation for the
one-vertex of the gEAA: 
\begin{eqnarray}
\partial_{t}\bar{\Gamma}_{k}^{(1)}[\bar{A}] & = & -\frac{1}{2}\textrm{Tr}\, G_{k}[0,\bar{A}]\left(\Gamma_{k}^{(2;1)}[0,\bar{A}]+R_{k}^{(1)}[\bar{A}]\right)G_{k}[0,\bar{A}]\partial_{t}R_{k}[\bar{A}]\nonumber \\
 &  & +\frac{1}{2}\textrm{Tr}\, G_{k}[0,\bar{A}]\partial_{t}R_{k}^{(1)}[\bar{A}]\,.\label{gauge_30}
\end{eqnarray}
Note that since $\partial_{t}\Gamma_{k}^{(0;1)}[0;\bar{A}]=\partial_{t}\bar{\Gamma}_{k}^{(1)}[\bar{A}]$,
equation (\ref{gauge_30}) is the same as the second equation in (\ref{gauge_27}),
but with $\varphi=0$. A further derivative of (\ref{gauge_30}) with
respect to $\bar{A}_{\mu}$ gives the flow equation for the two-vertex
of the gEAA:
\begin{eqnarray}
\partial_{t}\bar{\Gamma}_{k}^{(2)} & = & \textrm{Tr}\, G_{k}\left(\Gamma_{k}^{(2;1)}+R_{k}^{(1)}\right)G_{k}\left(\Gamma_{k}^{(2;1)}+R_{k}^{(1)}\right)G_{k}\partial_{t}R_{k}\nonumber \\
 &  & -\frac{1}{2}\textrm{Tr}\, G_{k}\left(\Gamma_{k}^{(2;2)}+R_{k}^{(2)}\right)G_{k}\partial_{t}R_{k}\nonumber \\
 &  & -\textrm{Tr}\, G_{k}\left(\Gamma_{k}^{(2;1)}+R_{k}^{(1)}\right)G_{k}\partial_{t}R_{k}^{(1)}+\frac{1}{2}\textrm{Tr}\, G_{k}\partial_{t}R_{k}^{(2)}\,.\label{gauge_31}
\end{eqnarray}
As for (\ref{gauge_30}), this equation is equal to the last equation
in (\ref{gauge_28}) if we set $\varphi=0$ and use $\partial_{t}\Gamma_{k}^{(0;2)}[0;\bar{A}]=\partial_{t}\bar{\Gamma}_{k}^{(2)}[\bar{A}]$.
As previously stated, the terms coming from functional derivatives
of the cutoff kernel (that are present in the background formalism
but not in the non-background one) are crucial in preserving the covariant
character of the flow of the gEAA and its vertices.

As for the bEAA, we can perform a Taylor expansion of the gEAA analogous
to (\ref{gauge_29}) and define the zero-field proper-vertices
\begin{equation}
\bar{\gamma}_{k,x_{1}...x_{n}}^{(n)}\equiv\bar{\Gamma}_{k}^{(n)}[0]_{x_{1}...x_{n}}\,,\label{gauge_31.1}
\end{equation}
to turn the hierarchy of flow equations for the proper-vertices of
the gEAA into an infinite system of coupled integro-differential equations
for the vertices $\bar{\gamma}_{k}^{(m)}$.

\subsubsection{Compact form}

We introduce now a compact notation to rewrite the flow equations
for the proper-vertices just derived. If we introduce the formal operator
\begin{equation}
\tilde{\partial}_{t}=(\partial_{t}R_{k}-\eta_{k}R_{k})\frac{\partial}{\partial R_{k}}\,,\label{gauge_32.1}
\end{equation}
where $\eta_{k}$ is the multiplet matrix of anomalous dimensions,
we can rewrite the flow equation for the bEAA (\ref{gauge_10}) as:
\begin{equation}
\partial_{t}\Gamma_{k}[\varphi;\bar{A}]=\frac{1}{2}\textrm{Tr}\, G_{k}[\varphi;\bar{A}]\partial_{t}R_{k}[\bar{A}]=-\frac{1}{2}\textrm{Tr}\,\tilde{\partial}_{t}\log G_{k}[\varphi;\bar{A}]\,.\label{gauge_32}
\end{equation}
In (\ref{gauge_32}) we used the following simple relations:
\[
\tilde{\partial}_{t}G_{k}=-G_{k}\partial_{t}R_{k}G_{k}\qquad\qquad\tilde{\partial}_{t}\log G_{k}=G_{k}^{-1}\tilde{\partial}_{t}G_{k}=G_{k}\partial_{t}R_{k}\,.
\]
In this way, we can rewrite the flow equation for the one-vertices
of the bEAA (\ref{gauge_27}) in the following compact form:
\begin{eqnarray}
\partial_{t}\Gamma_{k}^{(1;0)}[\varphi;\bar{A}] & = & \frac{1}{2}\textrm{Tr}\,\tilde{\partial}_{t}\left\{ \Gamma_{k}^{(3;0)}[\varphi;\bar{A}]G_{k}[\varphi;\bar{A}]\right\} \nonumber \\
\partial_{t}\Gamma_{k}^{(0;1)}[\varphi;\bar{A}] & = & \frac{1}{2}\textrm{Tr}\,\tilde{\partial}_{t}\left\{ \left(\Gamma_{k}^{(2;1)}[\varphi;\bar{A}]+R_{k}^{(1)}[\bar{A}]\right)G_{k}[\varphi;\bar{A}]\right\} \,,\label{gauge_33}
\end{eqnarray}
while the flow equations for the two-vertices of the bEAA (\ref{gauge_28})
read now:
\begin{eqnarray}
\partial_{t}\Gamma_{k}^{(2;0)} & = & -\frac{1}{2}\textrm{Tr}\,\tilde{\partial}_{t}\left\{ \Gamma_{k}^{(3;0)}G_{k}\,\Gamma_{k}^{(3;0)}G_{k}\right\} +\frac{1}{2}\textrm{Tr}\,\tilde{\partial}_{t}\left\{ \Gamma_{k}^{(4;0)}G_{k}\right\} \nonumber \\
\partial_{t}\Gamma_{k}^{(1;1)} & = & -\frac{1}{2}\textrm{Tr}\,\tilde{\partial}_{t}\left\{ \left(\Gamma_{k}^{(2;1)}+R_{k}^{(1)}\right)G_{k}\,\Gamma_{k}^{(3;0)}G_{k}\right\} +\frac{1}{2}\textrm{Tr}\,\tilde{\partial}_{t}\left\{ \Gamma_{k}^{(3;1)}G_{k}\right\} \nonumber \\
\partial_{t}\Gamma_{k}^{(0;2)} & = & -\frac{1}{2}\textrm{Tr}\,\tilde{\partial}_{t}\left\{ \left(\Gamma_{k}^{(2;1)}+R_{k}^{(1)}\right)G_{k}\left(\Gamma_{k}^{(2;1)}+R_{k}^{(1)}\right)G_{k}\right\} \nonumber \\
 &  & +\frac{1}{2}\textrm{Tr}\,\tilde{\partial}_{t}\left\{ \left(\Gamma_{k}^{(2;2)}+R_{k}^{(2)}\right)G_{k}\right\} \,.\label{gauge_34}
\end{eqnarray}
This compact form turns out to be very useful since the flow equations
(\ref{gauge_33}) and (\ref{gauge_34}) contain fewer terms than their
counter-parts (\ref{gauge_27}) and (\ref{gauge_28}), and are thus
much more manageable when employed in actual computations. The same
reasoning applies to all subsequent equations of the hierarchy and
extend to the flow equations for the zero-field proper-vertices $\gamma_{k}^{(n;m)}$.
The flow equations for the proper-vertices of the gEAA are just those
for the bEAA evaluated at $\varphi=0$ and we won't repeat them here.
We will see in the next section how the flow equation for the zero-field
proper-vertices just defined can be turned into a powerful computational
device to perform computations in the bEAA framework.

\subsection{Diagrammatic and momentum space techniques}

In this section we introduce a useful diagrammatic representation
to organize the various contributions to the flow equations for the
zero-field proper-vertices $\gamma_{k}^{(n;m)}$ and we expose the
momentum space rules that enables us to calculate these contributions
explicitly. As we said before, all these results are valid for any
theory with local gauge invariance. For this reason we try to maintain
our notation as general as possible.
\begin{figure}
\centering{}\includegraphics[scale=0.9]{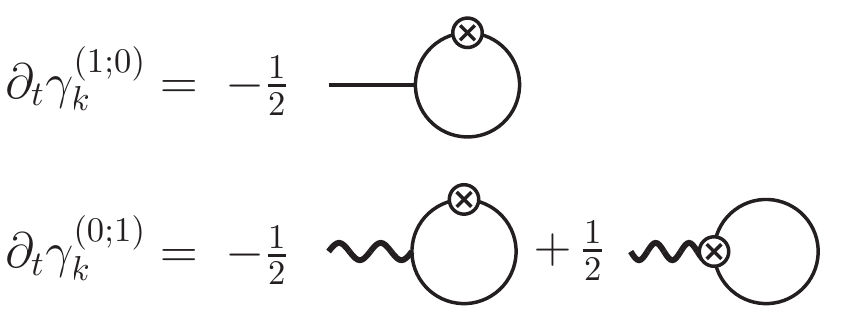}\caption{Diagrammatic representation of the flow equations for $\partial_{t}\gamma_{k}^{(1;0)}$
and $\partial_{t}\gamma_{k}^{(0;1)}$ as given in (\ref{gauge_27}).}
\end{figure}

As we will see, there are some non-trivial technical steps that have
to be made in order to write explicit momentum space flow equations
for zero-field proper-vertices with some background legs, i.e. containing
the vertices $\gamma^{(n;m)}$ with $m>0$. This issue is related
to the dependence of the cutoff kernel $R_{k}[\bar{A}]$ on the cutoff
operator which is constructed using the background field. The functional
derivatives of the cutoff kernel $R_{k}^{(m)}[\bar{A}]$ can be calculated
as terms of a Taylor series expansion of the cutoff kernel with respect
to the background field. We will perform this expansion using the
perturbative expansion for the (un-traced) heat kernel developed in
\cite{Codello_Zanusso_2012} and reviewed in appendix A.

\subsubsection{Diagrammatic rules}

We start to introduce the diagrammatic rules used to represent the
hierarchy of flow equations for the zero-field proper-vertices. In
particular, we show how to represent equations (\ref{gauge_27}) and
(\ref{gauge_28}) graphically. The virtue of diagrammatic techniques
is that they allow to switch from coordinate space to momentum space
straightforwardly.

When considering vertices with background lines, it is useful to introduce
the ``tilde'' bEAA%
\footnote{\noindent Note that $\tilde{\Gamma}_{k}[\varphi;\bar{A}]$ is the
actual Legendre transform of the functional generator of connected
correlation functions $W_{k}[J;\bar{A}]$.%
} defined by: 
\begin{equation}
\tilde{\Gamma}_{k}[\varphi;\bar{A}]=\Gamma_{k}[\varphi;\bar{A}]+\Delta S_{k}[\varphi;\bar{A}]\,,\label{gauge_D_1.01}
\end{equation}
and to define the related ``tilde'' zero-field proper-vertices
\begin{equation}
\tilde{\gamma}_{k}^{(n;m)}=\gamma_{k}^{(n;m)}+\Delta S_{k}^{(n;m)}[0;0]\,.\label{gauge_D_1.02}
\end{equation}
Obviously, for every $n$ the relation $\tilde{\gamma}_{k}^{(n;0)}=\gamma_{k}^{(n;0)}$
holds. Hence, if we want, we can re-phrase the flow equations for
the zero-field proper-vertices solely in terms of the tilde vertices
$\tilde{\gamma}^{(n;m)}$.

We represent the zero-field regularized propagator $G_{k}[0;0]$ with
an internal continuous line, the cutoff insertions $\partial_{t}R_{k}[0]$
are indicated with a crossed circle and the zero-field proper-vertices
$\tilde{\gamma}_{k}^{(n;m)}$ are represented as vertices with $n$
external continuous lines (fluctuation legs) and $m$ external thick
wavy lines (background legs). Note that $\Delta S_{k}^{(n;m)}[0;0]=0$
if $n>2$ since the cutoff action is quadratic in the fluctuation
fields. This diagrammatic rules are summarized graphically as follows:
\begin{figure}
\centering{}\includegraphics[scale=0.9]{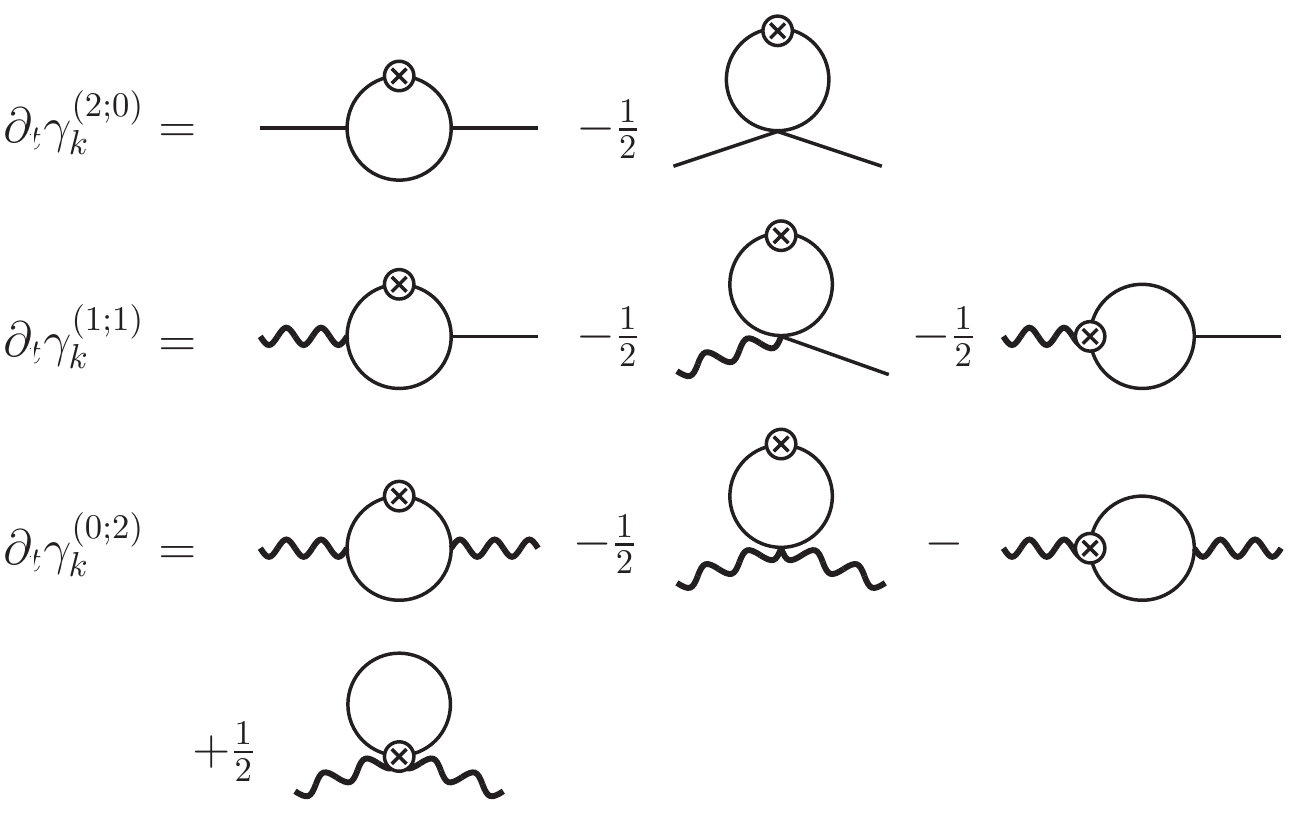}\caption{Diagrammatic representation of the flow equations for the vertices
$\partial_{t}\gamma_{k}^{(2;0)}$, $\partial_{t}\gamma_{k}^{(1;1)}$
and $\partial_{t}\gamma_{k}^{(0;2)}$ as in equation (\ref{gauge_28}).}
\end{figure}

\begin{center}
\includegraphics[scale=0.8]{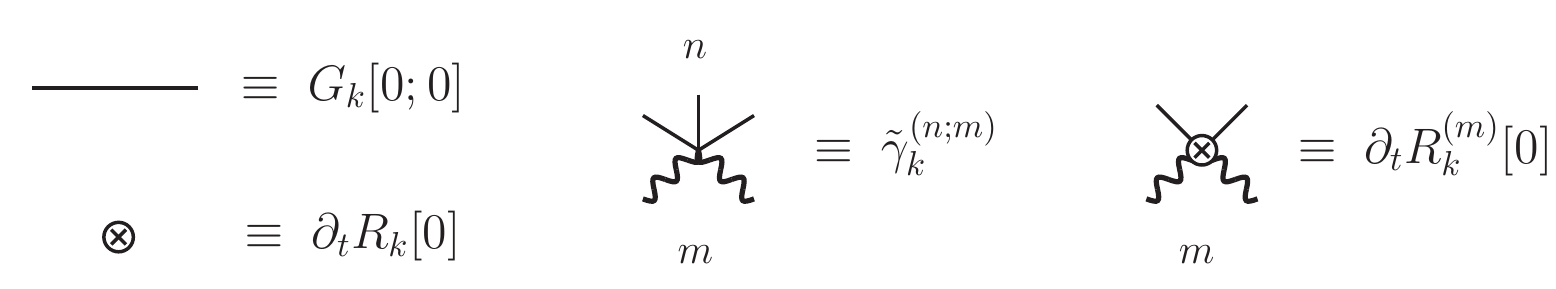}
\par\end{center}

\noindent Finally, to every closed loop we associate a coordinate
or a momentum%
\footnote{We define $\int_{x}\equiv\int d^{d}x$ and $\int_{q}\equiv\int\frac{d^{d}q}{(2\pi)^{d}}$.%
} $\Omega\int_{q}$ integral ($\Omega$ is the space-time volume),
together with the factor $\partial_{t}R_{k}-\eta R_{k}$. Here the
anomalous dimension $\eta_{k}$ pertains to the fields present in
the cutoff action. The application of these diagrammatic rules to
the flow equations (\ref{gauge_27}) for the zero-field one-vertices
$\partial_{t}\gamma_{k}^{(1;0)}$ and $\partial_{t}\gamma_{k}^{(0;1)}$
gives the representation of Figure 1, while the flow equations (\ref{gauge_28})
for the zero-field two-vertices $\partial_{t}\gamma_{k}^{(2;0)}$,
$\partial_{t}\gamma_{k}^{(1;1)}$ and $\partial_{t}\gamma_{k}^{(0;2)}$
can be represented as in Figure 2.

As explained in the previous section, it is sometimes useful to work
with the set of flow equations for the bEAA (\ref{gauge_33}) and
(\ref{gauge_34}) written employing the formal operator $\tilde{\partial}_{t}$
defined in (\ref{gauge_32.1}). In this case there is no explicit
insertion of the cutoff term $\partial_{t}R_{k}[0]$ in the loops,
but to every loop is now associated an integration together with the
action of this formal operator, i.e. $\int_{x}\tilde{\partial}_{t}$
in coordinate space or $\Omega\int_{q}\tilde{\partial}_{t}$ in momentum
space. In this way, the flow equations (\ref{gauge_33}) for the zero-field
one-vertices $\partial_{t}\gamma_{k}^{(1;0)}$ and $\partial_{t}\gamma_{k}^{(0;1)}$
can be represented as in Figure 3, while equations (\ref{gauge_34})
for the zero-field two-vertices $\partial_{t}\gamma_{k}^{(2;0)}$,
$\partial_{t}\gamma_{k}^{(1;1)}$ and $\partial_{t}\gamma_{k}^{(0;2)}$
can be represented as in Figure 4.

Note that in this last representation, the flow equations for the
different one-vertices or two-vertices assume a more symmetric form
with respect to each other. This reflects in a computational advantage,
especially when considering the flow equation for the two-vertex $\gamma_{k}^{(0;2)}$,
where the additional terms involving cutoff kernel vertices are accounted
for by the use of the tilde vertices and by the action of the operator
$\tilde{\partial}_{t}$. We will see in section 3.2.2 that this fact
is very useful.
\begin{figure}
\centering{}\includegraphics[scale=0.9]{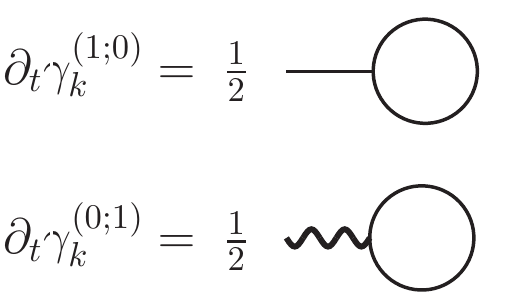}\caption{Diagrammatic representation of the flow equation for the one-vertices
$\partial_{t}\gamma_{k}^{(1;0)}$ and $\partial_{t}\gamma_{k}^{(0;1)}$
as given in (\ref{gauge_33}). }
\end{figure}

\subsubsection{Momentum space representation}

We are now in the position to write down the flow equations for the
zero-field proper-vertices $\gamma_{k}^{(n;m)}$ of the bEAA in momentum
space. The only non-standard step is to write the momentum space representation
of the terms involving functional derivatives of the cutoff kernel
$R_{k}[\bar{A}]$ with respect to the background field, i.e. the momentum
representation of the vertices $\gamma_{k}^{(2;1)}$ and $\gamma_{k}^{(2;2)}$.
The zero-field proper-vertex $\gamma_{k}^{(3;0)}$ is represented
graphically by the following diagram:

\begin{center}
\includegraphics[scale=0.9]{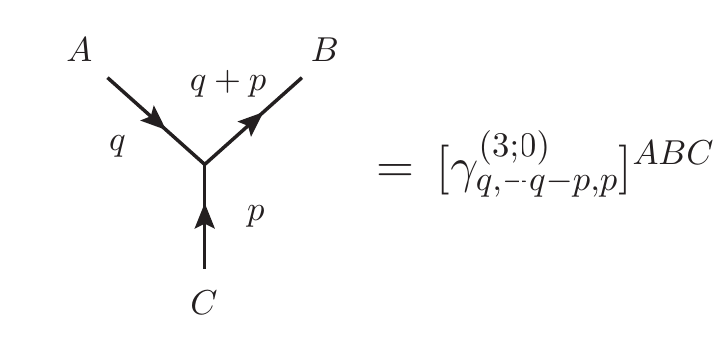}
\par\end{center}

\noindent Here we use capital letters to indicate general composite
indices, in the case of non-abelian gauge theories these have to be
interpreted as $A=a\,\alpha,B=b\,\beta,C=c\,\gamma$, while, for example,
in the gravitational context they have to interpreted as $A=\alpha\beta,B=\gamma\delta,C=\epsilon\kappa$.
Note that each index is associated with a momentum variable, so that
$A,B,C$ are the indices of the related momenta $q,p,-q-p$ respectively.
Note also that we always define ingoing momenta as being positive.
\begin{figure}
\centering{}\includegraphics[scale=0.9]{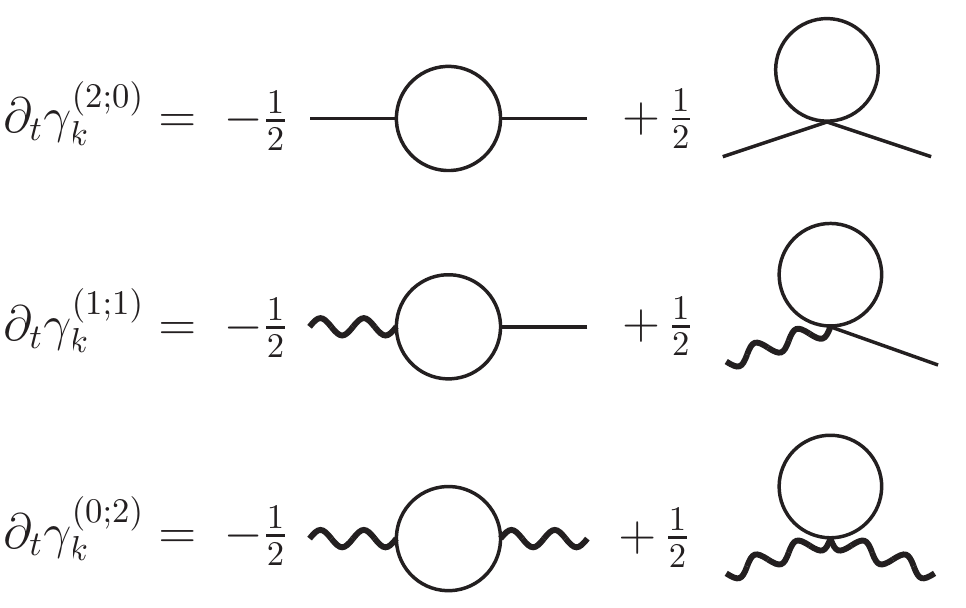}\caption{Diagrammatic representation of the flow equations for the two-vertices
$\partial_{t}\gamma_{k}^{(2;0)}$, $\partial_{t}\gamma_{k}^{(1;1)}$
and $\partial_{t}\gamma_{k}^{(0;2)}$ as in equation (\ref{gauge_34}).}
\end{figure}
 The two-fluctuations one-background zero-field proper-vertex $\tilde{\gamma}_{k}^{(2;1)}=\gamma_{k}^{(2;1)}+\Delta S_{k}^{(2;1)}[0;0]$
is represented graphically by the diagram:

\begin{center}
\includegraphics[scale=0.9]{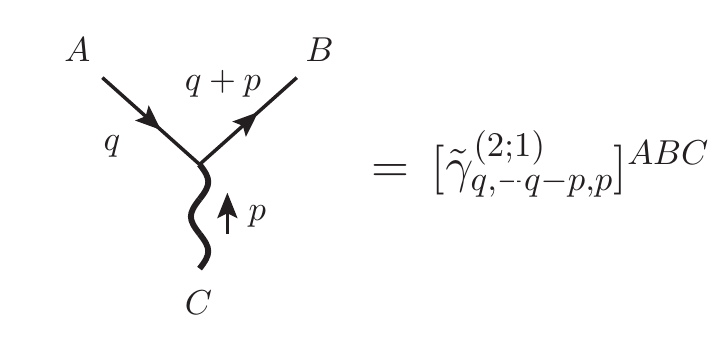}
\par\end{center}

\noindent The non-trivial technical step is to write the explicit
momentum space representation for the zero-field proper-vertex $\tilde{\gamma}_{k}^{(2;1)}$.
We start by stating the final result and we postpone the details of
the derivations to appendix A: 
\begin{equation}
[\tilde{\gamma}_{q,-q-p,p}^{(2;1)}]^{ABC}=[\gamma_{q,-q-p,p}^{(2;1)}]^{ABC}+[l_{q,-q-p,p}^{(2;1)}]^{ABC}R_{q+p,q}^{(1)}\,,\label{gauge_D_1.1}
\end{equation}
In (\ref{gauge_D_1.1}) we introduced the ``cutoff operator action''
$L[\varphi;\bar{A}]$, defined as that action whose Hessian with respect
to $\varphi$ is the cutoff operator and we defined its vertices as
\[
l_{x_{1}...x_{n}y_{1}...y_{m}}^{(n;m)}\equiv L^{(n;m)}[0;0]_{x_{1}...x_{n}y_{1}...y_{m}}\,.
\]
Furthermore,
\begin{equation}
R_{q+p,q}^{(1)}\equiv\frac{R_{q+p}-R_{q}}{(q+p)^{2}-q^{2}}\label{gauge_D_1}
\end{equation}
represents the first finite-difference derivative of the cutoff shape
function $R_{q}\equiv R_{k}(q^{2})$. The momentum space representation
(\ref{gauge_D_1.1}) is of crucial importance since it gives, together
with the generalization to higher vertices given in the following,
access to the computational use of the flow equations for the zero-field
proper-vertices $\gamma_{k}^{(n;m)}$. The four-fluctuation vertex
$\gamma_{k}^{(4;0)}$ is represented graphically as:

\begin{center}
\includegraphics[scale=0.9]{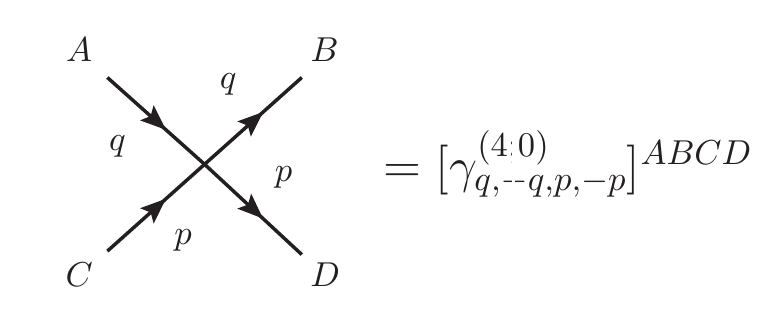}
\par\end{center}

\noindent Note that here we are giving the four-vertex only for a
particular combination of momenta, which is not the most general one,
since this will be the case that we will use in section 3. The two-fluctuations
two-backgrounds vertex $\tilde{\gamma}_{k}^{(2;2)}=\gamma_{k}^{(2;2)}+\Delta S_{k}^{(2;2)}[0;0]$
is represented instead by the following diagram:

\begin{center}
\includegraphics[scale=0.9]{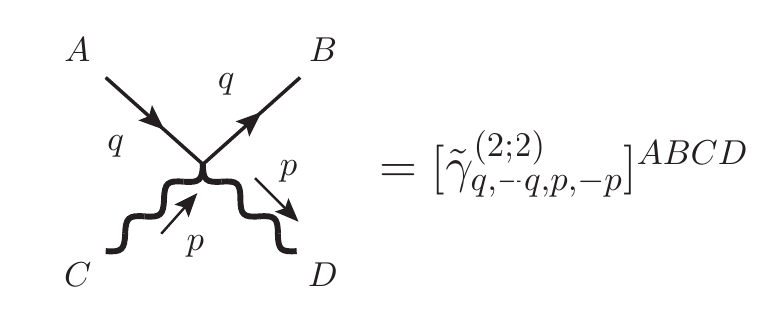}
\par\end{center}

\noindent The momentum space representation of $\tilde{\gamma}_{k}^{(2;2)}$,
derived in appendix B, turns out to be:
\begin{eqnarray}
[\tilde{\gamma}_{q,-q,p,-p}^{(2;2)}]^{ABCD} & = & [\gamma_{q,-q,p,-p}^{(2;2)}]^{ABC}+[l_{q,-q,p,-p}^{(2;2)}]^{ABCD}R_{q}'\nonumber \\
 &  & +[l_{q,-q-p,p}^{(2;1)}]^{ABM}[l_{q+p,-q,-p}^{(2;1)}]^{MCD}R_{q+p,q}^{(2)}\,.\label{gauge_D_1.2}
\end{eqnarray}
Note that now the cutoff operator action $L[\varphi;\bar{A}]$ enters
both as a four-vertex and as a product of two three-vertices, and
that this time we need to consider the second finite-difference derivative
of the cutoff shape function $R_{q}$:
\begin{equation}
R_{q+p,q}^{(2)}\equiv\frac{2}{(q+p)^{2}-q^{2}}\left[\frac{R_{q+p}-R_{q}}{(q+p)^{2}-q^{2}}-R_{q}'\right]\,.\label{gauge_D_2}
\end{equation}
The version of (\ref{gauge_D_1.2}) with general momenta is given
in appendix B. Equations (\ref{gauge_D_1.1}) and (\ref{gauge_D_1.2})
are the example of the kind of relations needed to obtain the explicit
momentum space representation of the flow equations for the zero-field
proper-vertices of the bEAA. With the technique of appendix B one
can find the explicit representation for all the other zero-field
proper-vertices $\tilde{\gamma}_{k}^{(n;m)}$.

The finite-difference derivatives (\ref{gauge_D_1}) and (\ref{gauge_D_2})
can be expanded for small external momenta as follows:
\begin{eqnarray}
R_{q+p,q}^{(1)} & = & R_{q}'+p\cdot q\, R_{q}''+\frac{1}{2}p^{2}\, R_{q}''+\frac{2}{3}(p\cdot q)^{2}\, R_{q}^{(3)}+O(p^{3})\nonumber \\
R_{q+p,q}^{(2)} & = & R_{q}''+\frac{2}{3}p\cdot q\, R_{q}^{(3)}+\frac{1}{3}p^{2}\, R_{q}^{(3)}+\frac{1}{3}(p\cdot q)^{2}\, R_{q}^{(4)}+O(p^{3})\,.\label{gauge_D_3}
\end{eqnarray}
As we will see in section 3, the ``correction terms'' in (\ref{gauge_D_3}),
proportional to $p$, $p^{2}$, or higher, are those needed to make
the flow of the zero-field proper-vertices $\bar{\gamma}_{k}^{(m)}=\gamma_{k}^{(0;m)}$
fully transverse, as they should be by construction.

We are now ready to write the flow equations for the zero-field proper-vertices
of the bEAA in their full momentum space form. We will present solely
the flow equations for the two-vertices, equations (\ref{gauge_28})
and (\ref{gauge_34}), since it is from these equations that in the
next section we will extract the beta functions of the gauge coupling.
All other equations can be derived using the rules stated in the previous
subsection and their generalizations.

We find the following momentum space representation for the first
equation in (\ref{gauge_28}), describing the flow of the zero-field
fluctuation-fluctuation two-vertex: 
\begin{eqnarray}
[\partial_{t}\gamma_{p,-p}^{(2;0)}]^{AB} & = & \Omega\int_{q}(\partial_{t}R_{q}-\eta R_{q})[G_{q}]^{12}[\gamma_{q,-q-p,p}^{(3;0)}]^{2A3}[G_{q+p}]^{34}[\gamma_{q+p,-q,-p}^{(3;0)}]^{4B5}[G_{q}]^{51}\nonumber \\
 &  & -\frac{1}{2}\Omega\int_{q}(\partial_{t}R_{q}-\eta R_{q})[G_{q}]^{12}[\gamma_{q,-q,p,-p}^{(4;0)}]^{2AB3}[G_{q}]^{31}\,.\label{gauge_D_4}
\end{eqnarray}
In (\ref{gauge_D_4}) and in the following relations $\eta$ is the
multiplet matrix of anomalous dimensions of the fluctuation fields
in $\varphi$. Note also that we are using the generalized notation
for the indices introduced before and we use just integers to denote
dummy indices. With respect to the first equation in Figure 2, the
first line in (\ref{gauge_D_4}) is the contribution from the first
diagram, while the second line is the contribution from the second
one. The second equation in (\ref{gauge_28}), describing the flow
of the fluctuation-background zero-field two-vertex, takes the following
form:
\begin{eqnarray}
[\partial_{t}\gamma_{p,-p}^{(1;1)}]^{AB} & = & \Omega\int_{q}(\partial_{t}R_{q}-\eta R_{q})[G_{q}]^{12}[\tilde{\gamma}_{q,-q-p,p}^{(2;1)}]^{2A3}[G_{q+p}]^{34}[\gamma_{q+p,-q,-p}^{(3;0)}]^{4B5}[G_{q}]^{51}\nonumber \\
 &  & -\frac{1}{2}\Omega\int_{q}(\partial_{t}R_{q}-\eta R_{q})[G_{q}]^{12}[\tilde{\gamma}_{q,-q,p,-p}^{(3;1)}]^{2AB3}[G_{q}]^{31}\nonumber \\
 &  & -\Omega\int_{q}[l_{q,-q-p,p}^{(2;1)}(\partial_{t}R_{q+p,q}^{(1)}-\eta R_{q+p,q}^{(1)})]^{4A1}\nonumber \\
 &  & \times[G_{q+p}]^{12}[\gamma_{q+p,-q,-p}^{(3;0)}]^{2B3}[G_{q}]^{34}\,.\label{gauge_D_5}
\end{eqnarray}
In (\ref{gauge_D_5}) there are now two tilde vertices since, referring
to the second equation in Figure 2, there is a background line attached
to, respectively, a three-vertex, a four-vertex and to the factor
$\partial_{t}R_{k}[\bar{A}]$. The contribution from these three diagrams
are respectively the first, second and third lines of (\ref{gauge_D_5}).
The last equation of (\ref{gauge_28}) takes the following form:
\begin{eqnarray}
[\partial_{t}\gamma_{p,-p}^{(0;2)}]^{AB} & = & \Omega\int_{q}(\partial_{t}R_{q}-\eta R_{q})[G_{q}]^{12}[\tilde{\gamma}_{q,-q-p,p}^{(2;1)}]^{2A3}[G_{q+p}]^{34}[\tilde{\gamma}_{q+p,-q,-p}^{(2;1)}]^{4B5}[G_{q}]^{51}\nonumber \\
 &  & -\frac{1}{2}\Omega\int_{q}(\partial_{t}R_{q}-\eta R_{q})[G_{q}]^{12}[\tilde{\gamma}_{q,-q,p,-p}^{(2;2)}]^{2AB3}[G_{q}]^{31}\nonumber \\
 &  & -\Omega\int_{q}[l_{q,p,-q-p}^{(2;1)}(\partial_{t}R_{q+p,q}^{(1)}-\eta R_{q+p,q}^{(1)})]^{1A2}[G_{q+p}]^{23}[\tilde{\gamma}_{q+p,-q,-p}^{(2;1)}]^{3B4}[G_{q}]^{41}\nonumber \\
 &  & +\frac{1}{2}\Omega\int_{q}\left\{ [l_{q,p,-p,-q}^{(2;2)}(\partial_{t}R_{q}'-\eta R_{q}')]^{1AB2}[G_{q}]^{21}\right.\nonumber \\
 &  & \left.+[l_{q,-q-p,p}^{(2;1)}]^{1A3}[l_{q+p,-q,-p}^{(2;1)}]^{3B2}(\partial_{t}R_{q+p,q}^{(2)}-\eta R_{q+p,q}^{(2)})[G_{q}]^{21}\right\} \,.\label{gauge_D_6}
\end{eqnarray}
Equation (\ref{gauge_D_6}) represents the flow of the zero-field
background-background two-vertex of the bEAA and thus every term is
written in terms of the tilde vertices and of the cutoff action $L[\varphi;\bar{A}]$.
As we explained earlier, these equations are very general and can
be adapted to every theory with local gauge symmetry.

In terms of the compact representation introduced in subsection 2.2.2
using the formal operator $\tilde{\partial}_{t}$ defined in (\ref{gauge_32.1}),
the flow equations for the zero-field two-vertices of the bEAA are
given in equation (\ref{gauge_34}). These are represented graphically
in Figure 4 and all three have the same overall structure. The flow
of the zero-field fluctuation-fluctuation two-vertex has the following
momentum space representation:
\begin{eqnarray}
[\partial_{t}\gamma_{p,-p}^{(2;0)}]^{AB} & = & -\frac{1}{2}\Omega\int_{q}\tilde{\partial}_{t}\left\{ [\gamma_{q,-q-p,p}^{(3;0)}]^{4A1}[G_{q+p}]^{12}[\gamma_{q+p,-q,-p}^{(3;0)}]^{2B3}[G_{q}]^{34}\right\} \nonumber \\
 &  & +\frac{1}{2}\Omega\int_{q}\tilde{\partial}_{t}\left\{ [\gamma_{q,-q,p,-p}^{(4;0)}]^{1AB2}[G_{q}]^{21}\right\} \,.\label{gauge_D_7}
\end{eqnarray}
The second equation in (\ref{gauge_34}), shown in Figure 4, expresses
the flow of the zero-field fluctuation-background two-vertex and differs
from (\ref{gauge_D_7}) in two tilde vertices:
\begin{eqnarray}
[\partial_{t}\gamma_{p,-p}^{(1;1)}]^{AB} & = & -\frac{1}{2}\Omega\int_{q}\tilde{\partial}_{t}\left\{ [\tilde{\gamma}_{q,-q-p,p}^{(2;1)}]^{4A1}[G_{q+p}]^{12}[\gamma_{q+p,-q,-p}^{(3;0)}]^{2B3}[G_{q}]^{34}\right\} \nonumber \\
 &  & +\frac{1}{2}\Omega\int_{q}\tilde{\partial}_{t}\left\{ [\tilde{\gamma}_{q,-q,p,-p}^{(3;1)}]^{1AB2}[G_{q}]^{21}\right\} \,.\label{gauge_D_8}
\end{eqnarray}
Finally, the compact form for the flow of the zero-field background-background
two-vertex is:
\begin{eqnarray}
[\partial_{t}\gamma_{p,-p}^{(0;2)}]^{AB} & = & -\frac{1}{2}\Omega\int_{q}\tilde{\partial}_{t}\left\{ [\tilde{\gamma}_{q,-q-p,p}^{(2;1)}]^{4A1}[G_{q+p}]^{12}[\tilde{\gamma}_{q+p,-q,-p}^{(2;1)}]^{2B3}[G_{q}]^{34}\right\} \nonumber \\
 &  & +\frac{1}{2}\Omega\int_{q}\tilde{\partial}_{t}\left\{ [\tilde{\gamma}_{q,-q,p,-p}^{(2;2)}]^{1AB2}[G_{q}]^{21}\right\} \,.\label{gauge_D_9}
\end{eqnarray}
Note that in equation (\ref{gauge_D_9}) all zero-field proper-vertices
are tilde vertices. As already said, equation (\ref{gauge_D_9}),
as equation (\ref{gauge_D_6}), represents also the flow of the zero-field
proper-vertex $\bar{\gamma}_{k}^{(2)}$ of the gEAA since we have
that $\partial_{t}\bar{\gamma}_{k}^{(2)}=\partial_{t}\gamma_{k}^{(0;2)}$.

Thus the flow equation for the zero-field proper-vertices of the bEAA
are formally as those of the standard EAA when written in terms of
the formal operator $\tilde{\partial}_{t}$ but with tilde vertices
in place of the standard vertices. All the non-trivial dependence
on the cutoff kernel is in this way hidden in the dependence of the
tilde vertices on it. This turns out to be a very useful property
in actual computations. One can thus naively draw the diagrams as
in the non-background formalism with the only caveat that the vertices
are tilde vertices and that these can be represented using the rules
derived in subsection 2.3.1. Remember also that $\tilde{\gamma}^{(n;0)}=\gamma^{(n;0)}$
and $\bar{\gamma}^{(m)}=\gamma^{(0;m)}.$ Clearly, these are only
the first equations of the respective hierarchies and the results
exposed in this section are valid for all the subsequent equations
of the hierarchy, for both the zero-field proper-vertices of the bEAA
and of the gEAA.

Equations (\ref{gauge_D_4}-\ref{gauge_D_9}), together with the momentum
space rules of subsection 2.3.1, are the main result of this paper.
A lot of information about the flow of both the bEAA and of the gEAA
can be extracted already from the flow of the zero-field two-vertices
described by these equations.

The results of this section, when combined with the flow equations
of the previous one, constitute the basis for a concrete framework
in which all truncations of the bEAA which are analytic in
the fields can be treated. In particular, one can consider a ``curvature
expansion'' where the gEAA is expanded in powers of the (generalized)
curvatures, and where the running is encoded in $k$--dependent structure
functions%
\footnote{See \cite{Codello_2010} for a first one--loop application in the
context of gravity.%
}. As we said, the methods presented in this section can be extended
to gravity, non-linear sigma models and membranes.

Finally, we remark that truncations like $\bar{\Gamma}_{k}[A]=\int W(F^{2})$,
where $W$ is an arbitrary function of the invariant $F^2\equiv F_{\mu\nu}F^{\mu\nu}$,
in non-abelian gauge theories \cite{ReuterWetterich_1994c}, or like
$\bar{\Gamma}_{k}[g]=\int\sqrt{g}f(R)$, in gravity \cite{Codello_Percacci_Rahmede_2008},
cannot be treated with the present method and one needs to resort
to other techniques to treated them.
To be more precise, one can treat only polynomial approximations to these functions since we ultimately set the background gauge field, or the background metric, to zero.

\newpage

\section{An application: non-abelian gauge theories}

In this section we apply the formalism developed in section 2 to non-abelian
gauge theories with gauge group $SU(N)$. We derive the beta function
for the gauge coupling and we show how standard results obtained by
heat kernel methods can be recovered in our new framework. Then we
calculate the anomalous dimensions of the fluctuation and ghost fields,
$\eta_{a,k}$ and $\eta_{c,k}$, and we use them to ``close'' the
beta function of $g_{k}$, obtaining new RG improved forms for this.
Finally, we study the phenomenology associated to these RG flows.

\subsection{Truncation ansatz}

As a simple application of the formalism, we consider a truncation
ansatz for the gEAA which is the RG improvement of the classical action:
\begin{equation}
\bar{\Gamma}_{k}[A]=I[A]\equiv\frac{1}{4}\int d^{d}x\, F_{\mu\nu}^{a}F^{a\mu\nu}\,.\label{gauge_A_1}
\end{equation}
The quantum gauge field can be expressed in terms of the background
and fluctuation fields as follows:
\begin{equation}
A_{\mu}=Z_{\bar{A},k}^{1/2}\bar{A}_{\mu}+Z_{a,k}^{1/2}a_{\mu}\,,\label{gauge_A_2}
\end{equation}
where we rescaled the gauge fields to account for their renormalization.
$Z_{a,k}$ is the running wave-function renormalization of the fluctuation
field while $Z_{\bar{A},k}$ is the running wave-function renormalization
of the background field. In the background field formalism the gauge
coupling is related to the wave-function renormalization of the background
gauge field \cite{Abbott_1981}: 
\begin{equation}
g_{k}=Z_{\bar{A},k}^{-1/2}\,.\label{gauge_A_4}
\end{equation}
It is thus sufficient to determine the scale dependence of $Z_{\bar{A},k}$
to find the running of $g_{k}$. One defines the anomalous dimension
of the background field:
\begin{equation}
\eta_{\bar{A},k}=-\partial_{t}\log Z_{\bar{A},k}=-g_{k}^{2}\partial_{t}Z_{\bar{A},k}\,;\label{gauge_A_2.1}
\end{equation}
then the beta function of the gauge coupling is:
\begin{equation}
\partial_{t}g_{k}=\frac{1}{2}\eta_{\bar{A},k}g_{k}\,.\label{gauge_A_2.2}
\end{equation}
Inserting (\ref{gauge_A_2}) into equation (\ref{gauge_V_2}) from
appendix B gives the following expansion for the gEAA (\ref{gauge_A_1}):
\begin{eqnarray}
\bar{\Gamma}_{k}[Z_{\bar{A},k}^{1/2}\bar{A}+Z_{a,k}^{1/2}a] & = & Z_{\bar{A},k}\bar{\Gamma}_{k}[\bar{A}]+Z_{\bar{A},k}^{1/2}Z_{a,k}^{1/2}\int d^{d}x\,\bar{F}^{a\mu\nu}(\bar{D}_{\mu}a_{\nu})^{a}+\nonumber \\
 &  & +\frac{1}{2}Z_{a,k}\int d^{d}x\, a_{\mu}^{a}\left[(-\bar{D}^{2})^{ab}g^{\mu\nu}+2f^{abc}F^{c\mu\nu}+\bar{D}^{ac\mu}\bar{D}^{cb\nu}\right]a_{\nu}^{b}+\nonumber \\
 &  & +g_{k}Z_{a,k}^{3/2}f^{abc}\int d^{d}x\,(\bar{D}_{\mu}a_{\nu})^{a}\, a_{\mu}^{b}a_{\nu}^{c}+\nonumber \\
 &  & +\frac{1}{4}g_{k}^{2}Z_{a,k}^{2}f^{abc}f^{ade}\int d^{d}x\, a^{b\mu}a^{c\nu}a_{\mu}^{d}a_{\nu}^{e}\,.\label{gauge_A_3}
\end{eqnarray}
We used (\ref{gauge_A_4}) to set $g_{k}Z_{\bar{A},k}^{1/2}=1$ in
the covariant derivatives and in the second term of the second line.
Quantities with a bar are constructed with the background gauge field
of (\ref{gauge_A_3}). We approximate the rEAA to be the sum of the
bare background gauge-fixing action (\ref{gauge_5}) and of the bare
background ghost action (\ref{gauge_6}) with rescaled fluctuation
and ghost fields\footnote{Again we used $g_{k}Z_{\bar{A},k}^{1/2}=1$.}:
\begin{equation}
\hat{\Gamma}_{k}[Z_{a,k}^{1/2}a,Z_{c,k}^{1/2}c,Z_{c,k}^{1/2}\bar{c};Z_{\bar{A},k}^{1/2}\bar{A}]=Z_{a,k}S_{gf}[a;\bar{A}]+Z_{c,k}S_{gh}[a,c,\bar{c};\bar{A}]\,.\label{gauge_A_3.1}
\end{equation}
With these definitions our truncation comprises three running couplings
$\{g_{k},Z_{a,k},Z_{c,k}\}$, or three anomalous dimensions $\{\eta_{\bar{A},k},\eta_{a,k},\eta_{c,k}\}$,
where the anomalous dimensions of the fluctuation and ghost
fields are defined as:
\begin{equation}
\eta_{a,k}=-\partial_{t}\log Z_{a,k}\qquad\qquad\eta_{c,k}=-\partial_{t}\log Z_{c,k}\,.\label{gauge_A_3.2}
\end{equation}
We can say that $g_{k}$ parametrizes the RG evolution of the gEAA,
while $Z_{a,k}$ and $Z_{c,k}$ account for the influence of the RG
evolution of the full bEAA. This truncation is an example of \emph{bi--field
}truncation in the nomenclature of \cite{Reuter_Bi_Field} since the
ansatz has a non-trivial $k$--dependence in both $\bar{\Gamma}_{k}[\bar{A}]$
and $\hat{\Gamma}_{k}[a,\bar{c},c;\bar{A}]$.
More general bi--field truncations will include, for example, terms like the fluctuation gluon mass. Even if these truncations can be treated with the present methods we will not treat these cases for the moment.

\subsection{Derivation of the beta functions}

In this subsection we calculate the anomalous dimensions of the background,
fluctuation and ghost fields. In subsection 3.2.1 we review the standard
heat kernel calculation of $\eta_{\bar{A},k}$, while in subsection
3.2.2 we re-derive it by the methods presented in section 2. In subsection
3.2.3 we use the methods of section 2 to calculate $\eta_{a,k}$ and
$\eta_{c,k}$.

\subsubsection{Heat kernel calculation of $\eta_{\bar{A},k}$}

We review how to calculate the beta function $\partial_{t}Z_{\bar{A},k}$
of the wave-function renormalization of the background field, using
the standard method based on the local heat kernel expansion \cite{Gies_2006,Reuter_Wetterich_1994a}.

If we take a scale derivative of the expansion (\ref{gauge_A_3})
and we evaluate at $a_{\mu}=0$, we find:
\begin{equation}
\partial_{t}\bar{\Gamma}_{k}[Z_{\bar{A},k}^{1/2}\bar{A}]=\partial_{t}Z_{\bar{A},k}\,\frac{1}{4}\int d^{d}x\,\bar{F}_{\mu\nu}^{a}\bar{F}^{a\mu\nu}\,.\label{gauge_ZA_7}
\end{equation}
From (\ref{gauge_ZA_7}) we see that to extract $\partial_{t}Z_{\bar{A},k}$
we need to consider the flow equation (\ref{gauge_15}) for the gEAA,
which we rewrite here for convenience 
\begin{eqnarray}
\partial_{t}\bar{\Gamma}_{k}[\bar{A}] & = & \frac{1}{2}\textrm{Tr}\left(\Gamma_{k}^{(2,0,0,0)}[0,0,0;\bar{A}]+\Delta S_{k}^{(2,0,0;0)}[0,0,0;\bar{A}]\right)^{-1}\partial_{t}\Delta S_{k}^{(2,0,0;0)}[0,0,0;\bar{A}]\nonumber \\
 &  & -\textrm{Tr}\left(\Gamma_{k}^{(0,1,1,0)}[0,0,0;\bar{A}]+\Delta S_{k}^{(0,1,1;0)}[0,0,0;\bar{A}]\right)^{-1}\partial_{t}\Delta S_{k}^{(0,1,1;0)}[0,0,0;\bar{A}]\,,\label{gauge_ZA_1}
\end{eqnarray}
and extract, from the functional traces on rhs, the coefficient of
the invariant $\frac{1}{4}\int\bar{F}^{2}$.

As a first step to evaluate the rhs of (\ref{gauge_ZA_1}), we make
the following choices for the cutoff kernels of the fluctuation and
ghost fields:
\begin{eqnarray}
\Delta S_{k}^{(2,0,0;0)}[0,0,0;\bar{A}] & = & Z_{a,k}R_{k}[\bar{A}]\nonumber \\
\Delta S_{k}^{(0,1,1;0)}[0,0,0;\bar{A}] & = & Z_{c,k}R_{k}^{gh}[\bar{A}]\,.\label{gauge_ZA_1.1}
\end{eqnarray}
Note that, consistently with the previous discussion, we introduced
in the respective cutoff kernels (\ref{gauge_ZA_1.1}) the wave-function
renormalization of the fluctuation fields $Z_{a,k}$ and $Z_{c,k}$.
We specify the differential operators that will be the arguments of
the cutoff kernels $R_{k}[\bar{A}]$ and $R_{k}^{gh}[\bar{A}]$ in
a moment.

Second, we calculate the Hessians present in the denominators of the
traces on the rhs of the flow equation (\ref{gauge_ZA_1}). Using
the general decomposition of the bEAA given in (\ref{gauge_6.2}),
these can be written as follows:
\begin{eqnarray}
\Gamma_{k}^{(2,0,0;0)}[a,\bar{c},c;\bar{A}] & = & \bar{\Gamma}_{k}^{(2)}[\bar{A}+a]+\hat{\Gamma}_{k}^{(2,0,0;0)}[a,\bar{c},c;\bar{A}]\nonumber \\
\Gamma_{k}^{(0,1,1;0)}[a,\bar{c},c;\bar{A}] & = & \hat{\Gamma}_{k}^{(0,1,1;0)}[a,\bar{c},c;\bar{A}]\,.\label{gauge_ZA_3}
\end{eqnarray}
From (\ref{gauge_A_3}) one can read-off the Hessian of the gEAA:
\begin{equation}
\bar{\Gamma}_{k}^{(2)}[\bar{A}+a]^{ab\,\mu\nu}=Z_{a,k}\left[(-D^{2})^{ab}g^{\mu\nu}+2f^{abc}F^{c\mu\nu}+D^{ac\mu}D^{cb\nu}\right]\,,\label{gauge_ZA_3.01}
\end{equation}
where $(-D^{2})^{ab}\equiv-D_{\mu}^{ac}D^{cb\mu}$. Note that we can
write the second term as $2f^{abc}F^{c\mu\nu}=-2(F^{\mu\nu})^{ab}$.
The Hessians of the rEAA (\ref{gauge_A_3.1}) are just the Hessians
of the bare background gauge-fixing and background ghost actions:
\begin{eqnarray}
\hat{\Gamma}_{k}^{(2,0,0;0)}[a,\bar{c},c;\bar{A}] & = & Z_{a,k}S_{gf}^{(2;0)}[a;\bar{A}]\nonumber \\
\hat{\Gamma}_{k}^{(0,1,1;0)}[a,\bar{c},c;\bar{A}] & = & Z_{c,k}S_{gh}^{(0,1,1;0)}[a,\bar{c},c;\bar{A}]\,,\label{gauge_ZA_3.1}
\end{eqnarray}
from (\ref{gauge_5}) and (\ref{gauge_6}) we can easily read off
the following forms:
\begin{eqnarray}
\hat{\Gamma}_{k}^{(2,0,0;0)}[a,\bar{c},c;\bar{A}]^{ab\,\mu\nu} & = & -\frac{1}{\alpha}Z_{a,k}\,\bar{D}^{ac\mu}\bar{D}^{cb\nu}\nonumber \\
\hat{\Gamma}_{k}^{(0,1,1;0)}[a,\bar{c},c;\bar{A}]^{ab} & = & -Z_{c,k}\,\bar{D}_{\mu}^{ac}D^{cb\mu}\,.\label{gauge_ZA_4}
\end{eqnarray}
In the flow equation (\ref{gauge_ZA_1}) we need the Hessian (\ref{gauge_ZA_3})
evaluated at $a_{\mu}=\bar{c}=c=0$; combining (\ref{gauge_ZA_3.01})
and (\ref{gauge_ZA_4}) we find:
\begin{eqnarray}
\Gamma_{k}^{(2,0,0;0)}[0,0,0;\bar{A}]^{ab\,\mu\nu} & = & Z_{a,k}\left[(-\bar{D}^{2})^{ab}g^{\mu\nu}+2f^{abc}\bar{F}^{c\mu\nu}+\left(1-\frac{1}{\alpha}\right)\bar{D}^{ac\mu}\bar{D}^{cb\nu}\right]\nonumber \\
\Gamma_{k}^{(0,1,1;0)}[0,0,0;\bar{A}]^{ab} & = & Z_{c,k}\delta^{ab}(-\bar{D}^{2})\,.\label{gauge_ZA_5}
\end{eqnarray}
From now on we set the gauge-fixing parameter to $\alpha=1$ in order
to make the first Hessian in (\ref{gauge_ZA_5}) proportional to the
Laplace-type differential operator
\begin{equation}
(D_{T}^{\mu\nu})^{ab}=(-D^{2})^{ab}g^{\mu\nu}-2(F^{\mu\nu})^{ab}\,.\label{gauge_ZA_5.1}
\end{equation}
We need now to choose the cutoff operator, the eigenvalues of which,
we compare to the RG scale $k$ to separate the fast field modes from
the slow field modes. Without introducing any running coupling in
the cutoff action, apart for the wave-function renormalization of
the fluctuation fields, there are two possible choices in the gauge
sector. Looking back at equation (\ref{gauge_ZA_5}), we see that
we can take as cutoff operator the covariant Laplacian $-\bar{D}^{2}$
or the full Laplace-type differential operator $\bar{D}_{T}$. We
call cutoff actions constructed in this way, respectively, type I
and type II. In the ghost sector there is no such freedom and we choose
as cutoff operator, in both cases, the covariant Laplacian $-\bar{D}^{2}$.

We start by considering the type II case. In view of (\ref{gauge_ZA_5})
we can rewrite the flow equation (\ref{gauge_ZA_1}) as follows:
\begin{equation}
\partial_{t}\bar{\Gamma}_{k}[\bar{A}]=\frac{1}{2}\textrm{Tr}_{1c}\frac{\partial_{t}R_{k}(\bar{D}_{T})-\eta_{a,k}R_{k}(\bar{D}_{T})}{\bar{D}_{T}+R_{k}(\bar{D}_{T})}-\textrm{Tr}_{0c}\frac{\partial_{t}R_{k}(-\bar{D}^{2})-\eta_{c,k}R_{k}(-\bar{D}^{2})}{-\bar{D}^{2}+R_{k}(-\bar{D}^{2})}\,,\label{gauge_ZA_6}
\end{equation}
where we emphasized that the traces are also over space-time as well
as color indices. We proceed by employing the local heat kernel expansion
(for the operators $\bar{D}_{T}$ and $-\bar{D}^{2}$) to expand the
traces on the rhs side of equation (\ref{gauge_ZA_6}) in terms of
gauge invariant operators. The functional trace of a Laplace-type
second order differential operator $\mathcal{O}=-D^{2}+U$ can be
expanded as: 
\begin{equation}
\textrm{Tr}\, h(\mathcal{O})=\frac{1}{(4\pi)^{d/2}}\sum_{n=0}^{\infty}B_{2n}(\Delta)Q_{\frac{d}{2}-n}[h]\,,\label{gauge_ZA_6.1}
\end{equation}
where the first heat kernel coefficients $B_{2n}(\mathcal{O})$ are:
\begin{eqnarray}
B_{0}(\mathcal{O}) & = & \int d^{d}x\,\textrm{tr}\,1\nonumber \\
B_{2}(\mathcal{O}) & = & \int d^{d}x\,\textrm{tr}\, U\nonumber \\
B_{4}(\mathcal{O}) & = & \int d^{d}x\,\textrm{tr}\left(\frac{1}{2}U^{2}+\frac{1}{6}D^{2}U+\frac{1}{12}[D_{\mu},D_{\nu}][D^{\mu},D^{\nu}]\right)\,,\label{gauge_ZA_6.2}
\end{eqnarray}
and the $Q$-functionals are defined as:
\begin{equation}
Q_{n}[h]=\left\{ \begin{array}{ccc}
\frac{1}{\Gamma(n)}\int_{0}^{\infty}dz\, z^{n-1}h(z) &  & n>0\\
(-1)^{n}h^{(n)}(0) &  & n\leq0
\end{array}\right..\label{gauge_ZA_6.3}
\end{equation}
For more details on the applications of the heat kernel expansion
to the calculation of functional traces see \cite{Codello_Percacci_Rahmede_2009}.
The invariant $\frac{1}{4}\int\bar{F}^{2}$ is contained in the heat
kernel coefficients $B_{4}(\bar{D}_{T})$ and $B_{4}(-\bar{D}^{2})$;
a use of equation (\ref{gauge_ZA_6.1}) gives the following:
\begin{eqnarray}
\left.\partial_{t}\bar{\Gamma}_{k}[\bar{A}]\right|_{\int\bar{F}^{2}} & = & \frac{1}{(4\pi)^{d/2}}\left\{ \frac{1}{2}B_{4}(\bar{D}_{T})Q_{\frac{d}{2}-2}\left[\left(\partial_{t}R_{k}-\eta_{a,k}R_{k}\right)G_{k}\right]\right.\nonumber \\
 &  & \left.-B_{4}(-\bar{D}^{2})Q_{\frac{d}{2}-2}\left[\left(\partial_{t}R_{k}-\eta_{c,k}R_{k}\right)G_{k}\right]\right\} \,,\label{gauge_ZA_7.1}
\end{eqnarray}
where we introduced the regularized propagator 
\begin{equation}
G_{k}(z)=\frac{1}{z+R_{k}(z)}\,.\label{gauge_ZA_7.2}
\end{equation}
For the differential operator $\bar{D}_{T}$ we find the following
heat kernel coefficient%
\footnote{Note that $\textrm{tr}\, U^{2}\equiv U^{ab\mu\nu}U_{\nu\mu}^{ba}$.%
}:
\begin{eqnarray}
B_{4}(\bar{D}_{T}) & = & \int d^{d}x\left\{ \frac{1}{2}\textrm{tr}\, U^{2}+\frac{1}{12}\textrm{tr}\,[D_{\mu},D_{\nu}][D^{\mu},D^{\nu}]\right\} \nonumber \\
 & = & \int d^{d}x\left\{ \frac{1}{2}\left(2f^{abc}\bar{F}^{c\mu\nu}\right)\left(2f^{bad}\bar{F}_{\nu\mu}^{d}\right)+\frac{1}{12}\left(-f^{abc}\bar{F}_{\mu\nu}^{c}\right)\left(-f^{bad}\bar{F}_{\mu\nu}^{d}\right)\right\} \nonumber \\
 & = & \frac{24-d}{12}N\int d^{d}x\,\bar{F}_{\mu\nu}^{a}\bar{F}^{a\mu\nu}\,,\label{gauge_ZA_8}
\end{eqnarray}
where we used $U^{ab\mu\nu}=2f^{abc}\bar{F}^{c\mu\nu}$, $[\bar{D}_{\mu},\bar{D}_{\nu}]^{ab}=-f^{abc}\bar{F}_{\mu\nu}^{c}$
and $f^{abc}f^{abd}=N\delta^{cd}$. The ghost operator $-\bar{D}^{2}$,
i.e. the Laplacian, has instead the following heat kernel coefficient:
\begin{equation}
B_{4}(-\bar{D}^{2})=\int d^{d}x\left\{ \frac{1}{12}\textrm{tr}\,[D_{\mu},D_{\nu}][D^{\mu},D^{\nu}]\right\} =-\frac{N}{12}\int d^{d}x\,\bar{F}_{\mu\nu}^{a}\bar{F}^{a\mu\nu}\,.\label{gauge_ZA_9}
\end{equation}
Inserting (\ref{gauge_ZA_8}) and (\ref{gauge_ZA_9}) in (\ref{gauge_ZA_7.1}),
comparing with (\ref{gauge_ZA_7}) and using (\ref{gauge_A_2.1})
finally gives:
\begin{equation}
\eta_{\bar{A},k}=-\frac{g_{k}^{2}N}{(4\pi)^{d/2}}\left\{ \frac{24-d}{6}Q_{\frac{d}{2}-2}\left[\left(\partial_{t}R_{k}-\eta_{a,k}R_{k}\right)G_{k}\right]+\frac{1}{3}Q_{\frac{d}{2}-2}\left[\left(\partial_{t}R_{k}-\eta_{c,k}R_{k}\right)G_{k}\right]\right\} \,.\label{gauge_ZA_10}
\end{equation}
Equation (\ref{gauge_ZA_10}) gives the type II anomalous dimension
of the background field, within the truncation specified by equations
(\ref{gauge_A_3}) and (\ref{gauge_A_3.1}), in the gauge $\alpha=1$,
in arbitrary dimension and for general cutoff shape function $R_{k}(z)$.

When instead we employ the type I cutoff, the gauge trace in the flow
equation (\ref{gauge_ZA_6}) becomes:
\begin{equation}
\frac{1}{2}\textrm{Tr}_{1c}\frac{\partial_{t}R_{k}(-\bar{D}^{2})-\eta_{a,k}R_{k}(-\bar{D}^{2})}{-\bar{D}^{2}-2\bar{F}+R_{k}(-\bar{D}^{2})}\,.\label{gauge_ZA_12}
\end{equation}
To collect all terms proportional to $\frac{1}{4}\int\bar{F}^{2}$
we expand the denominator in powers of the curvature:
\begin{eqnarray}
\left[-\bar{D}^{2}-2\bar{F}+R_{k}(-\bar{D}^{2})\right]_{\mu\nu}^{-1} & = & g_{\mu\nu}G_{k}(-\bar{D}^{2})+2\, G_{k}(-\bar{D}^{2})\bar{F}_{\mu\nu}\, G_{k}(-\bar{D}^{2})\nonumber \\
 &  & +4G_{k}(-\bar{D}^{2})\bar{F}_{\mu\alpha}\, G_{k}(-\bar{D}^{2})\bar{F}_{\;\nu}^{\alpha}\, G_{k}(-\bar{D}^{2})+O(\bar{F}^{3})\,.\label{gauge_ZA_13}
\end{eqnarray}
When we insert back the expansion (\ref{gauge_ZA_13}) in the trace
(\ref{gauge_ZA_12}), the third term we obtain is already proportional
to $\frac{1}{4}\int\bar{F}^{2}$:
\[
\left.\textrm{Tr}_{1c}\left\{ \left[\partial_{t}R_{k}(-\bar{D}^{2})-\eta_{a,k}R_{k}(-\bar{D}^{2})\right]G_{k}^{3}(-\bar{D}^{2})\bar{F}_{\mu\alpha}\bar{F}_{\;\nu}^{\alpha}\right\} \right|_{\int\bar{F}^{2}}=\qquad\qquad
\]
\begin{equation}
\qquad\qquad=\frac{1}{(4\pi)^{d/2}}Q_{\frac{d}{2}}\left[(\partial_{t}R_{k}-\eta_{a,k}R_{k})G_{k}^{3}\right]\left(N\int d^{d}x\,\bar{F}_{\mu\nu}^{a}\bar{F}^{a\mu\nu}\right)\,,\label{gauge_ZA_13.1}
\end{equation}
the second term we obtain is zero when traced over space-time indices,
while the first term we obtain generates a contribution proportional
to $\frac{1}{4}\int\bar{F}^{2}$ when expanded using the heat kernel:
\[
\left.\textrm{Tr}_{1c}\left[\left(\partial_{t}R_{k}(-\bar{D}^{2})-\eta_{a,k}R_{k}(-\bar{D}^{2})\right)G_{k}(-\bar{D}^{2})\right]\right|_{\int\bar{F}^{2}}=\qquad\qquad
\]
\begin{equation}
\qquad\qquad=\frac{1}{(4\pi)^{d/2}}Q_{\frac{d}{2}-2}\left[(\partial_{t}R_{k}-\eta_{a,k}R_{k})G_{k}\right]\left(-\frac{d}{12}N\int d^{d}x\,\bar{F}_{\mu\nu}^{a}\bar{F}^{a\mu\nu}\right)\,.\label{gauge_ZA_13.2}
\end{equation}
Note that with respect to (\ref{gauge_ZA_9}) the heat kernel coefficient
in (\ref{gauge_ZA_13.2}) contains an additional factor $d$, since
the trace is over spin one fields. Combining the contributions (\ref{gauge_ZA_13})
and (\ref{gauge_ZA_13.2}) gives the type I anomalous dimension of
the background field: 
\begin{eqnarray}
\eta_{\bar{A},k} & = & -\frac{g_{k}^{2}N}{(4\pi)^{d/2}}\left\{ -\frac{d}{6}Q_{\frac{d}{2}-2}\left[(\partial_{t}R_{k}-\eta_{a,k}R_{k})G_{k}\right]+8Q_{\frac{d}{2}}\left[(\partial_{t}R_{k}-\eta_{a,k}R_{k})G_{k}^{3}\right]\right.\nonumber \\
 &  & \left.+\frac{1}{3}Q_{\frac{d}{2}-2}\left[(\partial_{t}R_{k}-\eta_{c,k}R_{k})G_{k}\right]\right\} \,.\label{gauge_ZA_14}
\end{eqnarray}
As (\ref{gauge_ZA_10}), this relation is valid within the truncation
(\ref{gauge_A_3}) and (\ref{gauge_A_3.1}), in the gauge $\alpha=1$,
in arbitrary dimension and for general cutoff shape function $R_{k}(z)$.

Equation (\ref{gauge_ZA_10}) and (\ref{gauge_ZA_14}) are the main
results of this subsection and represent the RG running of the wave-function
renormalization of the background field in the two schemes, type I
and II. In subsection 3.3 we will specify the cutoff shape function
$R_{k}(z)$ in order to obtain explicit forms for the anomalous dimensions.

As a final comment, we remark that to calculate $\partial_{t}Z_{\bar{A},k}$
for general values of the gauge-fixing parameter $\alpha$ (possibly
scale dependent) using the heat kernel expansion, we need to be able
to evaluate functional traces containing the following non-minimal
operator:
\[
(D_{T}^{\mu\nu})^{ab}=(-D^{2})^{ab}g^{\mu\nu}+\left(1-\frac{1}{\alpha}\right)D^{ac\mu}D^{cb\nu}-2(F^{\mu\nu})^{ab}\,.
\]
A way to handle this is to use the re-summation technique proposed
in \cite{Benedetti_Groh_Machado_Saueressig_2010}; otherwise, one
can perform the calculation by decomposing the gauge field into its
spin components as was done in \cite{ReuterWetterich_1994c}.

\subsubsection{Calculation of $\eta_{\bar{A},k}$ from $\partial_{t}\bar{\gamma}_{k}^{(2)}$}

We show how to calculate the anomalous dimension of the background
field from the flow equation for the zero-field two-point function
$\bar{\gamma}_{k}^{(2)}$ of the gEAA employing the techniques introduced
in section 2.

In particular, we are going to work with the flow equation for $\bar{\gamma}_{k}^{(2)}$
in its compact form given in equation (\ref{gauge_D_9}). After the
field multiplet decomposition, this equation can be represented diagrammatically
as shown in Figure 5. In formulas, we can write the flow equation
as%
\footnote{In this subsection, as in the next we omit to write the scale index
$k$ on functionals and on couplings to simplify the notation.%
}:
\begin{eqnarray}
[\partial_{t}\bar{\gamma}_{p,-p}^{(2)}]^{mn\,\mu\nu} & = & -\frac{1}{2}\Omega\int_{q}\tilde{\partial}_{t}[a_{p,q}]^{mn\,\mu\nu}+\frac{1}{2}\Omega\int_{q}\tilde{\partial}_{t}[b_{p,q}]^{mn\,\mu\nu}+\Omega\int_{q}\tilde{\partial}_{t}[c_{p,q}]^{mn\,\mu\nu}\nonumber \\
 &  & -\Omega\int_{q}\tilde{\partial}_{t}[d_{p,q}]^{mn\,\mu\nu}\,.\label{gauge_ZA_22.1}
\end{eqnarray}
Here $\Omega=(2\pi)^{d}\delta(0)$ is a volume factor coming from
the momentum integrals. Tensor products in the integrands are:
\begin{figure}
\begin{centering}
\includegraphics[scale=0.9]{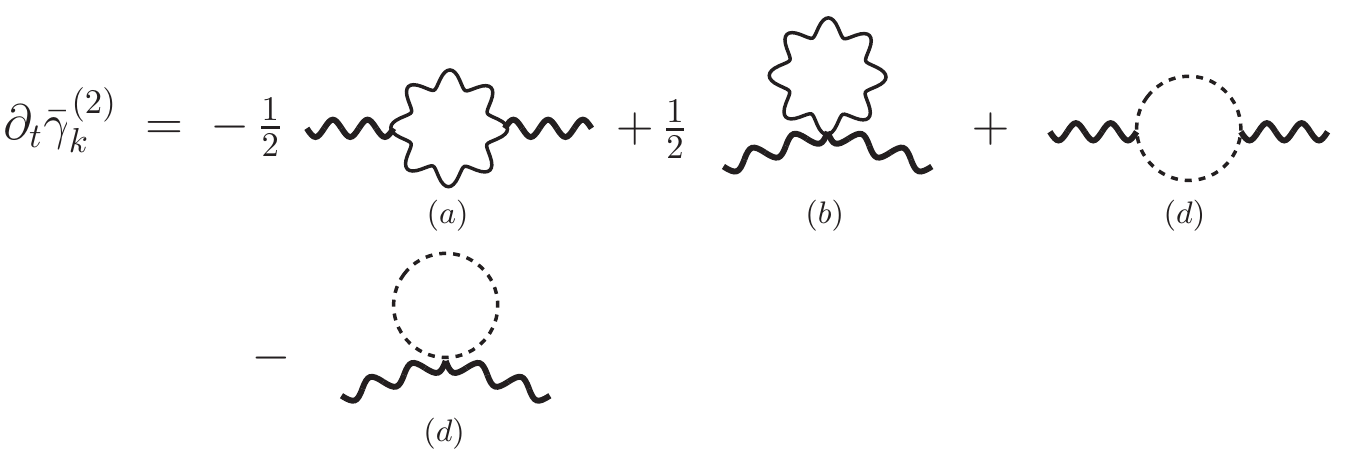}
\par\end{centering}

\caption{Diagrammatic representation of the compact form of the flow equation
for the zero-field two-point function $\bar{\gamma}_{k}^{(2)}$ of
the gEAA after the field multiplet decomposition within the truncation
specified by equations (\ref{gauge_A_3}) and (\ref{gauge_A_3.1}).
Thick wavy lines represent the background field $\bar{A}_{\mu}$,
light wavy lines represent the fluctuation field $a_{\mu}$, while
dotted lines represent the ghost fields $\bar{c}$ and $c$. }
\end{figure}
\begin{eqnarray}
[a_{p,q}]^{mn\,\mu\nu} & = & [\tilde{\gamma}_{q,-q-p,p}^{(2,0,0;1)}]^{abm\,\alpha\beta\mu}[G_{q+p}]^{bc\,\beta\gamma}[\tilde{\gamma}_{q+p,-q,-p}^{(2,0,0;1)}]^{cdn\,\gamma\delta\nu}[G_{q}]^{da\,\delta\alpha}\nonumber \\
{}[b_{p,q}]^{mn\,\mu\nu} & = & [\tilde{\gamma}_{q,-q,p,-p}^{(2,0,0;2)}]^{abmn\,\alpha\beta\mu\nu}[G_{q}]^{ba\,\beta\alpha}\nonumber \\
{}[c_{p,q}]^{mn\,\mu\nu} & = & [\tilde{\gamma}_{q,-q-p,p}^{(0,1,1;1)}]^{abm\,\mu}[G_{q+p}^{c}]^{bc}[\tilde{\gamma}_{q+p,-q,-p}^{(0,1,1;1)}]^{cdn\,\nu}[G_{q}^{c}]^{da}\nonumber \\
{}[d_{p,q}]^{mn\,\mu\nu} & = & [\tilde{\gamma}_{q,-q,p,-p}^{(0,1,1;2)}]^{abmn\,\mu\nu}[G_{q}^{c}]^{ba}\,.\label{gauge_ZA_22.2}
\end{eqnarray}
The ``tilde'' gauge vertices in (\ref{gauge_ZA_22.2}) are the following:
\begin{eqnarray}
[\tilde{\gamma}_{q,-q-p,p}^{(2,0,0;1)}]^{abm\,\alpha\beta\mu} & = & Z_{a}[I_{q,-q-p,p}^{(3)}]^{abm\,\alpha\beta\mu}+Z_{a}[S_{gf\, q,-q-p,p}^{(2;1)}]^{abm\,\alpha\beta\mu}\nonumber \\
 &  & +Z_{a}[l_{q,-q-p,p}^{(2;1)}]^{abm\,\alpha\beta\mu}R_{q+p,p}^{(1)}\,,\label{gauge_ZA_22.3}
\end{eqnarray}
(note that $\tilde{\gamma}_{q,-q-p,p}^{(2,0,0;1)}=\tilde{\gamma}_{q+p,-q,-p}^{(2,0,0;1)}$)
and:
\begin{eqnarray}
[\tilde{\gamma}_{q,-q,p,-p}^{(2,0,0;2)}]^{abmn\,\alpha\beta\mu\nu} & = & Z_{a}[I_{q,-q,p,-p}^{(4)}]^{abmn\,\alpha\beta\mu\nu}+Z_{a}[S_{gf\, q,-q,p,-p}^{(2;2)}]^{abmn\,\alpha\beta\mu\nu}\nonumber \\
 &  & +Z_{a}[l_{q,-q,p,-p}^{(2;2)}]^{abmn\,\alpha\beta\mu\nu}R'_{q}\nonumber \\
 &  & +Z_{a}[l_{q,-q-p,p}^{(2;1)}]^{abc\,\alpha\beta\gamma}[l_{q+p,-q,-p}^{(2;1)}]^{cmn\,\gamma\mu\nu}R_{q+p,q}^{(2)}\,.\label{gauge_ZA_22.4}
\end{eqnarray}
The ``tilde'' ghost vertices are instead:
\begin{equation}
[\tilde{\gamma}_{q,-q-p,p}^{(0,1,1;1)}]^{abm\,\mu}=Z_{c}[S_{gh\, q,-q-p,p}^{(0,1,1;1)}]^{abm\,\mu}+Z_{c}[l_{q,-q-p,p}^{(1,1;1)}]^{abm\,\mu}R_{q+p,q}^{(1)}\label{gauge_ZA_22.5}
\end{equation}
and
\begin{eqnarray}
[\tilde{\gamma}_{q,-q,p,-p}^{(0,1,1;2)}]^{abmn\,\mu\nu} & = & Z_{c}[S_{gh\, q,-q,p,-p}^{(0,1,1;2)}]^{abmn\,\mu\nu}+Z_{c}[l_{q,-q,p,-p}^{(1,1;2)}]^{abmn\,\mu\nu}R'_{q}\nonumber \\
 &  & +Z_{c}[l_{q,-q-p,p}^{(1,1;1)}]^{abc\,\mu}[l_{q+p,-q,-p}^{(1,1;1)}]^{cmn\,\nu}R_{q+p,q}^{(2)}\,.\label{gauge_ZA_22.6}
\end{eqnarray}
The momentum space representations of the vertices of the functionals
$I$, $S_{gf}$ and $S_{gh}$ appearing in these equations can be
found in appendix B. The vertices $l$ of the cutoff action will be
given after we make the cutoff action specifications.
We will soon see that these vertices,	 when reinserted in equation (\ref{gauge_ZA_22.1}), will correctly conspire to make the flow of $\partial_{t}\bar{\gamma}^{(2)}$ correctly transverse.

To proceed we first need to construct the regularized propagators
that enter the integrands in (\ref{gauge_ZA_22.2}). The fluctuation
field regularized propagator is:
\begin{equation}
[G_{p}^{-1}]^{\alpha\beta\, ab}=[\gamma_{p,-p}^{(2,0,0;0)}+R_{p}]^{\alpha\beta\, ab}\,.\label{gauge_P_0}
\end{equation}
Within our truncation, given by equations (\ref{gauge_A_3}) and (\ref{gauge_A_3.1}),
we have:
\begin{eqnarray}
[\gamma_{p,-p}^{(2,0,0;0)}]^{ab\,\alpha\beta} & = & [\bar{\gamma}_{p,-p}^{(2)}+\hat{\gamma}_{p,-p}^{(2,0,0;0)}]^{ab\,\alpha\beta}\nonumber \\
 & = & Z_{a}[I_{p,-p}^{(2)}]^{ab\,\alpha\beta}+Z_{a}[S_{gf\, p,-p}^{(2;0)}]^{ab\,\alpha\beta}\nonumber \\
 & = & \Omega\, Z_{a}\,\delta^{ab}\, p^{2}\left[(1-P)^{\alpha\beta}+\frac{1}{\alpha}P^{\alpha\beta}\right],\label{gauge_P_1}
\end{eqnarray}
where we used the two-vertices of $I$ and $S_{gf}$ from appendix
B and we introduced the longitudinal $P^{\alpha\beta}=p^{\alpha}p^{\beta}/p^{2}$
and transverse $(1-P)^{\alpha\beta}$ projectors. Inserting (\ref{gauge_P_1})
in (\ref{gauge_P_0}) gives the following form:
\begin{equation}
[G_{p}^{-1}]^{ab\,\alpha\beta}=\Omega\, Z_{a}\,\delta^{ab}\, p^{2}\,\left[(1-P)^{\alpha\beta}+\frac{1}{\alpha}P^{\alpha\beta}\right]+[R_{p}]^{ab\,\alpha\beta}\,.\label{gauge_P_2}
\end{equation}
We need now to specify the tensor structure of the cutoff kernel;
there are two basic choices:
\begin{eqnarray}
[R_{p}]^{ab\,\alpha\beta} & = & \Omega\, Z_{a}\,\delta^{ab}\left[(1-P)^{\alpha\beta}+\frac{1}{\alpha}P^{\alpha\beta}\right]R_{p}\label{gauge_P_3}\\
{}[R_{p}]^{ab\,\alpha\beta} & = & \Omega\, Z_{a}\,\delta^{ab}g^{\alpha\beta}R_{q}\,.\label{gauge_P_4}
\end{eqnarray}
Using the following relation that can be easily verified,
\[
\left[\mathbf{1}a+(\mathbf{1}-\mathbf{P})b+\mathbf{P}c\right]^{-1}=\frac{1}{a+b}(\mathbf{1}-\mathbf{P})+\frac{1}{a+c}\mathbf{P}\,,
\]
we can invert equation (\ref{gauge_P_2}) to obtain the explicit expression
for the regularized propagator. In the case that we are using the
cutoff kernel as defined in (\ref{gauge_P_3}), we find the form:
\begin{equation}
[G_{p}]^{ab\,\alpha\beta}=(\Omega\, Z_{a})^{-1}\left[\delta^{ab}\frac{1}{p^{2}+R_{p}}(1-P)^{\alpha\beta}+\delta^{ab}\frac{\alpha}{p^{2}+R_{p}}P^{\alpha\beta}\right]\,,\label{gauge_P_5}
\end{equation}
while in the case we are using the cutoff kernel as defined in (\ref{gauge_P_4}),
we get instead:
\begin{equation}
[G_{p}]^{ab\,\alpha\beta}=(\Omega\, Z_{a})^{-1}\left[\delta^{ab}\frac{1}{p^{2}+R_{p}}(1-P)^{\alpha\beta}+\delta^{ab}\frac{\alpha}{p^{2}+\alpha R_{p}}P^{\alpha\beta}\right]\,.\label{gauge_P_6}
\end{equation}
In this paper we will consider the choice corresponding to (\ref{gauge_P_4}),
since this is the minimal cutoff kernel choice we can make%
\footnote{In (\ref{gauge_P_3}), we are introducing in the cutoff kernel the
gauge-fixing parameter, if this is running it will give rise to additional
terms on the rhs of the flow equation, generated by $\partial_{t}R_{k}$
, proportional to $\partial_{t}\alpha_{k}$. A similar choice has
been made in \cite{Ellwanger_Hirsch_Weber_1996}.%
}. It's useful to define the transverse and longitudinal regularized
propagators:
\begin{equation}
G_{T,k}(z)=\frac{1}{z+R_{k}(z)}\qquad\qquad G_{L,k}(z)=\frac{\alpha}{z+\alpha R_{k}(z)}\,.\label{gauge_P_7}
\end{equation}
Note that the transverse regularized propagator is equal to the regularized
propagator defined in (\ref{gauge_ZA_7.2}), i.e. $G_{T,k}(z)=G_{k}(z)$.
We can now write (\ref{gauge_P_6}) as follows:
\begin{equation}
[G_{p}]^{ab\,\alpha\beta}=(\Omega\, Z_{a})^{-1}\left[\delta^{ab}(1-P)^{\alpha\beta}G_{p}^{T}+\delta^{ab}P^{\alpha\beta}G_{p}^{L}\right]\,.\label{gauge_P_7.3}
\end{equation}
Note that $G_{L,k}(z)=0$ if $\alpha=0$ and that $G_{L,k}(z)=G_{T,k}(z)=G_{k}(z)$
if $\alpha=1$; from now on we set $\alpha=1$ so that the tensor
structure of the regularized propagator is proportional to the identity.
The ghost regularized propagator, defined by
\[
[G_{p}^{-1}]^{ab}=[\gamma_{p,-p}^{(0,1,1;0)}+R_{p}]^{ab}\,,
\]
is easily obtained from
\begin{equation}
[\gamma_{p,-p}^{(0,1,1;0)}]^{ab}=Z_{c}[S_{gh\; p,-p}^{(0,1,1;0)}]^{ab}=\Omega\, Z_{c}\,\delta^{ab}p^{2}\,,\label{gauge_P_8}
\end{equation}
together with the minimal cutoff kernel choice
\begin{equation}
[R_{p}^{c}]^{ab}=\Omega\, Z_{c}\,\delta^{ab}R_{p}\,,\label{gauge_P_9}
\end{equation}
in the form:
\begin{equation}
[G_{p}^{c}]^{ab}=(\Omega\, Z_{c})^{-1}\delta^{ab}\frac{1}{p^{2}+R_{p}}=(\Omega\, Z_{c})^{-1}\delta^{ab}G_{p}\,.\label{gauge_P_10}
\end{equation}
This completes the construction of the regularized propagators needed
in the flow equation (\ref{gauge_ZA_22.1}).

We need now to specify the cutoff operator we employ to cutoff the
field modes. Remember from section 2 that the the cutoff operator
action, here $L[a;\bar{A}]$ in the gauge sector and $L[\bar{c},c;\bar{A}]$
in the ghost sector, is just the action whose Hessian is the cutoff
operator. As in the previous subsection, there are two basic options
in the gauge sector, which we called type I and type II, where the
cutoff operator is taken to be, respectively, the gauge Laplacian
$-\bar{D}^{2}$ or the operator $\bar{D}_{T}$.

We start to consider the type I case, where the gauge cutoff operator
action is the following:
\begin{equation}
L[a;\bar{A}]=\frac{1}{2}\int d^{d}x\,\bar{D}_{\mu}a_{\nu}\bar{D}^{\mu}a^{\nu}\,;\label{gauge_CA_1}
\end{equation}
one can easily check that the Hessian of this action is the gauge
Laplacian: 
\begin{equation}
L^{(2;0)}[0;\bar{A}]_{xy}^{ab\,\alpha\beta}=g^{\alpha\beta}\int d^{d}z\,\bar{D}_{z\mu}^{ac}\delta_{zx}\bar{D}_{z}^{cb\mu}\delta_{zy}=-g^{\alpha\beta}\bar{D}_{x\mu}^{ac}\bar{D}_{y}^{cb\mu}\delta_{xy}=(-\bar{D}_{x}^{2})^{ab}g^{\alpha\beta}\delta_{xy}\,.\label{gauge_CA_2}
\end{equation}
The ghost cutoff operator is the gauge Laplacian in both type I and
II cases; the ghost cutoff operator action is then simply%
\footnote{Note that $L[\bar{c},c;\bar{A}]=S_{gh}[0,\bar{c},c;\bar{A}]$.%
}:
\begin{equation}
L[\bar{c},c;\bar{A}]=\int d^{d}x\,\bar{D}_{\mu}\bar{c}\,\bar{D}^{\mu}c\,.\label{gauge_CA_2.1}
\end{equation}
Again we have $L^{(1,1;0)}[0,0;\bar{A}]_{xy}^{ab}=(-\bar{D}_{x}^{2})^{ab}\delta_{xy}$.
The vertices of the actions (\ref{gauge_CA_1}) and (\ref{gauge_CA_2.1})
can be found in appendix B.

Equation (\ref{gauge_ZA_22.1}) is a tensor equation from which we
can obtain two independent scalar equations after we contract it with
the projectors $(1-P)^{\mu\nu}$ and $P^{\mu\nu}$. We start by considering
the projection in the transverse direction; in particular, we find
the following results for the integrands (\ref{gauge_ZA_22.2}):
\begin{eqnarray}
(1-P)_{\mu\nu}[a_{p,q}]^{mn\,\mu\nu} & = & (\Omega\, Z_{a})^{-2}(1-P)_{\mu\nu}[\tilde{\gamma}_{q,-q-p,p}^{(2,0,0;1)}]^{abm\,\alpha\beta\mu}[\tilde{\gamma}_{q+p,-q,-p}^{(2,0,0;1)}]^{ban\,\beta\alpha\nu}G_{q}G_{q+p}\nonumber \\
 & = & N\delta^{mn}\left\{ 8(d-1)p^{2}-4dq^{2}(1-x^{2})\left(1+R_{q+p,q}^{(1)}\right)^{2}\right\} G_{q}G_{q+p}\nonumber \\
\nonumber \\
(1-P)_{\mu\nu}[b_{p,q}]^{mn\,\mu\nu} & = & (\Omega\, Z_{a})^{-1}(1-P)_{\mu\nu}[\tilde{\gamma}_{q,-q,p,-p}^{(2,0,0;2)}]^{aamn\,\alpha\alpha\mu\nu}G_{q}\nonumber \\
 & = & N\delta^{mn}\left\{ 2d(d-1)(1+R_{q}')+4dq^{2}(1-x^{2})R_{q+p,q}^{(2)}\right\} G_{q}\nonumber \\
\nonumber \\
(1-P)_{\mu\nu}[c_{p,q}]^{mn\,\mu\nu} & = & (\Omega\, Z_{c})^{-2}(1-P)_{\mu\nu}[\tilde{\gamma}_{q,-q-p,p}^{(0,1,1;1)}]^{abm\,\mu}[\tilde{\gamma}_{q+p,-q,-p}^{(0,1,1;1)}]^{ban\,\nu}G_{q}G_{q+p}\nonumber \\
 & = & N\delta^{mn}\left\{ 4q^{2}(1-x^{2})\left(1+R_{q+p,q}^{(1)}\right)^{2}\right\} G_{q}G_{q+p}\nonumber \\
\nonumber \\
(1-P)_{\mu\nu}[d_{p,q}]^{mn\,\mu\nu} & = & (\Omega\, Z_{c})^{-1}(1-P)_{\mu\nu}[\tilde{\gamma}_{q,-q,p,-p}^{(0,1,1;2)}]^{aamn\,\mu\nu}G_{q}\nonumber \\
 & = & N\delta^{mn}\left\{ 2(d-1)(1+R_{q}')+4q^{2}(1-x^{2})R_{q+p,q}^{(2)}\right\} G_{q}\,.\label{gauge_ZA_22.7}
\end{eqnarray}
In the first line of every contraction we used the fact that in the
gauge $\alpha=1$ the regularized propagators are proportional to
the identity, while in the second line of every contraction we performed
the tensor products between the ``tilde'' vertices. Here $x$ is
the cosine of the angle between the vectors $p$ and $q$. We evaluated
group factors as usual using $f^{abc}f^{abd}=N\delta^{cd}$. Note
also that all factors $\Omega\, Z_{a}$ and $\Omega\, Z_{c}$ simplified
in the final result.

The transverse component of the lhs of the flow equation (\ref{gauge_ZA_22.1})
is:
\begin{equation}
(1-P)_{\alpha\beta}[\partial_{t}\bar{\gamma}_{p,-p}^{(2)}]^{ab\,\alpha\beta}=\Omega\,\delta^{ab}(d-1)\partial_{t}Z_{\bar{A}}\, p^{2}\,;\label{gauge_ZA_23}
\end{equation}
inserting the contractions (\ref{gauge_ZA_22.7}) into the transverse
component of the flow equation (\ref{gauge_ZA_22.1}), gives the following
explicit relation:
\begin{eqnarray}
(d-1)\,\partial_{t}Z_{\bar{A}}\, p^{2} & = & -4(d-1)N\int_{q}\tilde{\partial}_{t}\left\{ G_{q}G_{q+p}\right\} \nonumber \\
 &  & -2Nd\int_{q}q^{2}(1-x^{2})\tilde{\partial}_{t}\left\{ G_{q}\left[G_{q+p}\left(1+R_{q+p,q}^{(1)}\right)^{2}-R_{q+p,q}^{(2)}\right]\right\} \nonumber \\
 &  & +Nd(d-1)\int_{q}\tilde{\partial}_{t}\left\{ G_{q}(1+R_{q}')\right\} \nonumber \\
 &  & +4N\int_{q}q^{2}(1-x^{2})\tilde{\partial}_{t}\left\{ G_{q}\left[G_{q+p}\left(1+R_{q+p,q}^{(1)}\right)^{2}-R_{q+p,q}^{(2)}\right]\right\} \nonumber \\
 &  & -2N(d-1)\int_{q}\tilde{\partial}_{t}\left\{ G_{q}(1+R_{q}')\right\} \,.\label{gauge_ZA_24}
\end{eqnarray}
In (\ref{gauge_ZA_24}) the first three lines can be traced back to
the gauge diagrams $(a)$ and $(b)$ of Figure 5, while the last two
lines come from the ghost diagrams $(c)$ and $(d)$. This equation
is the projected form of the flow equation for $\bar{\gamma}^{(2)}$
within the truncation of the bEAA we are considering and with the
functional trace explicitly evaluated. The rhs of (\ref{gauge_ZA_24})
contains contributions of arbitrary order in $p^{2}$: from the heat
kernel perspective this is equivalent to the re-summation of all contributions
proportional to the invariants of the form $\int\bar{F}_{\mu\nu}(-\bar{D}^{2})^{n}\bar{F}^{\mu\nu}$
present in the coefficients $B_{2n}$ for any $n\in\mathbb{N}$.

To extract the beta function $\partial_{t}Z_{\bar{A}}$ we need those
terms on the rhs of (\ref{gauge_ZA_24}) proportional to $p^{2}$.
The first integral in equation (\ref{gauge_ZA_24}) is needed only
in the limit $p\rightarrow0$, since its coefficient is already of
order $p^{2}$; it's easy to see that it can be rewritten as a $Q$-functional
in the following way:
\begin{eqnarray}
\int_{q}\tilde{\partial}_{t}\left\{ G_{q}G_{q+p}\right\}  & = & -2\int_{q}(\partial_{t}R_{q}-\eta R_{q})G_{q}^{2}G_{q+p}\nonumber \\
 & = & -\frac{2}{(4\pi)^{d/2}}Q_{\frac{d}{2}}\left[(\partial_{t}R_{k}-\eta R_{k})G_{k}^{3}\right]+O(p^{2})\,.\label{gauge_ZA_25}
\end{eqnarray}
The integrals in the second and fourth lines of (\ref{gauge_ZA_24})
have the same form; this is an important fact since it can be shown
that the following relation holds in arbitrary dimension:
\[
\int_{q}q^{2}(1-x^{2})\tilde{\partial}_{t}\left\{ G_{q}\left[G_{q+p}\left(1+R_{q+p,q}^{(1)}\right)^{2}-R_{q+p,q}^{(2)}\right]\right\} =\qquad\qquad\qquad\qquad\qquad
\]
\[
=\frac{d-1}{(4\pi)^{d/2}}\left\{ -\frac{1}{2}Q_{\frac{d}{2}-1}\left[(\partial_{t}R_{k}-\eta R_{k})G_{k}\right]+\frac{1}{12}Q_{\frac{d}{2}-2}\left[(\partial_{t}R_{k}-\eta R_{k})G_{k}\right]p^{2}\right.
\]
\begin{equation}
\qquad\qquad\qquad\left.-\frac{1}{120}Q_{\frac{d}{2}-3}\left[(\partial_{t}R_{k}-\eta R_{k})G_{k}\right]p^{4}\right\} +O(p^{6})\,.\label{gauge_ZA_27}
\end{equation}
Relation (\ref{gauge_ZA_27}) can be checked by inserting a sufficiently
smooth cutoff shape function and calculating both sides of it in a
Taylor expansion in $p^{2}$. Relations like (\ref{gauge_ZA_25})
and (\ref{gauge_ZA_27}) serve as a dictionary to transform the explicit
integrals on the rhs of the flow equation (\ref{gauge_ZA_24}) into
$Q$-functionals. With the aid of these relations we can extract from
the rhs of equation (\ref{gauge_ZA_24}) all the terms of order $p^{2}$
and a comparison with the lhs finally leads to the type I anomalous
dimension of the background field:
\begin{eqnarray*}
\eta_{\bar{A}} & = & -\frac{g^{2}N}{(4\pi)^{d/2}}\left\{ 8Q_{\frac{d}{2}}\left[(\partial_{t}R_{k}-\eta_{a}R_{k})G_{k}^{3}\right]-\frac{d}{6}Q_{\frac{d}{2}-2}\left[(\partial_{t}R_{k}-\eta_{a}R_{k})G_{k}\right]+\right.\\
 &  & \left.+\frac{1}{3}Q_{\frac{d}{2}-2}\left[(\partial_{t}R_{k}-\eta_{c}R_{k})G_{k}\right]\right\} \,,
\end{eqnarray*}
which is precisely equation (\ref{gauge_ZA_14}). With this result
we succeeded in showing that the $p^{2}$ terms of (\ref{gauge_ZA_24})
indeed correspond exactly to the coefficient of the $\frac{1}{4}\int\bar{F}^{2}$
term of the functional trace that we calculated previously using the
heat kernel expansion.

We consider now the type II cutoff. This means that in the above derivation
the gauge cutoff operator action has to be replaced by the following
action:
\begin{equation}
L[a;\bar{A}]=I[\bar{A}+a]+S_{gf}[a;\bar{A}]\,,\label{gauge_ZA_28}
\end{equation}
where the action $I[A]$ has been defined in (\ref{gauge_A_1}) and
the gauge fixing action in (\ref{gauge_5}). One can check that $L^{(2;0)}[0;\bar{A}]=\bar{D}_{T}$
as it should. Diagrams $(a)$ and $(b)$ now contract to:
\begin{eqnarray}
(1-P)_{\mu\nu}[a_{p,q}]^{mn\,\mu\nu} & = & N\delta^{mn}\left[8(d-1)p^{2}-4dq^{2}(1-x^{2})\right]\left(1+R_{q+p,q}^{(1)}\right)^{2}G_{q}G_{q+p}\nonumber \\
\nonumber \\
(1-P)_{\mu\nu}[b_{p,q}]^{mn\,\mu\nu} & = & N\delta^{mn}\left\{ 2d(d-1)(1+R_{q}')\right.\nonumber \\
 &  & \left.+\left[8(d-1)p^{2}-4dq^{2}(1-x^{2})\right]R_{q+p,q}^{(2)}\right\} G_{q}\,.\label{gauge_ZA_28.1}
\end{eqnarray}
In the ghost sector we don't make any changes. The first two terms
of equation (\ref{gauge_ZA_24}) are now replaced by:
\begin{equation}
-N\int_{q}\left[4(d-1)p^{2}+2dq^{2}(1-x^{2})\right]\tilde{\partial}_{t}\left\{ G_{q}\left[G_{q+p}\left(1+R_{q+p,q}^{(1)}\right)^{2}-R_{q+p,q}^{(2)}\right]\right\} \,.\label{gauge_ZA_29}
\end{equation}
By using the following relation, that can be checked as before,
\[
\int_{q}\tilde{\partial}_{t}\left\{ G_{q}\left[G_{q+p}\left(1+R_{q+p,q}^{(1)}\right)^{2}-R_{q+p,q}^{(2)}\right]\right\} =\qquad\qquad\qquad\qquad\qquad\qquad
\]
\begin{equation}
\qquad=\frac{1}{(4\pi)^{d/2}}\left\{ -Q_{\frac{d}{2}-2}\left[(\partial_{t}R_{k}-\eta R_{k})G_{k}\right]+\frac{1}{6}Q_{\frac{d}{2}-3}\left[(\partial_{t}R_{k}-\eta R_{k})G_{k}\right]\frac{p^{2}}{k^{2}}\right\} +O\left(\frac{p^{4}}{k^{4}}\right)\,,\label{gauge_ZA_29.1}
\end{equation}
to expand the rhs of (\ref{gauge_ZA_29}) one arrives at the expression
for the type II anomalous dimension of the background field:
\[
\eta_{\bar{A}}=-\frac{g^{2}N}{(4\pi)^{d/2}}\left\{ \frac{24-d}{6}Q_{\frac{d}{2}-2}\left[(\partial_{t}R_{k}-\eta_{a}R_{k})G_{k}\right]+\frac{1}{3}Q_{\frac{d}{2}-2}\left[(\partial_{t}R_{k}-\eta_{c}R_{k})G_{k}\right]\right\} \,.
\]
Again, as a further confirmation of our methods, we have re-derived
the result (\ref{gauge_ZA_10}) obtained before employing the heat
kernel expansion.

Along the same lines one can study the longitudinal equation obtained
from (\ref{gauge_ZA_22.1}) by contraction with $P^{\mu\nu}$ and
make an important check. In fact one finds, in both type I and II
cases, that the equation (in particular the rhs) turns out to be identically
zero; this is expected, since, by construction, the flow of the zero-field
proper-vertices $\bar{\gamma}^{(2)}$ of the gEAA is transverse. Another
check one can make is to control that the terms of order $p^{0}$
in equation (\ref{gauge_ZA_24}) cancel each other, since otherwise
they will generate a non-gauge invariant mass term for the background
field. Using (\ref{gauge_ZA_27}) and the following relation,
\begin{equation}
\int_{q}\tilde{\partial}_{t}\left\{ G_{q}(1+R_{q}')\right\} =-\frac{1}{(4\pi)^{d/2}}Q_{\frac{d}{2}-1}\left[(\partial_{t}R_{k}-\eta R_{k})G_{k}\right]\,,\label{gauge_ZA_30}
\end{equation}
to evaluate equation (\ref{gauge_ZA_24}) at $p=0$, gives:
\begin{eqnarray*}
 & -2Nd\left(-\frac{1}{2}\frac{d-1}{(4\pi)^{d/2}}Q_{\frac{d}{2}-1}\left[(\partial_{t}R_{k}-\eta_{a}R_{k})G_{k}\right]\right)+Nd(d-1)\left(-\frac{1}{(4\pi)^{d/2}}Q_{\frac{d}{2}-1}\left[(\partial_{t}R_{k}-\eta_{a}R_{k})G_{k}\right]\right)\\
 & +4N\left(-\frac{1}{2}\frac{d-1}{(4\pi)^{d/2}}Q_{\frac{d}{2}-1}\left[(\partial_{t}R_{k}-\eta_{c}R_{k})G_{k}\right]\right)-2N(d-1)\left(-\frac{1}{(4\pi)^{d/2}}Q_{\frac{d}{2}-1}\left[(\partial_{t}R_{k}-\eta_{c}R_{k})G_{k}\right]\right)\\
 & =\left[-2\left(-\frac{1}{2}\right)+(-1)\right]\frac{Nd(d-1)}{(4\pi)^{d/2}}Q_{\frac{d}{2}-1}\left[(\partial_{t}R_{k}-\eta_{a}R_{k})G_{k}\right]\\
 & +\left[4\left(-\frac{1}{2}\right)-(-2)\right]\frac{Nd(d-1)}{(4\pi)^{d/2}}Q_{\frac{d}{2}-1}\left[(\partial_{t}R_{k}-\eta_{c}R_{k})G_{k}\right]=0\,.
\end{eqnarray*}
As we see the $p^{0}$ contributions in both the gauge and ghost sectors
correctly cancel each other. The same can be checked in case of the
type II cutoff. We remark that these are nontrivial checks of the
formalism.

It this framework it is not difficult to relax the condition $\alpha=1$.
One does not need to change anything apart considering the regularized
propagator in its general form (\ref{gauge_P_7.3}) and using the
$\alpha$ dependent gauge-fixing vertices from appendix B. Work in
this direction will be material for a future studies \cite{Codello_in_preparation}. 

The aim of this subsection was to show how the method introduced in
section 2 can be used to reproduce the results obtained with the aid
of the local heat kernel expansion, and to show how the different
cutoff choices are reflected in this formalism. In this way we made
an important test of the method. We remark that this way of evaluating
the flow equation for the gEAA is very general and can be applied
to any general truncation ansatz, in particular to those that cannot
be treated with the aid of the local heat kernel expansion. We move
now to perform a new application of the technique by calculating $\eta_{a,k}$
and $\eta_{c,k}$ in the next subsection.

\subsubsection{Calculation of $\eta_{a,k}$ and $\eta_{c,k}$}

In this subsection we calculate the anomalous dimensions of the fluctuation
and ghost fields; these are related to the scale derivatives of wave-function
renormalization constants, $\partial_{t}Z_{a}$ and $\partial_{t}Z_{c}$,
by the relations:
\begin{equation}
\eta_{a}=-\partial_{t}\log Z_{a}\qquad\qquad\qquad\eta_{c}=-\partial_{t}\log Z_{c}\,.\label{gauge_Za_0}
\end{equation}
We will extract $\partial_{t}Z_{a}$ and $\partial_{t}Z_{c}$ form,
respectively, the flow equations for the zero-field two-point functions
$\gamma^{(2,0,0;0)}$ and $\gamma^{(0,1,1;0)}$ obtained from the
flow equation (\ref{gauge_D_4}) of section 2 after performing the
field multiplet decomposition.
\begin{figure}
\begin{centering}
\includegraphics[scale=0.9]{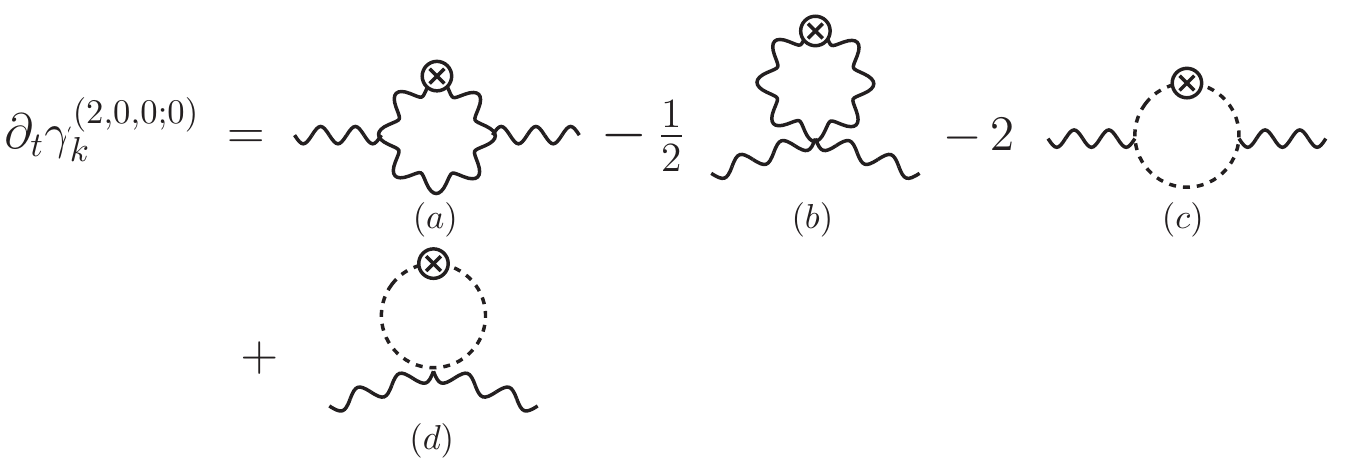}
\par\end{centering}

\caption{Diagrammatic representation of the flow equation for the zero-field
proper-vertex $\gamma_{k}^{(2,0,0;0)}$ of the bEAA used to calculate
$\partial_{t}Z_{a,k}$.}
\end{figure}

The flow equation for $\gamma^{(2,0,0;0)}$, within the truncation
represented by equations (\ref{gauge_A_3}) and (\ref{gauge_A_3.1}),
is given in Figure 6. In formulas this equation can be written as
follows:
\begin{eqnarray}
[\partial_{t}\gamma_{p,-p}^{(2,0,0;0)}]^{mn\,\mu\nu} & = & \Omega\int_{q}(\partial_{t}R_{q}-\eta_{a}R_{q})\,[a_{p,q}]^{mn\,\mu\nu}-\frac{1}{2}\Omega\int_{q}(\partial_{t}R_{q}-\eta_{a}R_{q})\,[b_{p,q}]^{mn\,\mu\nu}\nonumber \\
 &  & -2\Omega\int_{q}(\partial_{t}R_{q}-\eta_{c}R_{q})\,[c_{p,q}]^{mn\,\mu\nu}\,,\label{gauge_Za_2.10}
\end{eqnarray}
where the tensor products entering it are:
\begin{eqnarray}
[a_{p,q}]^{mn\,\mu\nu} & = & [G_{q}]^{ab\,\alpha\beta}[\gamma_{q,-q-p,p}^{(3,0,0;0)}]^{bcm\,\beta\gamma\mu}[G_{q+p}]^{cd\,\gamma\delta}[\gamma_{q+p,-q,-p}^{(3,0,0;0)}]^{den\,\delta\epsilon\nu}[G_{q}]^{ea\,\epsilon\alpha}\nonumber \\
{}[b_{p,q}]^{mn\,\mu\nu} & = & [G_{q}]^{ab\,\alpha\beta}[\gamma_{q,-q,p,-p}^{(4,0,0;0)}]^{bcmn\,\beta\gamma\mu\nu}[G_{q}]^{ca\,\gamma\alpha}\nonumber \\
{}[c_{p,q}]^{mn\,\mu\nu} & = & [G_{q}^{c}]^{ab}[\gamma_{p,q,-q-p}^{(1,1,1;0)}]^{mbc\,\mu}[G_{q}^{c}]^{cd}[\gamma_{p,q+p,-q}^{(1,1,1;0)}]^{nde\,\nu}[G_{q}^{c}]^{ea}\,,\label{gauge_Za_2.12}
\end{eqnarray}
and the vertices the following:
\begin{eqnarray}
[\gamma_{p_{1},p_{2},p_{3}}^{(3,0,0;0)}]^{abm\,\alpha\beta\mu} & = & gZ_{a}^{3/2}[I_{p_{1},p_{2},p_{3}}^{(3)}]^{abm\,\alpha\beta\mu}\nonumber \\
{}[\gamma_{p_{1},p_{2},p_{3},p_{4}}^{(4,0,0;0)}]^{abmn\,\alpha\beta\mu\nu} & = & g^{2}Z_{a}[I_{p_{1},p_{2},p_{3},p_{4}}^{(4)}]^{abmn\,\alpha\beta\mu\nu}\nonumber \\
{}[\gamma_{p_{1},p_{2},p_{3}}^{(1,1,1;0)}]^{mab\,\mu} & = & Z_{c}^{1/2}[S_{gh\, p_{1},p_{2},p_{3}}^{(1,1,1;0)}]^{mab\,\mu}\,.\label{gauge_Za_2.13}
\end{eqnarray}
Diagram $(d)$ in Figure 6 is identically zero, since in our truncation
there is no term bilinear both in the ghost and in the gauge fields,
and we omitted to write it in (\ref{gauge_Za_2.10}). The vertices
of the actions in (\ref{gauge_Za_2.13}) are given in appendix B.

To deal with scalar equations we project equation (\ref{gauge_Za_2.10})
using the orthogonal projectors $(1-P)^{\mu\nu}$ and $P^{\mu\nu}$.
In this way we obtain two independent equations describing, respectively,
the flow of the transverse and longitudinal components of $\gamma^{(2,0,0;0)}$.
We will use the transverse equation to extract $\partial_{t}Z_{a}$.
We mention here that the longitudinal equation can be used to extract
the running of $\alpha$, if one considers a truncation with running
gauge parameter \cite{Codello_in_preparation}. Applying the transverse
projector to the lhs of the flow equation (\ref{gauge_Za_2.10}) and
using (\ref{gauge_P_1}) gives:

\begin{equation}
(1-P)_{\mu\nu}[\partial_{t}\bar{\gamma}_{p,-p}^{(2)}+\partial_{t}\hat{\gamma}_{p,-p}^{(2,0,0;0)}]^{mn\,\mu\nu}=\Omega\,\delta^{mn}(d-1)\,\partial_{t}Z_{a}\, p^{2}\,,\label{gauge_Za_1}
\end{equation}
where we used the trace $(1-P)_{\alpha}^{\alpha}=d-1$. Equation (\ref{gauge_Za_1})
shows that we can extract $\partial_{t}Z_{a}$ as the coefficient
of the $p^{2}$ term of the transverse projection of equation (\ref{gauge_Za_2.10}).
Acting on the first integrand in (\ref{gauge_Za_2.12}) with the transverse
projector gives:
\begin{eqnarray}
(1-P)_{\mu\nu}[a_{p,q}]^{mn\,\mu\nu} & = & g^{2}\,\Omega\, Z_{a}(-N\delta^{mn})\left\{ -5(d-1)p^{2}-2(d-1)p\cdot q\right.\nonumber \\
 &  & \left.+2\left[(2d-3)x^{2}-3d+4\right]q^{2}\right\} (G_{q})^{2}G_{q+p}\,.\label{gauge_Za_2.2}
\end{eqnarray}
The group factor here is $f^{amb}f^{bna}=-N\delta^{mn}$, while, as
before, the variable $x$ is the cosine of the angle between $p$
and $q$. The transverse contribution from the second integrand in
(\ref{gauge_Za_2.12}) is:
\begin{equation}
(1-P)_{\mu\nu}[b_{p,q}]^{mn\,\mu\nu}=g^{2}\,\Omega\, Z_{a}(-2N\delta^{mn})\left\{ -(d-1)^{2}\right\} (G_{q})^{2}\,,\label{gauge_Za_2.3}
\end{equation}
here the group factor is $f^{abm}f^{anb}+f^{abn}f^{amb}=-2N\delta^{mn}$.
From (\ref{gauge_Za_2.12}) we find the transverse contribution from
the ghost diagram $(c)$: 
\begin{equation}
(1-P)_{\mu\nu}[c_{p,q}]^{mn\,\mu\nu}=g^{2}\,\Omega\, Z_{a}(-N\delta^{mn})\left\{ -(1-x^{2})q^{2}\right\} (G_{q})^{2}G_{q+p}\,,\label{gauge_Za_2.4}
\end{equation}
where the group factor is $f^{abm}f^{ban}=-N\delta^{mn}$.

Note that each integrand (\ref{gauge_Za_2.2}-\ref{gauge_Za_2.4})
(or diagram) is proportional to $g^{2}Z_{a}$ since the fluctuation
three-vertex comes with a factor $gZ_{a}^{3/2}$, the four-vertex
with a factor $g^{2}Z_{a}^{2}$, while the regularized gauge propagators
come with a power of $Z_{a}^{-1}$ and a gauge cutoff kernel insertion
with a power of $Z_{a}$. In the ghost diagrams the three-vertex gives
a factor $gZ_{a}^{1/2}Z_{c}$, there is no four-vertex, the regularized
ghost propagator has a factor of $Z_{c}^{-1}$ and the ghost cutoff
kernel insertion has a power of $Z_{c}$. Also, all the space-time
volume factors $\Omega$ on both sides of equation (\ref{gauge_Za_2.10})
delete each other.

Once we insert equations (\ref{gauge_Za_2.2}), (\ref{gauge_Za_2.3})
and (\ref{gauge_Za_2.4}) back in (\ref{gauge_Za_2.10}) we obtain
the explicit flow of the vertex $\gamma_{p,-p}^{(2,0,0;0)}$, to all
orders in the external momenta $p$. The momentum integrals in (\ref{gauge_Za_2.10})
can be performed using spherical coordinates with the $z$-axis along
the vector $p$:
\begin{equation}
\int_{q}\rightarrow\frac{S_{d-1}}{(2\pi)^{d}}\int_{0}^{\infty}dq\, q^{d-1}\int_{-1}^{1}dx\left(1-x^{2}\right)^{\frac{d-3}{2}}\,,\label{gauge_Za_2.5}
\end{equation}
where $S_{d}=\frac{2\pi^{\frac{d}{2}}}{\Gamma(\frac{d}{2})}$ is the
volume of the $d$-dimensional sphere. We also change variable $z=q^{2}$
in the radial integral so that:
\begin{equation}
\int_{0}^{\infty}dq\, q^{d-1}\rightarrow\frac{1}{2}\int_{0}^{\infty}dz\, z^{\frac{d}{2}-1}\,.\label{gauge_Za_2.51}
\end{equation}
After expanding the transverse projection of equation (\ref{gauge_Za_2.10})
in powers of $p$ the $x$-integrals can be easily performed; from
the terms proportional to $p^{2}$ we can extract $\partial_{t}Z_{a}$
and from these we obtain the anomalous dimension of the fluctuation
field:
\begin{eqnarray}
\eta_{a} & = & \frac{g^{2}N}{(4\pi)^{d/2}}\left\{ -5Q_{\frac{d}{2}}\left[\left(\partial_{t}R_{k}-\eta_{a}R_{k}\right)G_{k}^{3}\right]-(3d-1)Q_{\frac{d}{2}+1}\left[\left(\partial_{t}R_{k}-\eta_{a}R_{k}\right)G_{k}^{2}G_{k}'\right]\right.\nonumber \\
 &  & -\left(3d-1\right)Q_{\frac{d}{2}+2}\left[\left(\partial_{t}R_{k}-\eta_{a}R_{k}\right)G_{k}^{2}G_{k}''\right]+Q_{\frac{d}{2}+1}\left[\left(\partial_{t}R_{k}-\eta_{c}R_{k}\right)G_{k}^{2}G_{k}'\right]\nonumber \\
 &  & \left.+Q_{\frac{d}{2}+2}\left[\left(\partial_{t}R_{k}-\eta_{c}R_{k}\right)G_{k}^{2}G_{k}''\right]\right\} \,,\label{gauge_Za_5}
\end{eqnarray}
where we wrote everything in terms of $Q$-functionals. Equation (\ref{gauge_Za_5})
is valid in arbitrary dimension and for every cutoff shape function,
and gives implicitly the anomalous dimension of the fluctuation field,
in the gauge $\alpha=1$, as part of a linear system for the variables
$\eta_{a}$ and $\eta_{c}$. In the following we derive an analogous
equation for the anomalous dimension $\eta_{c}$ of the ghost fields,
that together with (\ref{gauge_Za_5}), can be used to solve for $\eta_{a}$
and $\eta_{c}$ in function of $g$.

We now calculate the scale derivative of the ghost wave-function renormalization
$\partial_{t}Z_{c}$. The only term in our truncation (\ref{gauge_A_3.1})
that contains the ghost fields and the related wave-function renormalization
is the following:
\begin{equation}
Z_{c}\int d^{d}x\,\bar{D}_{\mu}\bar{c}\left(\bar{D}^{\mu}+gZ_{a}^{1/2}a^{\mu}\right)c\,.\label{gauge_gh_1}
\end{equation}
We can extract $\partial_{t}Z_{c}$ from the flow equation for the
ghost-ghost zero-field proper-vertex $\gamma^{(0,1,1;0)}$. This equation
is obtained from the flow equation (\ref{gauge_D_4}) for the zero-field
proper-vertex $\gamma^{(2;0)}$ after the field multiplet decomposition;
it is represented graphically in Figure 7 and reads as follows:
\begin{equation}
[\partial_{t}\gamma_{p,-p}^{(0,1,1;0)}]^{mn}=\Omega\int_{q}(\partial_{t}R_{q}-\eta_{a}R_{q})\,[e_{p,q}]^{mn}+\Omega\int_{q}(\partial_{t}R_{q}-\eta_{c}R_{q})\,[f_{p,q}]^{mn}\,,\label{gauge_gh_1.1}
\end{equation}
where the integrands are:
\begin{figure}
\begin{centering}
\includegraphics[scale=0.9]{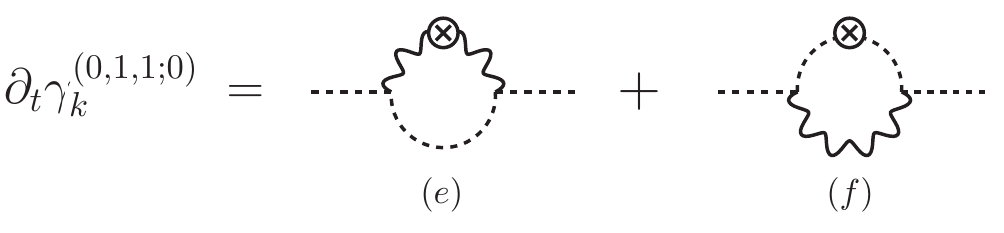}
\par\end{centering}

\caption{Graphical representation of the flow equation for the ghost-ghost
zero-field proper vertex $\gamma_{k}^{(0,1,1;0)}$ from (\ref{gauge_gh_1.1}).}
\end{figure}
\begin{eqnarray}
[e_{p,q}]^{mn} & = & [G_{q}]^{ab\,\alpha\beta}[\gamma_{q,p,-q-p}^{(1,1,1;0)}]^{bmc\,\beta}[G_{q+p}]^{cd}[\gamma_{-q,q+p,-p}^{(1,1,1;0)}]^{den\,\gamma}[G_{q}]^{ea\,\gamma\alpha}\nonumber \\
{}[f_{p,q}]^{mn} & = & [G_{q}]^{ab}[\gamma_{-q-p,,p,q}^{(1,1,1;0)}]^{bmc\,\gamma}[G_{q+p}]^{cd\,\gamma\delta}[\gamma_{q+p,-q,-p}^{(1,1,1;0)}]^{den\,\delta}[G_{q}]^{ea}\,.\label{gauge_gh_2.1}
\end{eqnarray}
The only vertex we need is thus the following:
\[
[\gamma_{p_{1},p_{2},p_{3}}^{(1,1,1;0)}]^{mab\,\mu}=gZ_{c}^{1/2}[S_{gh\, p_{1},p_{2},p_{3}}^{(1,1,1;0)}]^{mab\,\mu}\,,
\]
which can be found in appendix B. In equation (\ref{gauge_gh_1.1})
only the ghost-ghost-gluon vertex appears since the action (\ref{gauge_gh_1})
is at most trilinear in the fluctuation and ghost fields, thus the
second term in the flow equation (\ref{gauge_D_4}) vanishes. Inserting
(\ref{gauge_gh_1}) in the lhs of (\ref{gauge_gh_1.1}) gives: 
\begin{equation}
[\partial_{t}\gamma_{p,-p}^{(0,1,1;0)}]^{mn}=\Omega\,\delta^{mn}\,\partial_{t}Z_{c}\, p^{2}\,.\label{gauge_gh_2}
\end{equation}
As before we perform the tensor products in (\ref{gauge_gh_2.1})
and we evaluate the integrals in (\ref{gauge_gh_1.1}); we then extract
$\partial_{t}Z_{c}$ from the term of order $p^{2}$; after writing
everything in terms of $Q$-functionals, we obtain the anomalous dimension
of the ghost fields in the following form:
\begin{eqnarray}
\eta_{c} & = & \frac{g^{2}N}{(4\pi)^{d/2}}\left\{ -Q_{\frac{d}{2}}\left[(\partial_{t}R-\eta_{a}R)G_{k}^{3}\right]-Q_{\frac{d}{2}+1}\left[(\partial_{t}R-\eta_{a}R)G_{k}^{2}G_{k}'\right]\right.\nonumber \\
 &  & \left.+Q_{\frac{d}{2}+1}\left[(\partial_{t}R-\eta_{c}R)G_{k}^{2}G_{k}'\right]\right\} \,.\label{gauge_gh_7}
\end{eqnarray}
Equation (\ref{gauge_gh_7}) is the ghost anomalous dimensions in
the gauge $\alpha=1$ (within the truncation we are considering) and
is valid for general cutoff shape function and dimension. Equations
(\ref{gauge_Za_5}) and (\ref{gauge_gh_7}) are the main results of
this subsection.

\subsection{Beta functions and non-perturbative predictions}

In this subsection we study the different ways one has to ``close''
the beta function for the gauge coupling derived in the previous subsection
and we make a comparison with other approaches. Then we study some
possible phenomenology related to these RG flows.

\subsubsection{Beta functions}

To obtain an explicit form for the anomalous dimension of the background
field,
\begin{equation}
\eta_{\bar{A},k}=-\partial_{t}\log\, Z_{\bar{A},k}\,,\label{BF_0}
\end{equation}
given in equation (\ref{gauge_ZA_10}) for type II cutoff or in equation
(\ref{gauge_ZA_14}) for type I cutoff, we need to specify the cutoff
shape function $R_{k}(z)$. We will consider the theta-- or optimized--cutoff
shape function:
\begin{equation}
R_{k}(z)=Z_{k}(k^{2}-z)\theta(k^{2}-z)\,,\label{BF_0.01}
\end{equation}
(where $Z_{k}$ represents $Z_{a,k}$ or $Z_{c,k}$) since it allows
us to perform the integrals in the $Q$-functionals analytically for
every value of the dimension $d$. A simple integration gives the
following form:
\begin{equation}
Q_{n}[(\partial_{t}R_{k}-\eta R_{k})G_{k}^{m}]=\frac{k^{2(n+1-m)}}{\Gamma(n+2)}\left[2(n+1)-\eta\right]\label{BF_0.1}
\end{equation}
(where $\eta$ stands for $\eta_{a,k}$ or $\eta_{c,k}$). If we insert
(\ref{BF_0.1}) in the type I anomalous dimension (\ref{gauge_ZA_14}),
we find:
\begin{equation}
\eta_{\bar{A},k}=\frac{g_{k}^{2}N\, k^{d-4}}{(4\pi)^{d/2}\Gamma\left(\frac{d}{2}\right)}\left[-\frac{192-d(d-2)^{2}}{6d}+\frac{192-d^{2}(d+2)}{6d(d+2)}\eta_{a,k}+\frac{1}{3}\eta_{c,k}\right]\,;\label{BF_1}
\end{equation}
while if we insert (\ref{BF_0.1}) in the type II anomalous dimension
(\ref{gauge_ZA_10}) we get:
\begin{equation}
\eta_{\bar{A},k}=\frac{g_{k}^{2}N\, k^{d-4}}{(4\pi)^{d/2}\Gamma\left(\frac{d}{2}\right)}\left[-\frac{(26-d)(d-2)}{6}+\frac{24-d}{6}\eta_{a,k}+\frac{1}{3}\eta_{c,k}\right]\,.\label{BF_2}
\end{equation}
Note that only the type II anomalous dimension has the one--loop term
proportional to $26-d$, while type I becomes positive already at
$d\approx7.174$; for $d\rightarrow\infty$ both coefficients go as
$d^{2}/12$; at $d=2$ the type II coefficient is zero while the type
I is negative. See Figure 8 for a comparison.

Equation (\ref{BF_1}) and (\ref{BF_2}) show that the anomalous dimension
of the background field $\eta_{\bar{A},k}$ is determined by the anomalous
dimensions of the fluctuation and ghost fields
\begin{equation}
\eta_{a,k}=-\partial_{t}\log\, Z_{a,k}\qquad\qquad\eta_{c,k}=-\partial_{t}\log\, Z_{c,k}\,.\label{BF_3}
\end{equation}
This reflects the fact that the flow of the gEAA (here represented
by $\eta_{\bar{A},k}$) is not closed but depends on the flow of the
full bEAA (here represented by $\eta_{a,k}$ and $\eta_{c,k}$). The
anomalous dimensions of the fluctuation and of the ghost fields are
given, respectively, in equations (\ref{gauge_Za_5}) and (\ref{gauge_gh_7}).
The $Q$-integrals that we need now are:
\begin{figure}
\begin{centering}
\includegraphics[scale=0.8]{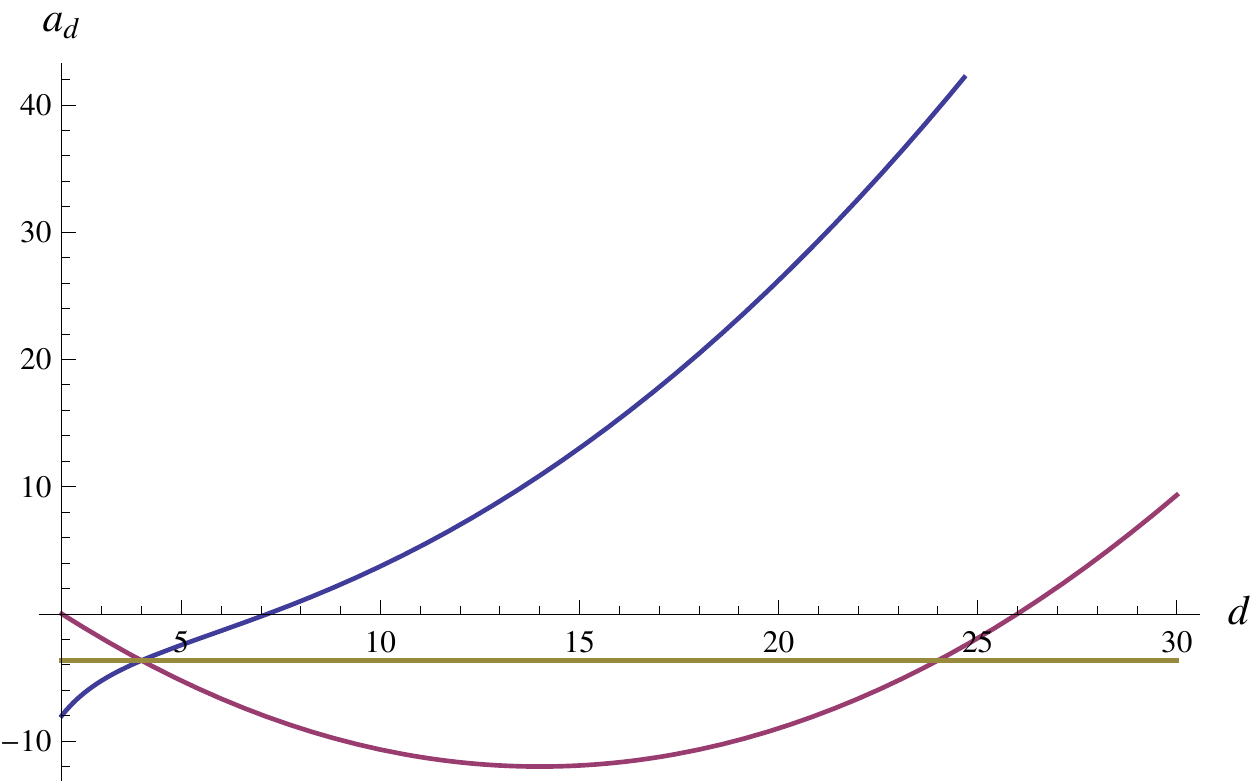}
\par\end{centering}

\caption{The one-loop coefficient $a$ in the cutoff schemes type I (lower
curve at $d=2$) and type II (upper curve at $d=2$). For reference
we plotted the line $-\frac{11}{3}$.}
\end{figure}
\begin{eqnarray}
Q_{n}[(\partial_{t}R_{k}-\eta R_{k})G_{k}^{m}G'_{k}] & = & 0\nonumber \\
Q_{n}[(\partial_{t}R_{k}-\eta R_{k})G_{k}^{m}G_{k}''] & = & -\frac{k^{2(n+1-m)}}{\Gamma(n)}\,;\label{BF_3.1}
\end{eqnarray}
with these we find the following system:
\begin{eqnarray}
\eta_{a,k} & = & \frac{g_{k}^{2}N\, k^{d-4}}{(4\pi)^{d/2}\Gamma\left(\frac{d}{2}\right)}\left[-\frac{8(d+6)}{d(d+2)}+\frac{20}{d(d+2)}\eta_{a,k}\right]\label{BF_4}\\
\eta_{c,k} & = & \frac{g_{k}^{2}N\, k^{d-4}}{(4\pi)^{d/2}\Gamma\left(\frac{d}{2}\right)}\left[-\frac{4}{d}+\frac{4}{d(d+2)}\eta_{a,k}\right]\,.\label{BF_5}
\end{eqnarray}
We want to remember to the reader that we are working in the gauge
$\alpha=1$ and that (\ref{BF_4}) and (\ref{BF_5}) generally depend
on the gauge-fixing parameter $\alpha$. Contrary to $\eta_{\bar{A},k}$,
the anomalous dimensions $\eta_{a,k}$ and $\eta_{c,k}$ are the same
for both cutoff types. Note also that the one--loop terms in (\ref{BF_4})
and (\ref{BF_5}) are always negative.

In the background field formalism the gauge coupling is related to
the wave-function renormalization of the background field by (\ref{gauge_A_4}),
i.e. $g_{k}=Z_{\bar{A},k}^{-1/2}$. Thus the beta function for the
gauge coupling can be obtained from the anomalous dimensions of the
background field by the following simple relation:
\begin{equation}
\partial_{t}g_{k}=\frac{1}{2}\eta_{\bar{A},k}g_{k}\,.\label{BF_6}
\end{equation}
One can now shift to dimensionless coupling $\tilde{g}_{k}=k^{\frac{d-4}{2}}g_{k}$
to find:
\begin{equation}
\partial_{t}\tilde{g}_{k}=\frac{1}{2}(d-4+\eta_{\bar{A},k})\tilde{g}_{k}\,.\label{BF_6.1}
\end{equation}
Even if the previous equations are valid in any dimension, in the
following we will consider the physical interesting case $d=4$ where
$g_{k}=\tilde{g}_{k}$.

Owning to the structure of both (\ref{BF_1}) and (\ref{BF_2}) we
can write the following general form for the background field anomalous
dimension in a given scheme:
\begin{equation}
\eta_{\bar{A},k}=\left(2a+b\,\eta_{a,k}+b'\,\eta_{c,k}\right)g_{k}^{2}\,,\label{BF_7}
\end{equation}
together with the following values for the coefficients in the two
cutoff schemes:
\begin{equation}
a_{I}=-\frac{11}{3}\frac{N}{(4\pi)^{2}}\qquad\qquad b_{I}=\frac{2}{3}\frac{N}{(4\pi)^{2}}\qquad\qquad b_{I}'=\frac{1}{3}\frac{N}{(4\pi)^{2}}\,,\label{BF_8}
\end{equation}

\begin{equation}
a_{II}=-\frac{11}{3}\frac{N}{(4\pi)^{2}}\qquad\qquad b_{II}=\frac{10}{3}\frac{N}{(4\pi)^{2}}\qquad\qquad b_{II}'=\frac{1}{3}\frac{N}{(4\pi)^{2}}\,.\label{BF_9}
\end{equation}
As expected we recover the scheme independent one--loop result $a_{I}=a_{II}=-\frac{11}{3}\frac{N}{(4\pi)^{2}}$.
The coefficients $b$ depend instead on the cutoff type; however,
these coefficients are gauge dependent and may have closer values
in another gauge such as the Landau $\alpha=0$. Note also that $b'_{I}=b'_{II}$
since the ghost contributions are the same in both schemes.

Since the anomalous dimensions of the fluctuation and ghost fields
(\ref{BF_4}) and (\ref{BF_5}), to lowest order, are proportional
to $g_{k}^{2}$, the only term on the lhs of (\ref{BF_7}) of order
$g_{k}^{2}$ is the first. To this order we recover the standard one--loop
beta function of the gauge coupling:
\begin{equation}
\partial_{t}g_{k}=a\, g_{k}^{3}+O(g_{k}^{5})=-\frac{11}{3}\frac{N}{(4\pi)^{2}}g_{k}^{3}+O(g_{k}^{5})\,.\label{BF_10}
\end{equation}
Since the beta function is negative at small coupling, non-abelian
gauge theories are asymptotically free in $d=4$ \cite{Gross_Wilczek_1973}.
From this we see that the one--loop approximation is equivalent to
set $\eta_{a,k}=\eta_{c,k}=0$ in (\ref{BF_7}). For a discussion
of the physical mechanism behind (\ref{BF_10}) asymptotic freedom
we refer to \cite{Reuter_Nink_2012}.

In the general case, to determine $\eta_{\bar{A},k}$ from equation
(\ref{BF_7}) we need to calculate or specify the anomalous dimensions
$\eta_{a,k}$ and $\eta_{c,k}$ as functions of $g_{k}$. As we said,
this is the actual manifestation of the fact that the flow of the
gEAA ($g_{k}$) is not closed but is given in terms of the flow of
the bEAA ($\eta_{a,k}$ and $\eta_{c,k}$). We discuss now two different
approximations that allow us to obtain a closed form for $\eta_{\bar{A},k}$
and thus an improved form for the beta function of the gauge coupling.

The first approximation, or improvement, consists in identifying the
anomalous dimension of the background field with the anomalous dimension
of the fluctuation field, and by setting the anomalous dimension of
the ghost fields to zero:
\begin{equation}
\eta_{\bar{A},k}=\eta_{a,k}\qquad\qquad\eta_{c,k}=0\,.\label{BF_11}
\end{equation}
Previous applications of the bEAA to non-abelian gauge theories \cite{Reuter_Wetterich_1994a,ReuterWetterich_1994c}
used this identification.
As we will see in a moment, this identification becomes exact in the supersymmetric limit.
If we insert now (\ref{BF_11}) in (\ref{BF_7})
we find the following linear system for the variable $\eta_{\bar{A},k}$:
\begin{equation}
\eta_{\bar{A},k}=\left(2a+b\,\eta_{\bar{A},k}\right)g_{k}^{2}\,,\label{BF_12}
\end{equation}
 which is easily solved to yield: 
\begin{equation}
\eta_{\bar{A},k}=\frac{2a}{1-b\, g_{k}^{2}}g_{k}^{2}\,,\label{BF_13}
\end{equation}
or, using (\ref{BF_6}), the following beta function for the gauge
coupling:
\begin{equation}
\partial_{t}g_{k}=\frac{a}{1-b\, g_{k}^{2}}g_{k}^{3}\,.\label{BF_14}
\end{equation}
The beta function (\ref{BF_14}) is a rational function of the gauge
coupling; this shows how the identification (\ref{BF_11}) implements
the re--summation of an infinite number of perturbative contributions.
One clearly sees the presence of a singularity in the beta function
(\ref{BF_14}) at $g_{k}^{2}=1/b$. We will see in the next subsection
a possible physical consequence of this fact. The beta function (\ref{BF_14})
in the two schemes is shown in Figure 9. In the spirit of \cite{Reuter_Bi_Field}
this is a single--field truncation.
\begin{figure}
\begin{centering}
\includegraphics[scale=0.9]{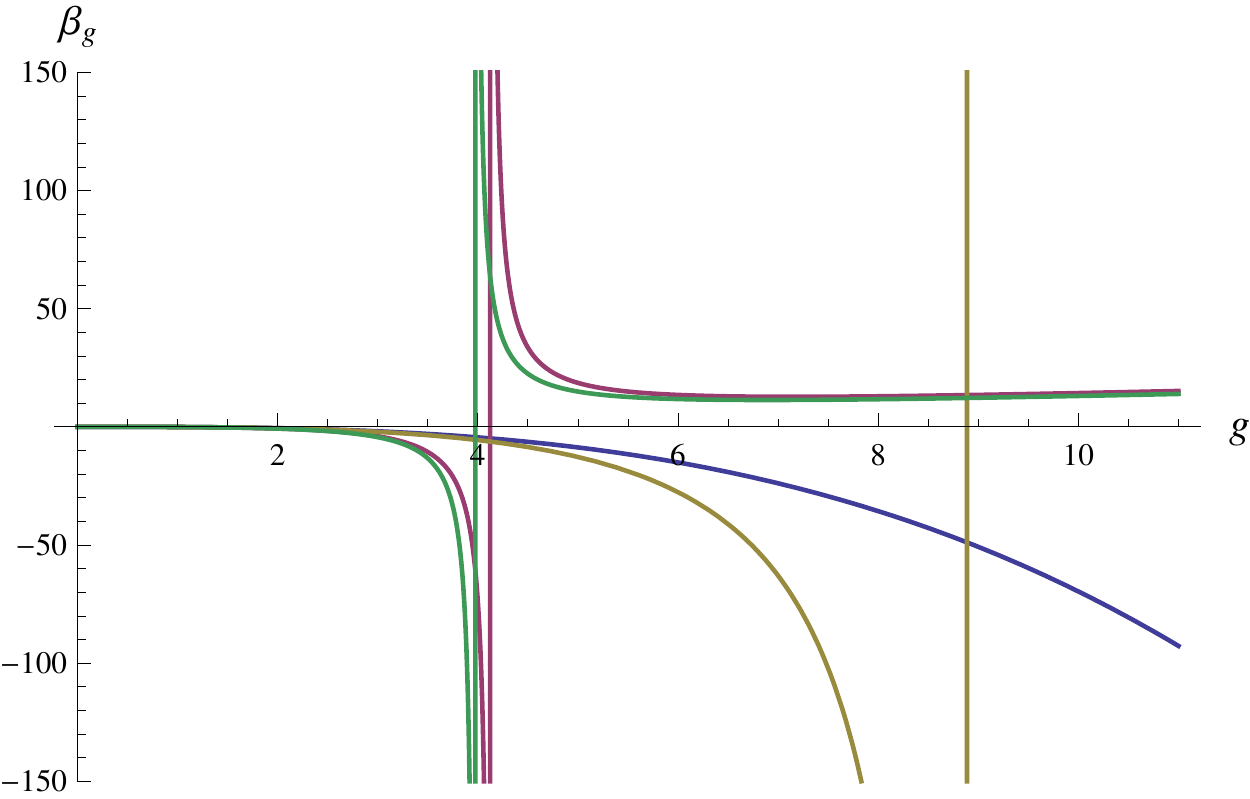}
\par\end{centering}

\caption{Single--field beta functions from (\ref{BF_14}). Asymptotes from
the left: type II, RS and type I beta functions. The one-loop beta
function (without asymptote) is shown for comparison. Note the similarity
between the type II and the RS beta functions.}
\end{figure}

Ryttov and Sannino (RS) \cite{Ryttov_Sannino_2008} proposed an ``all
orders'' beta function for the gauge coupling that has exactly the
structure found in (\ref{BF_14}), but with the following coefficients:
\begin{equation}
a_{RS}=-\frac{11}{3}\frac{N}{(4\pi)^{2}}\qquad\qquad b_{RS}=\frac{34}{11}\frac{N}{(4\pi)^{2}}\,,\label{BF_14.01}
\end{equation}
chosen in order to match the one-- and two--loop results for the beta
function upon Taylor expansion in $g_{k}$. The inspiration behind
the ansatz of RS comes from the knowledge of the exact beta function
found by Novikov-Shifman-Vainshtein-Zakharov (NSVZ) \cite{Novikov_Shifman_Vainshtein_Zakharov_1983}
in the supersymmetric $\mathcal{N}=1$ version of the model. The NSVZ
beta function is also of the form (\ref{BF_14}) but with coefficients
$a_{NSVZ}=-3\frac{N}{(4\pi)^{2}}$ and $b_{NSVZ}=2\frac{N}{(4\pi)^{2}}$.
RS postulated the existence of a perturbative massless scheme where
all the perturbative orders are reproduced by the choice (\ref{BF_14.01});
here we note the interesting fact that the RS beta function is very
similar to our type II beta function, as clearly shown in Figure 9.
We see that the bEAA formalism offers a framework where beta functions
of the form (\ref{BF_14}) can be obtained from first principles.
The linear relation for the anomalous dimension of the background
field postulated in \cite{Pica_Sannino_2011} to derive the RS beta
function, is here a consequence of the structure of the exact RG flow
equation for the gEAA. As a future work, it will be interesting to
reproduce the exact results for supersymmetric beta functions of \cite{Novikov_Shifman_Vainshtein_Zakharov_1983}
using functional RG methods.

When the coupling is small $g_{k}\ll1$ we can expand the beta function
(\ref{BF_14}) to find:
\begin{equation}
\partial_{t}g_{k}=a\, g_{k}^{3}+ab\, g_{k}^{5}+O\left(g_{k}^{7}\right)\,.\label{BF_14.1}
\end{equation}
As we already noticed, the one-loop contributions are scheme independent
$a_{I}=a_{II}=a_{RS}$ and equal to the perturbative result, while
the two--loop contributions are instead different. The RS coefficient
is, by construction, equal to the perturbative result $a_{RS}b_{RS}=-\frac{g_{k}^{5}}{(4\pi)^{4}}\frac{102}{9}N^{2}$
\cite{Abbott_1981}, while we find $a_{I}b_{I}=-\frac{g_{k}^{5}}{(4\pi)^{4}}\frac{22}{9}N^{2}$
and $a_{II}b_{II}=-\frac{g_{k}^{5}}{(4\pi)^{4}}\frac{110}{9}N^{2}$.
Note that the type I coefficient is $79\%$ smaller than the two--loop
coefficient, while type II coefficient is just $8\%$ bigger. These
two--loop coefficients are expected not to be equal to the perturbative
result, since this is scheme independent only in massless schemes,
while the bEAA formalisms implements a kind of mass dependent regularization
\cite{Gies_2006}.

The second way we can obtain a closed beta function for the gauge
coupling, is to first calculate the anomalous dimensions $\eta_{a,k}$
and $\eta_{c,k}$ and then reinsert them back in (\ref{BF_7}). This
means that we are considering the flow of the full bEAA which takes
place in the enlarged theory space of all functionals of the fluctuation
fields $a_{\mu}$, $\bar{c}$, $c$, and of the background field $\bar{A}_{\mu}$.
In the truncation we are considering, given by equations (\ref{gauge_A_3})
and (\ref{gauge_A_3.1}), this is the three-dimensional space parametrized
by the coupling constants $\{g_{k},Z_{a,k},Z_{c,k}\}$. In the spirit
of \cite{Reuter_Bi_Field} this is a bi--field truncation; see also
\cite{Rosten_2011} for a discussion of this enlarged class of truncations.
We have seen that the anomalous dimensions of the fluctuation field
and of the ghost fields are determined by a linear system; we will
solve this to obtain $\eta_{a,k}$ and $\eta_{c,k}$ as functions
of the gauge coupling $g_{k}$ and then we will reinsert them back
in equation (\ref{BF_7}) to obtain $\eta_{\bar{A},k}$; from this
we then obtain a new closed form for $\partial_{t}g_{k}$. 

In the most general case (compatible with our truncation), the anomalous
dimensions $\eta_{a,k}$ and $\eta_{c,k}$ are determined by the following
linear system;
\begin{eqnarray}
\eta_{a,k} & = & (A+B\,\eta_{a,k}+C\,\eta_{c,k})g_{k}^{2}\nonumber \\
\eta_{c,k} & = & (A'+B'\,\eta_{a,k}+C'\,\eta_{c,k})g_{k}^{2}\,.\label{BF_16}
\end{eqnarray}
The coefficients in (\ref{BF_16}) do not depend on the scheme I or
II, but they depend on the gauge and on $R_{k}(z)$. In particular,
setting $d=4$ in (\ref{BF_4}) and (\ref{BF_5}) gives the following
values for the coefficients:
\[
A=-\frac{10}{3}\frac{N}{(4\pi)^{2}}\qquad\qquad B=\frac{5}{6}\frac{N}{(4\pi)^{2}}\qquad\qquad C=0\,,
\]
\begin{equation}
A'=-\frac{N}{(4\pi)^{2}}\qquad\qquad B'=\frac{1}{6}\frac{N}{(4\pi)^{2}}\qquad\qquad C'=0\,.\label{BF_16.1}
\end{equation}
As a note, we mention that the scheme independent coefficients $A$
and $A'$ become, respectively, $-\frac{13-3\alpha}{3}$ and $-\frac{3-\alpha}{2}$
when $\alpha\neq1$ and the coefficients $C$ and $C'$ become non-zero.
A system analogous to (\ref{BF_16}) was studied within the non-background
EAA approach to non-abelian gauge theories in \cite{Ellwanger_Hirsch_Weber_1996}.
Considering that the anomalous dimensions in the rhs of (\ref{BF_16})
are at least of order $g_{k}^{2}$ we find, to lowest order, the following
forms:
\begin{eqnarray}
\eta_{a,k} & = & -\frac{10}{3}\frac{g_{k}^{2}N}{(4\pi)^{2}}+O(g_{k}^{4})\nonumber \\
\eta_{c,k} & = & -\frac{g_{k}^{2}N}{(4\pi)^{2}}+O(g_{k}^{4})\,.\label{BF_15.1}
\end{eqnarray}
The terms on the rhs of (\ref{BF_15.1}) are scheme independent and
agree with the ones of \cite{Ellwanger_Hirsch_Weber_1996}. To solve
the system (\ref{BF_16}) we rewrite it as a matrix equation:
\begin{equation}
\bar{\eta}_{k}=\left(\bar{A}+\mathbf{M}\,\bar{\eta}_{k}\right)g_{k}^{2}\,,\label{BF_17}
\end{equation}
where we defined:
\begin{equation}
\bar{\eta}_{k}=\left(\begin{array}{c}
\eta_{a,k}\\
\eta_{c,k}
\end{array}\right)\qquad\qquad\bar{A}=\left(\begin{array}{c}
A\\
A'
\end{array}\right)\qquad\qquad\mathbf{M}=\left(\begin{array}{cc}
B & C\\
B' & C'
\end{array}\right)\,.\label{BF_18}
\end{equation}
The linear system (\ref{BF_17}) is easily solved:
\begin{equation}
\bar{\eta}_{k}=\left(1-g_{k}^{2}\,\mathbf{M}\right)^{-1}\bar{A}\,,\label{BF_20}
\end{equation}
or more explicitly:
\begin{eqnarray}
\eta_{a,k} & = & \frac{A(1-C'g_{k}^{2})+ACg_{k}^{2}}{(1-Bg_{k}^{2})(1-C'g_{k}^{2})+B'Cg_{k}^{4}}g_{k}^{2}\nonumber \\
\eta_{c,k} & = & \frac{A'(1-Bg_{k}^{2})+AB'g_{k}^{2}}{(1-Bg_{k}^{2})(1-C'g_{k}^{2})+B'Cg_{k}^{4}}g_{k}^{2}\,.\label{BF_21}
\end{eqnarray}
We can now turn back to the beta function for the gauge coupling and
``close'' it by inserting the functions (\ref{BF_21}) in (\ref{BF_7}).
Under the condition $C=C'=0$, we find the following RG improved result
for the beta function of the gauge coupling:
\begin{figure}
\begin{centering}
\includegraphics[scale=0.9]{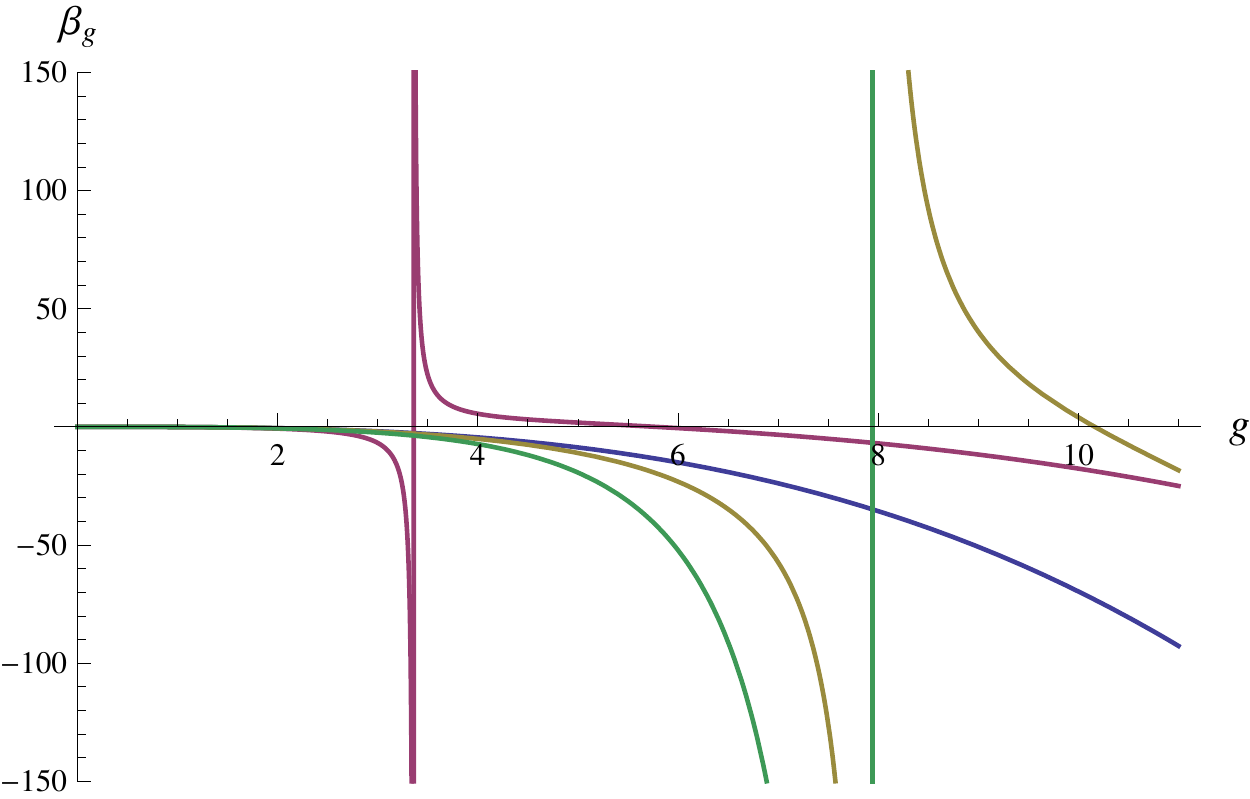}
\par\end{centering}

\caption{Bi--field beta functions from (\ref{BF_22}). Asymptotes from the left:
Padè, type II and type I beta functions. Now the type I and type II
beta function have the asymptotes in the same place. The one--loop
beta function (without asymptote) is shown for comparison.}
\end{figure}
\begin{equation}
\partial_{t}g_{k}=\frac{a+c\, g_{k}^{2}+d\, g_{k}^{4}}{1-b\, g_{k}^{2}}g_{k}^{3}\,,\label{BF_22}
\end{equation}
with coefficients:
\begin{equation}
c=\frac{1}{2}(bA+b'A'-2aB)\qquad\qquad d=\frac{b'}{2}(AB'-A'B)\,.\label{BF_23}
\end{equation}
Using the values (\ref{BF_8}) and (\ref{BF_16.1}) we find:
\[
c_{I}=\frac{16}{9}\frac{N^{2}}{(4\pi)^{4}}\qquad\qquad d_{I}=\frac{5}{108}\frac{N^{3}}{(4\pi)^{6}}\,,
\]
\begin{equation}
c_{II}=-\frac{8}{3}\frac{N^{2}}{(4\pi)^{4}}\qquad\qquad d_{II}=\frac{5}{108}\frac{N^{3}}{(4\pi)^{6}}\,.\label{BF_24}
\end{equation}
Note that only the coefficients $c$ are different in the two schemes.
The beta function (\ref{BF_22}) clearly shows the influence of the
fluctuation couplings, here represented by $Z_{a,k}$ and $Z_{c,k}$,
on the flow of the gauge coupling $g_{k}$. We remember that the anomalous
dimensions we used to obtain the closed form for the beta function
of $g_{k}$ was calculated in the gauge $\alpha=1$, in a future work
\cite{Codello_in_preparation} we will study the dependence on the
gauge-fixing parameter, considering in particular the Landau gauge\emph{
$\alpha=0$. }

Beta functions of the form (\ref{BF_22}) are obtained as Padé approximations
of the perturbative beta function \cite{Gockeler_Horsley_Irving_Pleiter_Rakow_Schierholz_Stuben_2006}.
By employing data at three--loops one finds a beta function of the
form (\ref{BF_22}) with coefficients given by: 
\[
a_{\textrm{Padè}}=-b_{1}\qquad\qquad b_{\textrm{Padè}}=\frac{b_{2}}{b_{1}}\qquad\qquad c_{\textrm{Padè}}=-b_{1}+\frac{b_{0}b_{2}}{b_{1}}\qquad\qquad d_{\textrm{Padè}}=0\,,
\]
where $b_{1},b_{2},b_{3}$ are the one--, two-- and three--loop coefficients.
The bi--field beta functions are compared with the Padè re-summed one
in Figure 10. When making this comparisons one has to remember that
starting from the three--loop coefficients these become scheme dependent
even when employing massless renormalization schemes. Finally, the
coefficient $d$ becomes non-zero when one considers data at four--loops,
thus our beta function (\ref{BF_22}) contains re-summed information
up to four--loops.

\subsubsection{Flow and physical observables}

We now integrate the beta functions derived in the previous subsection.
We will be able to do these steps analytically. We start to consider
the single--field beta function (\ref{BF_14}); a simple integration
of this gives:
\begin{figure}
\begin{centering}
\includegraphics[scale=0.85]{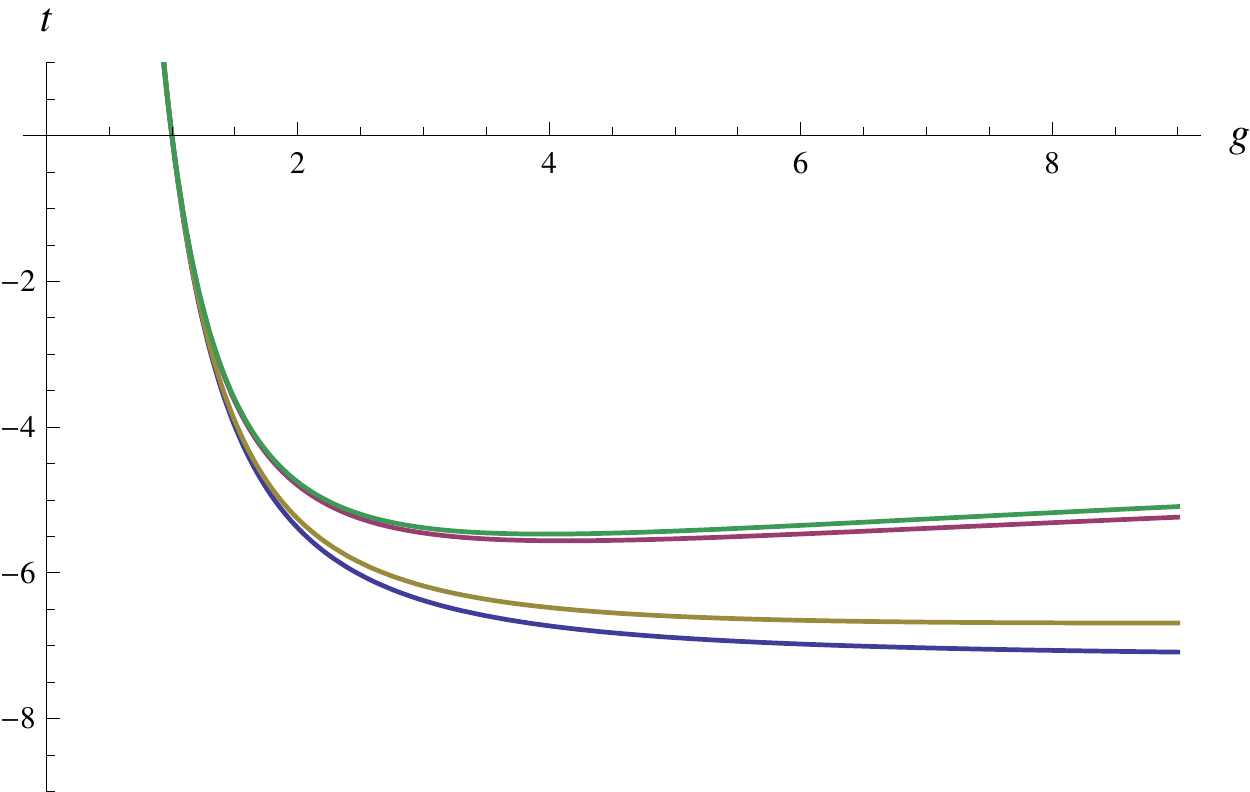}
\par\end{centering}

\caption{Implicit plot for the single--field flow of $g_{k}$ from (\ref{IF_2}).
From top: RS, type II, type I and one-loop flows. }
\end{figure}
\begin{equation}
t=-\frac{1}{2a}\left(\frac{1}{g_{k}^{2}}-\frac{1}{g_{0}^{2}}\right)-\frac{b}{a}\log\frac{g_{k}}{g_{0}}\,,\label{IF_2}
\end{equation}
where $t=\log\frac{k}{k_{0}}$, with $k_{0}$ a reference scale where
$g_{k}=g_{k_{0}}$. Equation (\ref{IF_2}) is transcendental and cannot
be inverted analytically except in the one--loop case $b=0$. In this
last case one obtains:
\begin{figure}
\begin{centering}
\includegraphics[scale=0.85]{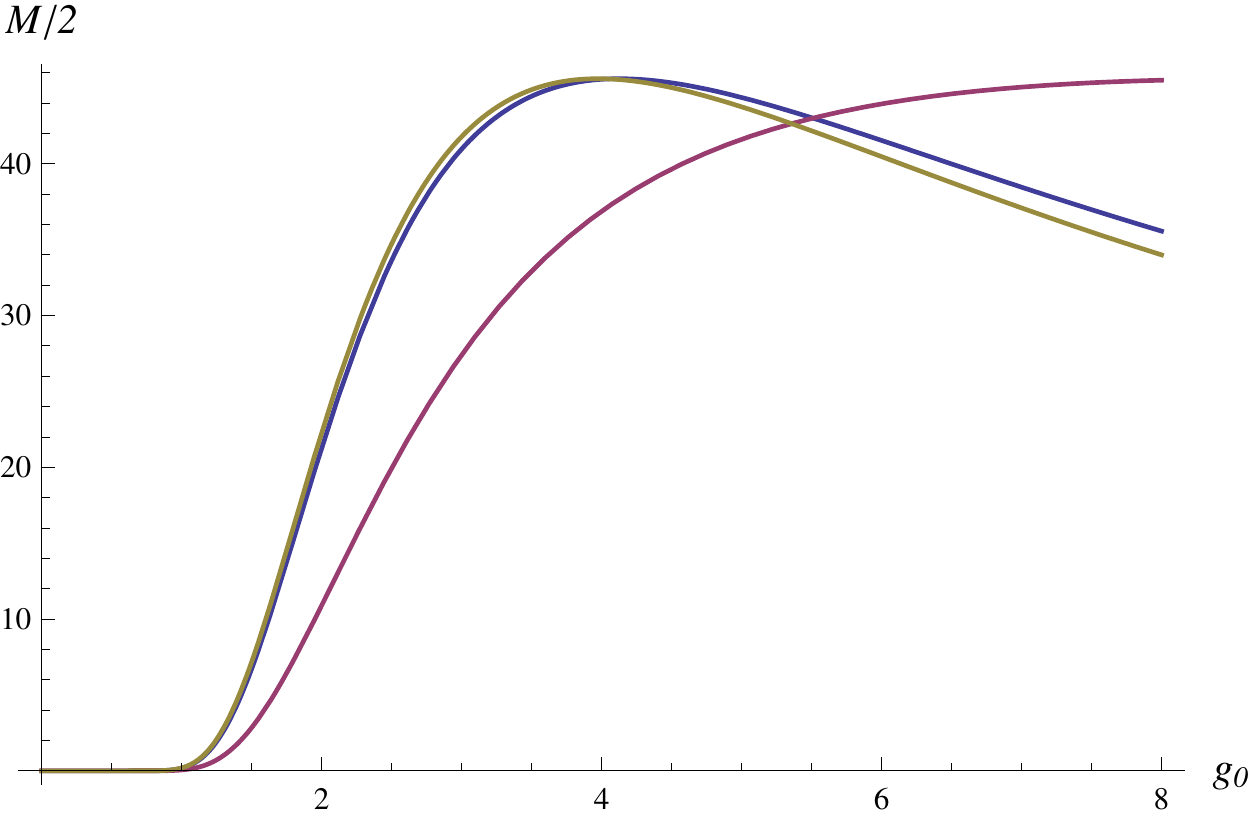}
\par\end{centering}

\caption{The mass gap as a function of $g_{0}$ (with $k_{0}=91.19$ GeV) for
the single--field flows compared to the flow proposed by RS from (\ref{IF_2.3}).
From top-left the RS, type II and type I curves.}
\end{figure}
\begin{equation}
g_{k}^{2}=\frac{g_{0}^{2}}{1-2a\log\frac{k}{k_{0}}}\,.\label{IF_2.2}
\end{equation}
Since $a=-\frac{11}{3}\frac{N}{(4\pi)^{2}}<0$, in the UV limit $k\rightarrow\infty$
the theory is asymptotically free, i.e. $g_{k}\rightarrow0$. 

When $b\neq0$ we can make an implicit plot of $t(g_{k})$ using (\ref{IF_2});
the result for the three cases (type I, type II and RS) together with
the one--loop case are shown in Figure 11. One sees that the type
II and RS flows are very similar. The improved flows appear to be
convex functions (contrary to the perturbative result) and have a
minimum at the value where the beta function (\ref{BF_14}) has the
pole, i.e. at $g_{k}^{2}=\frac{1}{b}$. Note the important fact that
even if the beta functions have a pole, the flow they generate is
instead smooth. But one cannot invert $t(g_{k})$ to obtain $g_{k}(t)$
for every value of $t\in\mathbb{R}$. If we choose to be continuously
connected to the asymptotic free regime, when we lower the RG scale
toward the IR we reach the point $t(b^{-1/2})$ after which we cannot
proceed further, i.e. the curve $g_{k}(t)$ cannot be extended.
This signals that our single coupling truncation breaks down (and one needs to consider more sophisticated truncations to proceed further toward the IR \cite{Ellwanger_Hirsch_Weber_1998}).
Following \cite{Sannino_Schechter_2010} we can interpret $t(b^{-1/2})$ as the point
where bound states are formed and identify the relative mass scale
$M$ as (twice) the mass gap. With this identification, using (\ref{IF_2}),
we find the following relation for the mass gap as a function of $k_{0}$
and $g_{0}$:
\begin{equation}
M=k_{0}\left(g_{0}^{2}b\right)^{\frac{b}{2a}}e^{\frac{1}{2a}\left(\frac{1}{g_{0}^{2}}-b\right)}\,.\label{IF_2.3}
\end{equation}
This estimate for the mass gap is plotted in Figure 12. For $k_{0}=91.19$
GeV and $N=3$ one finds $M_{I}=0.61$ GeV and $M_{II}=1.79$ GeV,
to be compared with the estimate $M_{RS}=1.69$ GeV of \cite{Sannino_Schechter_2010}.
As for the beta functions, the type II estimate and the one based
on the RS beta function are similar, while a strong scheme dependence
is observed in the type I case.

In the same way one can integrate the bi--field beta function (\ref{BF_22})
to obtain the following analytical relation:
\begin{figure}
\begin{centering}
\includegraphics[scale=0.85]{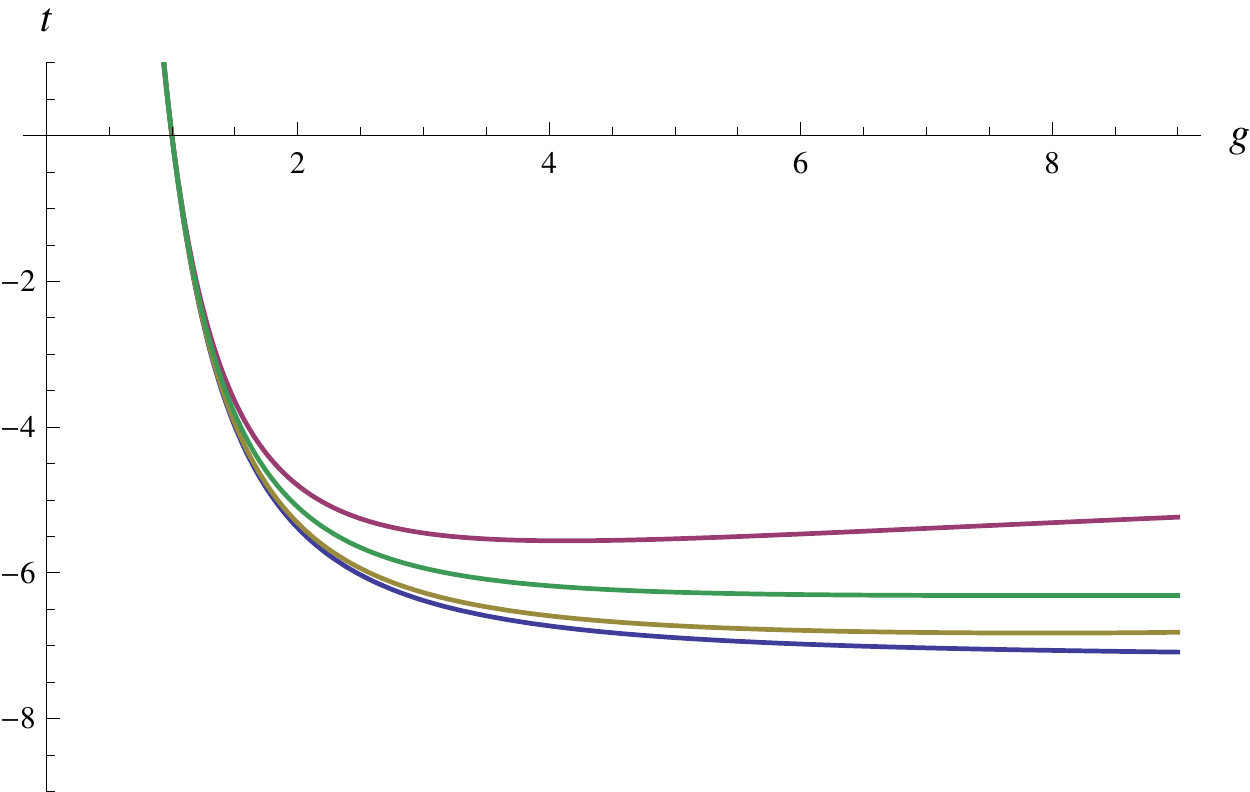}
\par\end{centering}

\caption{Implicit plot for the bi--field flow of $g_{k}$ from (\ref{IF_4}).
From the top: RS, type II, type I and one-loop flows. Note that the
type I and II are closer than in Figure 11.}
\end{figure}
\begin{eqnarray}
t & = & -\frac{1}{2a}\left(\frac{1}{g_{k}^{2}}-\frac{1}{g_{0}^{2}}\right)-\frac{ab+c}{a^{2}}\log\frac{g_{k}}{g_{0}}+\frac{ab+c}{4a^{2}}\log\frac{a+cg_{k}^{2}+dg_{k}^{4}}{a+cg_{0}^{2}+dg_{0}^{4}}\nonumber \\
 &  & +\frac{abc-2ad+c^{2}}{4a^{2}\sqrt{c^{2}-4ad}}\log\frac{\left(c+2dg_{k}^{2}-\sqrt{c^{2}-4ad}\right)\left(c+2dg_{0}^{2}+\sqrt{c^{2}-4ad}\right)}{\left(c+2dg_{k}^{2}+\sqrt{c^{2}-4ad}\right)\left(c+2dg_{0}^{2}-\sqrt{c^{2}-4ad}\right)}\,.\label{IF_4}
\end{eqnarray}
The implicit plot of (\ref{IF_4}) is shown in Figure 13. Obviously
if we set $c=d=0$ we recover (\ref{IF_2}). As before, we can estimate
the mass gap $M$, the results is now:
\begin{eqnarray}
M & = & k_{0}\left(g_{0}^{2}b\right)^{\frac{ab+c}{2a^{2}}}e^{\frac{1}{2a}\left(\frac{1}{g_{0}^{2}}-b\right)}\left(\frac{a+c/b+d/b^{2}}{a+cg_{0}^{2}+dg_{0}^{4}}\right)^{\frac{ab+c}{4a^{2}}}\nonumber \\
 &  & \times\left[\frac{\left(c+2d/b-\sqrt{c^{2}-4ad}\right)\left(c+2dg_{0}^{2}+\sqrt{c^{2}-4ad}\right)}{\left(c+2d/b+\sqrt{c^{2}-4ad}\right)\left(c+2dg_{0}^{2}-\sqrt{c^{2}-4ad}\right)}\right]^{\frac{abc-2ad+c^{2}}{4a^{2}\sqrt{c^{2}-4ad}}}\,.\label{IF_5}
\end{eqnarray}
Evaluating (\ref{IF_5}) for the different cutoff choices gives Figure
14. For $k_{0}=91.19$ GeV and $N=3$ one now finds the estimates
$M_{I}=0.54$ GeV and $M_{II}=0.85$ GeV, which are nearer to each
other but smaller than the RS or the lattice predictions.
\begin{figure}
\begin{centering}
\includegraphics[scale=0.85]{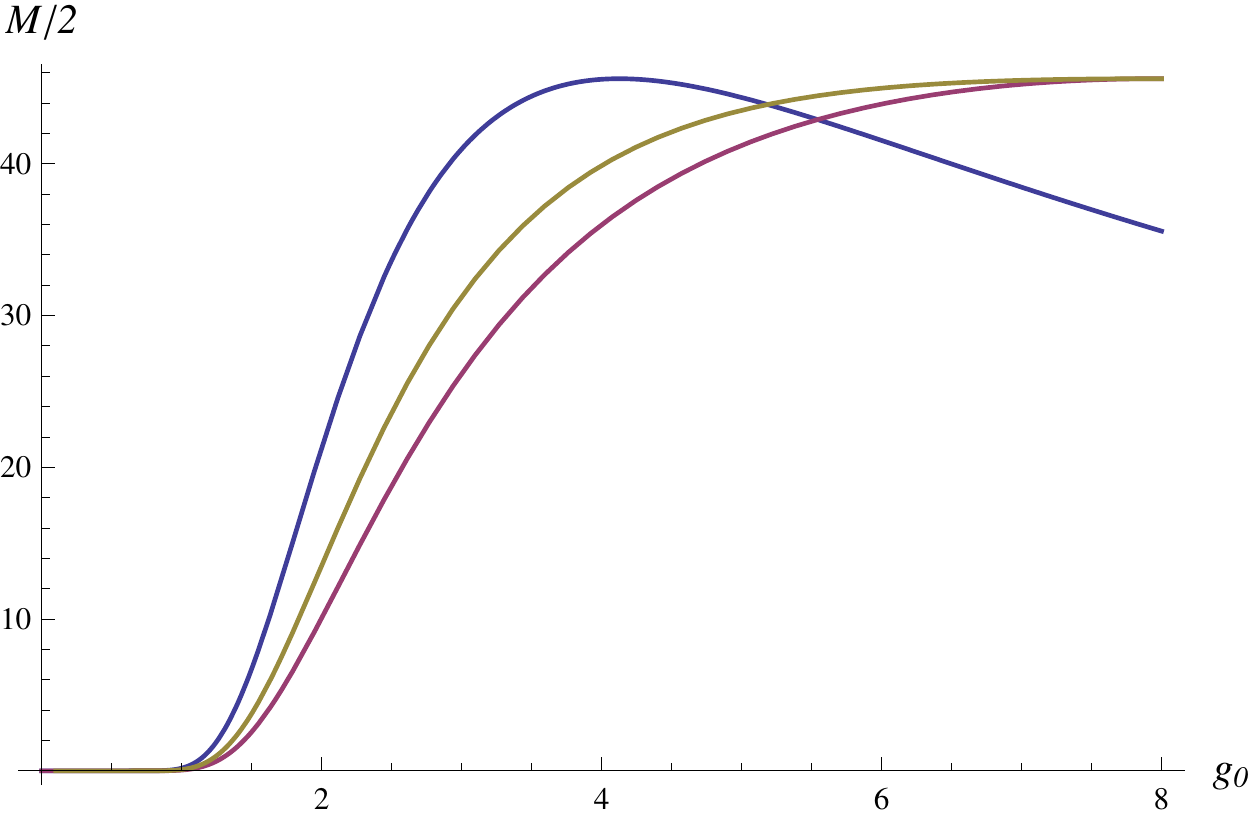}
\par\end{centering}

\caption{The mass gap $M$ as a function of $g_{0}$ (with $k_{0}=91.19$ GeV)
for the (from top-left) RS and bi--field flows.}
\end{figure}

To conclude, we discuss the RG scale dependence of a physical observable
$\mathcal{O}_{k}(g_{k})$. In particular, if this is RG invariant
$\frac{d}{dt}\mathcal{O}_{k}(g_{k})=0$, then it's flow is governed
by the beta function $\beta(g_{k})=\partial_{t}g_{k}$ of the coupling
constant. If the observable has dimension $d_{\mathcal{O}}$ the we
can set $\mathcal{O}_{k}(g_{k})=k^{d_{\mathcal{O}}}\tilde{\mathcal{O}}(g_{k})$
and obtain:
\begin{equation}
d_{\mathcal{O}}\tilde{\mathcal{O}}(g_{k})+\beta(g_{k})\partial_{g}\tilde{\mathcal{O}}(g_{k})=0\,.\label{NP_1}
\end{equation}
It's easy to integrate (\ref{NP_1}) to find:
\begin{equation}
\mathcal{O}_{k}=\mathcal{O}_{k_{0}}e^{-d_{\mathcal{O}}\int_{g_{0}}^{g_{k}}\frac{dg}{\beta(g)}}\,,\label{NP_2}
\end{equation}
where $t=\int_{g_{0}}^{g_{k}}\frac{dg}{\beta(g)}$. If we insert (\ref{IF_2})
into (\ref{NP_2}) we find:
\begin{equation}
\mathcal{O}_{k}=\mathcal{O}_{k_{0}}\left(\frac{g_{k}}{g_{0}}\right)^{d_{\mathcal{O}}\frac{b}{a}}e^{\frac{d_{\mathcal{O}}}{2a}\left(\frac{1}{g_{k}^{2}}-\frac{1}{g_{0}^{2}}\right)}\,.\label{NP_3}
\end{equation}
One notices that for any value of $d_{\mathcal{O}}$ the observable
is a smooth function of $g_{k}$ even if the beta function has a pole.
A similar, but more complex relation, can be obtained starting from
(\ref{IF_4}):
\begin{eqnarray}
\mathcal{O}_{k} & = & \mathcal{O}_{k_{0}}\left(\frac{g_{k}}{g_{0}}\right)^{d_{\mathcal{O}}\frac{ab+c}{a^{2}}}e^{\frac{d_{\mathcal{O}}}{2a}\left(\frac{1}{g_{k}^{2}}-\frac{1}{g_{0}^{2}}\right)}\left(\frac{a+cg_{k}^{2}+dg_{k}^{4}}{a+cg_{0}^{2}+dg_{0}^{4}}\right)^{d_{\mathcal{O}}\frac{ab+c}{4a^{2}}}\nonumber \\
 &  & \times\left[\frac{\left(c+2dg_{k}^{2}-\sqrt{c^{2}-4ad}\right)\left(c+2dg_{0}^{2}+\sqrt{c^{2}-4ad}\right)}{\left(c+2dg_{k}^{2}+\sqrt{c^{2}-4ad}\right)\left(c+2dg_{0}^{2}-\sqrt{c^{2}-4ad}\right)}\right]^{d_{\mathcal{O}}\frac{abc-2ad+c^{2}}{4a^{2}\sqrt{c^{2}-4ad}}}\,.\label{NP_4}
\end{eqnarray}
Even in this case observables are smooth functions of the coupling
constant. The simplest observable is the invariant scale $M$; we
can recover (\ref{IF_5}) by setting $d_{\mathcal{O}}=1$, $g_{k}^{2}=1/b$,
$\mathcal{O}_{k}=M$ and $\mathcal{O}_{k_{0}}=k_{0}$ in (\ref{NP_4}).
Other possible observables have different canonical dimensions, but
in general observables turn out to be smooth functions of the coupling
constant in spite of the singular behavior of the RG improved (or
re-summed) beta functions. Thus this kind of beta functions are consistent
approximations to the RG flow that can be applied to the study of
the IR physics of non-abelian gauge theories.

\newpage

\section{Discussion and future perspectives}

In this paper we introduced a method that enables the projection of
the RG flow equations of a large new class of truncations of the background
effective average action (bEAA). Our method is based on the explicit
momentum space representation of the hierarchy of flow equations satisfied
by the proper-vertices of the bEAA. The key step in our construction
is the determination of the explicit momentum space form of vertices
with background legs, since these are related to functional derivatives
of the cutoff action and must be treated with care. We showed how
these vertices indeed have a simple form related to the vertices of
the cutoff action, the action defined by having the cutoff operator
as Hessian, and contains finite difference derivatives of the cutoff
shape function. We also gave simple diagrammatic rules that provide
a representation of the flow equations for the proper-vertices of
the bEAA that can be used in actual computations. Our method is very
general and can be applied to a variety of new interesting truncations
of the bEAA, with applications to non-abelian gauge theories, nonlinear
sigma models, gravity and membranes; or any other theory characterized
by local gauge symmetries.

As a first application, we studied a bi--field truncation of the bEAA
for $SU(N)$ non-abelian gauge theories. Employing bi--field truncations
we can explore the dependency of the bEAA on all its arguments: fluctuation
and background fields. We considered the three-dimensional subspace
of theory space parametrized by the three anomalous dimensions $\{\eta_{\bar{A},k},\eta_{a,k},\eta_{c,k}\}$.
We showed how all the results obtained with the aid of the local heat
kernel expansion, usually employed in this framework, are reproduced
by our technique. The structure of the exact flow equation for the
bEAA implies that $\eta_{\bar{A},k}$ is linearly related to both
$\eta_{a,k}$ and $\eta_{c,k}$; in the background field method, the
beta function of the gauge coupling is related to $\eta_{\bar{A},k}$,
thus to obtain $\partial_{t}g_{k}$ one needs $\eta_{a,k}$ and $\eta_{c,k}$.
Usually one imposes $\eta_{a,k}=\eta_{\bar{A},k}$ and $\eta_{c,k}=0$,
in order to treat the problem by using only a single--field truncation.
Instead, we calculated $\eta_{a,k}$ and $\eta_{c,k}$ directly and
showed how they can be determined as a function of $g_{k}$ by solving
a linear system; in this way we obtained a new RG improved form for
the beta function of the gauge coupling. We then discussed some phenomenological
implications of the flow so obtained, which has similarities with
re-summed perturbative beta functions. As a future extension of this
study, one can consider more general bi--field truncations by allowing
the gauge fluctuation mass and gauge-fixing parameter to run \cite{Codello_in_preparation}.

In another application one can study bi--field truncations in the
context of quantum gravity and check how Asymptotic Safety, up to
now observed in single--field truncations, is robust to the bi--field
extension \cite{Codello_D'Odorico_Pagani_2013}. 

Still, as work done in the non-background formalism \cite{Ellwanger_Hirsch_Weber_1998}
has shown, the non-trivial part of the bEAA has a non-local structure.
With the method presented here we can project a large class of non-local
truncations. In a gauge invariant formalism, the finite part of the
effective action is encoded in ``structure functions'', which are
generally functions of the covariant Laplacian, and act on local invariants.
As an example, in non-abelian gauge theories, one can consider the
following ansatz for the gauge invariant part of the bEAA:
\begin{equation}
\bar{\Gamma}_{k}[A]=\int d^{d}x\, F_{\mu\nu}^{a}f_{k}(-D^{2})F^{a\mu\nu}+O(F^{3})\,,\label{C_1}
\end{equation}
where $f_{k}$ is a running structure function which encodes non-trivial
physical information. By applying our algorithm we can obtain a RG
flow equation for the running structure function, which will be of
the form: 
\begin{equation}
\partial_{t}f_{k}=\mathcal{F}_{R_{k}}^{d}\left(g_{k},f_{k},f_{k}',f_{k}''\right)\,.\label{C_1.1}
\end{equation}
By following the flow of the running structure function $f_{k}$ from
the UV, where the theory is asymptotically free, down to the IR, we
can obtain the full finite non-local effective action, from which
we can extract relevant physical information. Similar applications
can be made in the context of gravity, extending the results obtained
using the non-local heat kernel expansion \cite{Codello_2010}, in
the context of non-linear sigma models or in the context membrane theory.

For these reasons, among others, the computational strategy based
on the hierarchy of flow equations for the zero-field proper-vertices
of the bEAA, when combined with the momentum space rules that we derived,
is a promising tool for further studies of the bEAA in its full generality.

\section*{Acknowledgments}

I would like to thank M. Reuter for stimulating discussions and O.
Zanusso, M. Demmel, R. Percacci for careful reading the manuscript.

\newpage

\appendix

\section{Momentum space representation of background vertices}

In this appendix we give the details of the derivation of the momentum
space reprentation of vertices of the bEAA with background legs.

\subsection{Perturbative expansion of the heat kernel}

We shortly review the perturbative expansion for the heat kernel as
developed in \cite{Codello_Zanusso_2012}, to which we refer for more
details. This expansion will be used in the next subsection to derive
the momentum space representation of background vertices.

The heat kernel $K^{s}(x,y)$ satisfies the following partial differential
equation with boundary condition:
\begin{equation}
\left(\partial_{s}+\Delta_{x}\right)K^{s}(x,y)=0\qquad\qquad K^{0}(x,y)=\delta(x-y)\,,\label{HK_1}
\end{equation}
where $\Delta=-D_{\mu}D^{\mu}+U$ is the Laplacian operator constructed
using the covariant derivatives $D_{\mu}$ and $U$ is a potential
term. We decompose the the Laplacian as the sum of a ``non-interacting''
Laplacian $-\partial^{2}$ and of an interaction $V$ in the following
way:
\begin{equation}
\Delta=-\partial^{2}+V\,.\label{HK_PT_1}
\end{equation}
The potential $V$ contains $U$ and all terms proportional to the
gauge connection $A_{\mu}$. Two examples are the flat space Laplacian
$\Delta=-\partial^{2}+U$, where simply $V=U$, or the abelian gauge
Laplacian where $V$ contains all terms that vanish for $A_{\mu}=0$:
\[
\Delta=-D_{\mu}D^{\mu}=-\left(\partial_{\mu}+iA_{\mu}\right)\left(\partial^{\mu}+iA^{\mu}\right)=-\partial^{2}\underbrace{-2A_{\mu}\partial^{\mu}-\partial_{\mu}A^{\mu}+A_{\mu}A^{\mu}}_{V}\,.
\]

To derive the perturbative series for the heat kernel we need first
to calculate the heat kernel%
\footnote{Here and in the following we use the compact notation $K_{xy}^{s}\equiv K^{s}(x,y)$
and $\delta_{xy}\equiv\delta^{(d)}(x-y)$. We also define $\int_{x}\equiv\int d^{d}x$
and $\int_{q}\equiv\int\frac{d^{d}q}{(2\pi)^{d}}$.%
} $K_{0,xy}^{s}$ of the operator $-\partial^{2}$, around which we
will perform the expansion. From equation (\ref{HK_1}) we see that
it satisfies the following equation with boundary condition:
\begin{equation}
\left(\partial_{s}-\partial_{x}^{2}\right)K_{0,xy}^{s}=0\qquad\qquad K_{0,xy}^{0}=\delta_{xy}\,;\label{HK_PT_2}
\end{equation}
the solution of (\ref{HK_PT_2}) is the standard Gaussian:
\begin{equation}
K_{0,xy}^{s}=\frac{1}{(4\pi s)^{d/2}}e^{-\frac{(x-y)^{2}}{4s}}\,.\label{HK_PT_5}
\end{equation}
Equation (\ref{HK_PT_5}) is the solution around which we will construct
the perturbative expansion. Note that the heat kernel (\ref{HK_PT_5})
satisfies:
\begin{equation}
K_{0,xy}^{s_{1}+s_{2}}=\int_{z}K_{0,xz}^{s_{1}}K_{0,zy}^{s_{2}}\,.\label{HK_PT_5.1}
\end{equation}
To derive the perturbative expansion of $K_{xy}^{s}$ around $K_{0,xy}^{s}$
we define the operator%
\footnote{Not to be confused with the potential $U$.%
} $U_{xy}^{s}=\int_{z}K_{0,xz}^{-s}K_{zy}^{s}$. Using (\ref{HK_1})
and (\ref{HK_PT_2}) we find it satisfies the following equation:
\begin{eqnarray}
\partial_{s}U_{xy}^{s} & = & \int_{z}\left[\partial_{s}K_{0,xz}^{-s}K_{zy}^{s}+K_{0,xz}^{-s}\partial_{s}K_{zy}^{s}\right]\nonumber \\
 & = & \int_{z}\left[K_{0,xz}^{-s}(-\partial_{z}^{2})K_{zy}^{s}-K_{0,xz}^{-s}\Delta_{z}K_{zy}^{s}\right]\nonumber \\
 & = & -\int_{z}K_{0,xz}^{-s}V_{z}K_{zy}^{s}\nonumber \\
 & = & -\int_{zw}K_{0,xz}^{-s}V_{z}K_{0,zw}^{s}U_{wy}^{s}\,.\label{HK_PT_6}
\end{eqnarray}
We know that equation (\ref{HK_PT_6}) is solved by Dyson's series:
\[
U_{xy}^{s}=T\exp\left\{ -\int_{0}^{s}dt\,\int_{z}K_{0,xz}^{-t}V_{z}K_{0,zy}^{t}\right\} \,,
\]
where the exponential is time-ordered with respect to $s$. Using
(\ref{HK_PT_5.1}) we immediately find:
\begin{equation}
K_{xy}^{s}=\int_{z}K_{0,xz}^{s}\, T\exp\left\{ -\int_{0}^{s}dt\,\int_{w}K_{0,zw}^{-t}V_{w}K_{0,wy}^{t}\right\} \,.\label{HK_PT_7}
\end{equation}
Rescaling the integration variable in (\ref{HK_PT_7}) as $t\rightarrow t/s$
and using (\ref{HK_PT_5.1}) gives the final formula for the perturbative
expansion of the heat kernel:
\begin{eqnarray}
K_{xy}^{s} & = & \int_{z}K_{0,xz}^{s}\, T\exp\left\{ -s\int_{0}^{1}dt\int_{w}K_{0,zw}^{-st}V_{w}K_{0,wy}^{st}\right\} \nonumber \\
 & = & K_{0,xy}^{s}-s\int_{0}^{1}dt\,\int_{z}K_{0,xz}^{s(1-t)}V_{z}\, K_{0,zy}^{st}+\nonumber \\
 &  & +s^{2}\int_{0}^{1}dt_{1}\int_{0}^{t_{1}}\, dt_{2}\,\int_{zw}K_{0,xz}^{s(1-t_{1})}V_{z}\, K_{0,zw}^{s(t_{1}-t_{2})}V_{w}\, K_{0,wy}^{st_{2}}+O(V^{3})\,.\label{HK_PT_8}
\end{eqnarray}
In the following section we will use this relation to find an explicit
momentum space representation for the functional derivatives of the
cutoff action.

\subsection{Derivation of momentum space representation}

In this section we derive the momentum space representation of background
vertices, in particular we show how to obtain relations (\ref{gauge_D_1.1})
and (\ref{gauge_D_1.2}) used in section 2.3 to obtain the momentum
space representation of the flow equations for the zero-field proper-vertices
of the bEAA.

We need to calculate the momentum space representation of the cutoff
vertices $\Delta S_{k}^{(2;1)}[0;0]$ and $\Delta S_{k}^{(2;2)}[0;0]$.
From the definition of the cutoff action (\ref{gauge_2}) we see that%
\footnote{In the gravitational case one must additionally care about factors
of $\sqrt{g}$.%
}:
\begin{equation}
\Delta S_{k}^{(2;0)}[0;\bar{A}]_{xy}^{AB}=\int_{zw}\delta_{xz}R_{k}[\bar{A}]_{zw}^{AB}\delta_{wy}=R_{k}[\bar{A}]_{xy}^{AB}\,.\label{gauge_MSR_1}
\end{equation}
The cutoff kernel $R_{k}[\bar{A}]$ is a function of the cutoff operator
$L^{(2;0)}[0;\bar{A}]$, constructed as the Hessian of the cutoff
operator action $L[a;\bar{A}]$. For example, if the cutoff operator
is the gauge Laplacian, then:
\[
L^{(2;0)}[0;\bar{A}]_{xy}=\int_{z}\bar{D}_{z\mu}\delta_{zx}\bar{D}_{z}^{\mu}\delta_{zy}=-\int_{z}\delta_{zx}\bar{D}_{z\mu}\bar{D}_{z}^{\mu}\delta_{zy}=-\bar{D}_{\mu x}\bar{D}^{\mu y}\delta_{xy}=-\bar{D}_{x}^{2}\delta_{xy}\,.
\]
Thus, after making a Laplace transform, we can write the cutoff kernel
in terms of the heat kernel of the cutoff operator%
\footnote{For simplicity we consider $R_{k}^{AB}$ having tensor structure proportional
to $\delta^{AB}$.%
}: 
\begin{equation}
R_{k}[\bar{A}]_{xy}^{AB}=\int_{0}^{\infty}ds\,\tilde{R}_{k}(s)\, K^{s}[\bar{A}]_{xy}^{AB}\,,\label{gauge_MSR_2}
\end{equation}
where the heat kernel can be written in terms of the Hessian of the
cutoff operator action:
\begin{equation}
K^{s}[\bar{A}]=\exp\left\{ -sL^{(2;0)}[0;\bar{A}]\right\} \,.\label{gauge_MSR_3}
\end{equation}
Inserting (\ref{gauge_MSR_2}) in equation (\ref{gauge_MSR_1}) and
setting the background field to zero after differentiating with respect
to it one time, gives the following representation for the cutoff
vertex with one background leg in terms of the heat kernel:
\begin{equation}
\Delta S_{k}^{(2;1)}[0;0]_{xyz}^{ABC}=\left.\frac{\delta R_{k}[\bar{A}]_{xy}^{AB}}{\delta\bar{A}_{z}^{C}}\right|_{\bar{A}=0}=\int_{0}^{\infty}ds\,\tilde{R}_{k}(s)\left.\frac{\delta K^{s}[\bar{A}]_{xy}^{AB}}{\delta\bar{A}_{z}^{C}}\right|_{\bar{A}=0}\,.\label{gauge_MSR_4}
\end{equation}
We can now use the perturbative expansion for the heat kernel obtained
in the previous section, equation (\ref{HK_PT_8}), to write the last
term of (\ref{gauge_MSR_4}) as follows:
\begin{equation}
\int_{0}^{\infty}ds\,\tilde{R}_{k}(s)\left.\frac{\delta K^{s}[\bar{A}]_{xy}^{AB}}{\delta\bar{A}_{z}^{C}}\right|_{\bar{A}=0}=\int_{0}^{\infty}ds\,\tilde{R}_{k}(s)(-s)\int_{0}^{1}dt\, K_{0,xw}^{s(1-t)}L^{(2;1)}[0;0]_{wuz}^{ABC}K_{0,uy}^{st}\,.\label{gauge_MSR_5}
\end{equation}
In (\ref{gauge_MSR_5}) we omitted to explicitly write the coordinate
integrals and we wrote the flat space heat kernels as $K^{s}[0]_{xy}^{AB}=K_{0,xy}^{s}\delta^{AB}$,
where $K_{0,xy}^{s}$ is given in equation (\ref{HK_PT_5}). Going
to momentum space and inserting (\ref{gauge_MSR_5}) in (\ref{gauge_MSR_4})
gives the following representation for the cutoff vertex:
\begin{eqnarray}
\Delta S_{k}^{(2;1)}[0;0]_{p_{1},p_{2},p_{3}}^{ABC} & = & \int_{0}^{\infty}ds\,\tilde{R}_{k}(s)(-s)\int_{0}^{1}dt\, K_{0,p_{1}}^{s(1-t)}[l_{p_{1},p_{2},p_{3}}^{(2;1)}]^{ABC}K_{0,p_{2}}^{st}\nonumber \\
 & = & \int_{0}^{\infty}ds\,\tilde{R}_{k}(s)(-s)\int_{0}^{1}dt\, e^{-s(1-t)p_{1}^{2}}[l_{p_{1},p_{2},p_{3}}^{(2;1)}]^{ABC}e^{-stp_{2}^{2}}\nonumber \\
 & = & [l_{p_{1},p_{2},p_{3}}^{(2;1)}]^{ABC}\int_{0}^{1}dt\int_{0}^{\infty}ds\,\tilde{R}_{k}(s)(-s)e^{-s(1-t)p_{1}^{2}-stp_{2}^{2}}\,,\label{gauge_MSR_6}
\end{eqnarray}
where the cutoff operator vertices are defined as:
\[
l_{x_{1}...x_{n}y_{1}...y_{m}}^{(n;m)}\equiv L^{(n;m)}[0;0]_{x_{1}...x_{n}y_{1}...y_{m}}\,.
\]
Here we used the following simple momentum space representation for
the flat space heat kernel $K_{0,p}^{s}=e^{-sp^{2}}$. It is left
to evaluate the double integral in (\ref{gauge_MSR_6}). This can
be done with the aid of the $Q$-functionals (see the appendix A of
\cite{Codello_Percacci_Rahmede_2009}):
\begin{equation}
Q_{n}[h](z+a)=\int_{0}^{\infty}ds\,\tilde{h}(s)s^{-n}e^{-sa}\,,\label{gauge_MSR_6.01}
\end{equation}
\begin{equation}
Q_{n}[h](z+a)=\left\{ \begin{array}{ccc}
\frac{1}{\Gamma(n)}\int_{0}^{\infty}dz\, z^{n-1}h(z+a) &  & n>0\\
(-1)^{n}h^{(n)}(a) &  & n\leq0
\end{array}\right..\label{gauge_MSR_6.1}
\end{equation}
Using (\ref{gauge_MSR_6.1}) we find: 
\begin{eqnarray}
\int_{0}^{\infty}ds\,\tilde{R}_{k}(s)(-s)e^{-s(1-t)p_{1}^{2}-stp_{2}^{2}} & = & -Q_{-1}[R_{k}(z+s(1-t)p_{1}^{2}+stp_{2}^{2})]\nonumber \\
 & = & R_{k}'(s(1-t)p_{1}^{2}+stp_{2}^{2})\,.\label{gauge_MSR_7}
\end{eqnarray}
Now the parameter integral is easily evaluated: 
\begin{equation}
\int_{0}^{1}dt\, R_{k}'(s(1-t)p_{1}^{2}+stp_{2}^{2})=\frac{R_{k}(p_{2}^{2})-R_{k}(p_{1}^{2})}{p_{2}^{2}-p_{1}^{2}}\,.\label{gauge_MSR_8}
\end{equation}
If we introduce the first finite-difference derivative, defined as
\[
f_{p_{1},p_{2}}^{(1)}=\frac{f(p_{2}^{2})-f(p_{1}^{2})}{p_{2}^{2}-p_{1}^{2}}\,,
\]
we can finally write, for the cutoff vertex with one external background
leg (\ref{gauge_MSR_6}), the following momentum space representation:
\begin{equation}
\Delta S_{k}^{(2;1)}[0;0]_{p_{1},p_{2},p_{3}}^{ABC}=[l_{p_{1},p_{2},p_{3}}^{(2;1)}]^{ABC}R_{p_{1},p_{2}}^{(1)}\,.\label{gauge_MSR_9}
\end{equation}
We just need now to consider (\ref{gauge_MSR_9}) with the momentum
values $p_{1}=q$, $p_{2}=-q-p$ and $p_{3}=p$ to prove the relation
given in equation (\ref{gauge_D_1.1}):
\begin{equation}
\Delta S_{k}^{(2;1)}[0;0]_{q,-q-p,p}^{ABC}=[l_{q,-q-p,p}^{(2;1)}]^{ABC}R_{q+p,q}^{(1)}\,.\label{gauge_MSR_10}
\end{equation}
Along the same lines we can derive the momentum space representation
for the cutoff vertex with two external background legs. In place
of (\ref{gauge_MSR_4}) we have now
\begin{equation}
\Delta S_{k}^{(2;2)}[0;0]_{xyzw}^{ABCD}=\left.\frac{\delta^{2}R_{k}[\bar{A}]_{xy}^{AB}}{\delta\bar{A}_{w}^{D}\delta\bar{A}_{z}^{C}}\right|_{\bar{A}=0}=\int_{0}^{\infty}ds\,\tilde{R}_{k}(s)\left.\frac{\delta^{2}K^{s}[\bar{A}]_{xy}^{AB}}{\delta\bar{A}_{w}^{D}\delta\bar{A}_{z}^{C}}\right|_{\bar{A}=0}\,.\label{gauge_MSR_11}
\end{equation}
Using the perturbative expansion (\ref{HK_PT_8}) gives the following
expansion for the second functional derivative of the cutoff kernel:
\begin{eqnarray}
\left.\frac{\delta^{2}R_{k}[\bar{A}]_{xy}^{AB}}{\delta\bar{A}_{w}^{D}\delta\bar{A}_{z}^{C}}\right|_{\bar{A}=0} & = & \int_{0}^{\infty}ds\,\tilde{R}_{k}(s)\,\left.\frac{\delta^{2}K^{s}[\bar{A}]_{xy}^{AB}}{\delta\bar{A}_{w}^{D}\delta\bar{A}_{z}^{C}}\right|_{\bar{A}=0}\nonumber \\
 & = & \int_{0}^{\infty}ds\,\tilde{R}_{k}(s)(-s)\int_{0}^{1}dt\, K_{0,xu}^{s(1-t)}L^{(2;2)}[0;0]_{uvzw}^{ABCD}K_{0,vy}^{st}\nonumber \\
 &  & +2\int_{0}^{\infty}ds\,\tilde{R}_{k}(s)\, s^{2}\int_{0}^{1}dt_{1}\int_{0}^{t_{1}}dt_{2}\, K_{0,xu}^{s(1-t_{1})}L^{(2;1)}[0;0]_{uvz}^{AMB}\nonumber \\
 &  & \times K_{0,vr}^{s(t_{1}-t_{2})}L^{(2;1)}[0;0]_{rtw}^{CMD}K_{0,ty}^{st_{2}}\,.\label{gauge_MSR_12}
\end{eqnarray}
When we insert (\ref{gauge_MSR_12}) into (\ref{gauge_MSR_11}) and
shift to momentum space, the first contribution in (\ref{gauge_MSR_12}),
like in the previous case, takes the form:
\begin{equation}
[l_{p_{1},p_{2},p_{3},p_{4}}^{(2;2)}]^{ABCD}R_{p_{4},p_{1}}^{(1)}\,.\label{gauge_MSR_13}
\end{equation}
The second contribution takes instead the following form:
\[
[l_{p_{1},-p_{1}-p_{2},p_{2}}^{(2;1)}]^{AMB}[l_{p_{3},p_{1}+p_{2},p_{4}}^{(2;1)}]^{CMD}\qquad\qquad\qquad
\]
\begin{equation}
\qquad\qquad\qquad\times\int_{0}^{1}dt_{1}\int_{0}^{t_{1}}dt_{2}\int_{0}^{\infty}ds\,\tilde{R}_{k}(s)\, s^{2}\, e^{-s(1-t_{1})p_{1}^{2}-s(t_{1}-t_{2})(p_{1}+p_{2})^{2}-s_{2}t_{2}p_{4}^{2}}\,.\label{gauge_MSR_14}
\end{equation}
We can calculate the double integral in (\ref{gauge_MSR_14}) using
the properties of the $Q$-functionals as before:
\[
\int_{0}^{\infty}ds\,\tilde{R}_{k}(s)\, s^{2}\, e^{-s(1-t_{1})p_{1}^{2}-s(t_{1}-t_{2})(p_{1}+p_{2})^{2}-s_{2}t_{2}p_{4}^{2}}=\qquad\qquad
\]
\[
\qquad\qquad=Q_{-2}[R_{k}(z+s(1-t_{1})p_{1}^{2}+s(t_{1}-t_{2})(p_{1}+p_{2})^{2}+s_{2}t_{2}p_{4}^{2})]
\]
\begin{equation}
\qquad=R_{k}''(s(1-t_{1})p_{1}^{2}+s(t_{1}-t_{2})(p_{1}+p_{2})^{2}+s_{2}t_{2}p_{4}^{2})\,,\label{gauge_MSR_15}
\end{equation}
and
\[
2\int_{0}^{1}dt_{1}\int_{0}^{t_{1}}dt_{2}\, R_{k}''(s(1-t_{1})p_{1}^{2}+s(t_{1}-t_{2})(p_{1}+p_{2})^{2}+s_{2}t_{2}p_{4}^{2})=\qquad\qquad
\]
\begin{equation}
\qquad\qquad=\frac{2}{(p_{1}+p_{2})^{2}-p_{4}^{2}}\left[\frac{R_{k}((p_{1}+p_{2})^{2})-R_{k}(p_{1}^{2})}{(p_{1}+p_{2})^{2}-p_{1}^{2}}-\frac{R_{k}(p_{4}^{2})-R_{k}(p_{1}^{2})}{p_{4}^{2}-p_{1}^{2}}\right]\,.\label{gauge_MSR_16}
\end{equation}
Finally, inserting (\ref{gauge_MSR_16}) in (\ref{gauge_MSR_14})
and combining with (\ref{gauge_MSR_13}), gives the following momentum
space representation for (\ref{gauge_MSR_11}): 
\[
\Delta S_{k}^{(2;2)}[0;0]_{p_{1},p_{2},p_{3},p_{4}}^{ABCD}=[l_{p_{1},p_{2},p_{3},p_{4}}^{(2;2)}]^{ABCD}R_{p_{4},p_{1}}^{(1)}+\qquad\qquad
\]
\begin{equation}
\qquad\qquad+[l_{p_{1},-p_{1}-p_{2},p_{2}}^{(2;1)}]^{AMB}[l_{p_{3},p_{1}+p_{2},p_{4}}^{(2;1)}]^{CMD}\frac{2}{(p_{1}+p_{2})^{2}-p_{4}^{2}}\left[R_{p_{1}+p_{2},p_{1}}^{(1)}-R_{p_{4},p_{1}}^{(1)}\right]\,.\label{gauge_MSR_17}
\end{equation}
To recover relation (\ref{gauge_D_1.2}) we set $p_{1}=-p_{2}=q$
and $p_{3}=-p_{4}=p$ in (\ref{gauge_MSR_17}) so that:
\begin{equation}
\Delta S_{k}^{(2;2)}[0;0]_{q,-q,p,-p}^{ABCD}=[l_{q,-q,p,-p}^{(2;2)}]^{ABCD}R_{p}'+[l_{q,-q-p,p}^{(2;1)}]^{AMB}[l_{q+p,-q,-p}^{(2;1)}]^{CMD}R_{p+q,q}^{(2)}\,,\label{gauge_MSR_18}
\end{equation}
where $R_{p}'\equiv R'(p^{2})$. In (\ref{gauge_MSR_18}) we defined
the second finite-difference derivative of a function as:
\begin{equation}
f_{p+q,q}^{(2)}=\frac{2}{(p+q)^{2}-p^{2}}\left[f_{p+q,p}^{(1)}-f'_{p}\right]\,.\label{gauge_MSR_19}
\end{equation}
This concludes the derivation of relations (\ref{gauge_D_1.1}) and
(\ref{gauge_D_1.2}) needed in section 2.3 to write the explicit momentum
space representation for the flow equations of the zero-field proper-vertices
of the bEAA.

\newpage

\section{Basic relations for non-abelian gauge theories}

In this appendix we review our conventions and we collect the relevant
formulae for functional variations and derivatives that are used in
the paper.

\subsection{Definitions and conventions}

We consider the Lie groups $SU(N)$; we pick up a representation such
that the group elements are represented by matrices $R=e^{-\theta}$,
where $\theta=-i\theta^{a}t^{a}$ are the (infinitesimal) group parameters,
the indices $a,b,...$ run from one to $\dim SU(N)=N^{2}-1$ and the
$\dim R\times\dim R$ matrices $t^{a}$ are the generators of the
Lie algebra of $SU(N)$ in the given representation. The generators
satisfy the following commutation relations $\left[t^{a},t^{b}\right]=if^{abc}t^{c}$;
in a general representation the structure constants $f^{abc}$ are
antisymmetric in the first two indices $f^{abc}=-f^{bac}$. The covariant
derivative is defined as $D_{\mu}=\partial_{\mu}+gA_{\mu}$, where
$g$ is the gauge coupling constant and $A_{\mu}$ is the Lie algebra
valued connection. The components of the connection are defined by
$A_{\mu}=-iA_{\mu}^{a}t^{a}$ and one may write the covariant derivative
as:
\begin{equation}
D_{\mu}=\partial_{\mu}-iA_{\mu}^{a}t^{a}\,.\label{BNA_1.1}
\end{equation}
The gauge field strength $F_{\mu\nu}$ is the curvature of the gauge
connection; it can be defined as the commutator of covariant derivatives
acting on matter fields $\phi$, $[D_{\mu},D_{\nu}]\phi=F_{\mu\nu}\phi$,
with $F_{\mu\nu}=\partial_{\mu}A_{\nu}-\partial_{\nu}A_{\mu}+[A_{\mu},A_{\nu}]$
the field strength. In component form $F_{\mu\nu}=-iF_{\mu\nu}^{a}t^{a}$
and we have:
\begin{equation}
F_{\mu\nu}^{a}=\partial_{\mu}A_{\nu}^{a}-\partial_{\nu}A_{\mu}^{a}+g\, f^{abc}A_{\mu}^{b}A_{\nu}^{c}\,.\label{BNA_4}
\end{equation}
In the adjoint representation the structure constants are related
to the generators $f^{abc}=i(t_{\textrm{ad}}^{c})^{ab}$ and are completely
antisymmetric and we have $\dim\textrm{ad}=\dim SU(N)$. In this representation
the covariant derivative (\ref{BNA_1.1}) becomes:
\begin{equation}
D_{\mu}^{ab}=\partial_{\mu}\delta^{ab}-iA_{\mu}^{c}(t_{\textrm{ad}}^{c})^{ab}=\partial_{\mu}\delta^{ab}+g\, f^{acb}A_{\mu}^{c}\,.\label{BNA_5}
\end{equation}
We also have the following relation relating the commutator of the
covariant derivatives and the field strength $[D_{\mu},D_{\nu}]^{ab}=-f^{abc}F_{\mu\nu}^{c}$.

Under an infinitesimal gauge transformation $R\approx1-\theta$ matter
fields and the connection transform as:
\begin{equation}
\delta_{\theta}\phi=[\phi,\theta]\qquad\delta_{\theta}A_{\mu}=D_{\mu}\theta\qquad\delta_{\theta}F_{\mu\nu}=\left[F_{\mu\nu},\theta\right]\,.\label{BNA_6}
\end{equation}
In components we have instead%
\footnote{Note that $-iA_{\mu}^{c}(t_{\textrm{ad}}^{c})^{ab}\theta^{b}=i\theta^{c}(t_{\textrm{ad}}^{c})^{ab}A_{\mu}^{b}$.%
}:
\begin{equation}
\delta_{\theta}\phi^{i}=i\theta^{a}(t_{\textrm{R}}^{a})^{ij}\phi^{j}\qquad\delta_{\theta}A_{\mu}^{a}=D_{\mu}^{ab}\theta^{b}=\partial_{\mu}\theta^{a}+f^{abc}A_{\mu}^{b}\theta^{c}\qquad\delta_{\theta}F_{\mu\nu}^{a}=i\theta^{c}(t_{\textrm{ad}}^{c})^{ab}F_{\mu\nu}^{b}=f^{abc}F_{\mu\nu}^{b}\theta^{c}\,.\label{BNA_7}
\end{equation}
where $i,j,...$ to the dimension of the representation of the matter
fields.

\subsection{Variations and functional derivatives}

We consider the functional: 
\begin{equation}
I[A]=\frac{1}{4}\int d^{d}x\, F_{\mu\nu}^{a}F^{a\mu\nu}\,,\label{gauge_V_1}
\end{equation}
 which can be easily shown to be gauge invariant $\delta_{\theta}I[A]=0$.
Expanding (\ref{gauge_V_1}) around a background configuration $A_{\mu}=\bar{A}_{\mu}+a_{\mu}$
gives the following:
\begin{eqnarray}
I[\bar{A}+a] & = & I[\bar{A}]+\int d^{d}x\,\bar{F}^{a\mu\nu}(\bar{D}_{\mu}a_{\nu})^{a}\nonumber \\
 &  & +\frac{1}{2}\int d^{d}x\, a_{\mu}^{a}\left[(-\bar{D}^{2})^{ab}g^{\mu\nu}+2f^{abc}F^{c\mu\nu}+\bar{D}^{ac\mu}\bar{D}^{cb\nu}\right]a_{\nu}^{b}\nonumber \\
 &  & +g\, f^{abc}\int d^{d}x\,(\bar{D}_{\mu}a_{\nu})^{a}\, a_{\mu}^{b}a_{\nu}^{c}+\frac{1}{4}g^{2}f^{abc}f^{ade}\int d^{d}x\, a^{b\mu}a^{c\nu}a_{\mu}^{d}a_{\nu}^{e}\,.\label{gauge_V_2}
\end{eqnarray}
One may use $2f^{abc}F^{c\mu\nu}=2iF^{c\mu\nu}(t_{\textrm{ad}}^{c})^{ab}=-2F^{\mu\nu}$
to write the differential operator in the quadratic term of (\ref{gauge_V_2})
as $-\bar{D}^{2}g^{\mu\nu}-2F^{\mu\nu}+\bar{D}^{\mu}\bar{D}^{\nu}$.
Note that the background gauge-fixing action (\ref{gauge_5}) and
the background ghost action (\ref{gauge_6}) are already in their
varied form since they are by construction quadratic in the fields.

We now calculate the functional derivatives of the functional (\ref{gauge_V_1}),
of the gauge-fixing action (\ref{gauge_5}) and of the ghost action
(\ref{gauge_6}) that are needed in the flow equations for both the
bEAA and the gEAA. The Hessian of $I[A]$ at $A_{\mu}=0$ becomes:
\begin{equation}
[I_{p_{1},p_{2}}^{(2)}]^{\alpha\beta\, ab}=(2\pi)^{d}\delta_{p_{1}+p_{2}}\delta^{ab}\left[-g^{\alpha\beta}(p_{1}\cdot p_{2})+\frac{1}{2}\left(p_{1}^{\alpha}p_{2}^{\beta}+p_{1}^{\beta}p_{2}^{\alpha}\right)\right]\,.\label{gauge_FD_4}
\end{equation}
Note that in both (\ref{gauge_FD_4}), as in all the following formulas,
the indices and the momentum variables are related in the precise
order they appear. The third functional derivative of (\ref{gauge_V_1}),
which gives the gauge three-vertex, is:
\begin{equation}
[I_{p_{1},p_{2},p_{3}}^{(3)}]^{\alpha\beta\gamma\, abc}=(2\pi)^{d}\delta_{p_{1}+p_{2}+p_{3}}igf^{abc}\left[g^{\alpha\beta}(p_{2}-p_{1})^{\gamma}+g^{\beta\gamma}(p_{3}-p_{2})^{\alpha}+g^{\gamma\alpha}(p_{1}-p_{3})^{\beta}\right]\,.\label{gauge_FD_5}
\end{equation}
The fourth functional derivative of (\ref{gauge_V_1}), which gives
the gauge four-vertex, is:
\begin{eqnarray}
[I_{p_{1},p_{2},p_{3},p_{4}}^{(4)}]^{\alpha\beta\gamma\delta\, abcd} & = & (2\pi)^{d}\delta_{p_{1}+p_{2}+p_{3}+p_{4}}g^{2}\left[f^{eab}f^{ecd}\left(g^{\alpha\gamma}g^{\beta\delta}-g^{\alpha\delta}g^{\beta\gamma}\right)\right.\nonumber \\
 &  & +f^{eac}f^{ebd}\left(g^{\alpha\beta}g^{\gamma\delta}-g^{\alpha\delta}g^{\beta\gamma}\right)\nonumber \\
 &  & \left.+f^{ead}f^{ebc}\left(g^{\alpha\beta}g^{\gamma\delta}-g^{\alpha\gamma}g^{\beta\delta}\right)\right]\,.\label{gauge_FD_6}
\end{eqnarray}
We will use the vertices (\ref{gauge_FD_5}) and (\ref{gauge_FD_6})
to construct the zero-field proper vertices of the bEAA with both
fluctuation and background legs. This is possible, since for a gauge
invariant action as is (\ref{gauge_V_1}), the following property
holds:
\begin{equation}
\frac{\delta I[A]}{\delta A_{\mu}^{a}}=\frac{\delta I[\bar{A}+a]}{\delta a_{\mu}^{a}}=\frac{\delta I[\bar{A}+a]}{\delta\bar{A}_{\mu}^{a}}\,.\label{gauge_FD_7}
\end{equation}

Next, we need the vertices coming from the gauge-fixing action (\ref{gauge_5}),
which in component form reads: 
\begin{equation}
S_{gf}[a;\bar{A}]=\frac{1}{2\alpha}\int d^{d}x\,(\partial_{\mu}a^{a\mu}+g\, f^{abc}\bar{A}_{\mu}^{b}a^{c\mu})(\partial_{\nu}a^{a\nu}+g\, f^{ade}\bar{A}_{\nu}^{d}a^{e\nu})\,,\label{gauge_FD_7.1}
\end{equation}
The second functional derivative of (\ref{gauge_FD_7.1}) with respect
to the fluctuation field gives rise to: 
\begin{equation}
[S_{gf\; p,-p}^{(2;0)}]^{\alpha\beta\, ab}=-(2\pi)^{d}\delta_{p_{1}+p_{2}+p_{3}}\frac{1}{\alpha}\delta^{ab}p_{1}^{\alpha}p_{2}^{\beta}\,.\label{gauge_FD_8}
\end{equation}
The mixed functional derivatives give the fluctuation-fluctuation-background
vertex:
\begin{equation}
[S_{gf\; p_{1},p_{2},p_{3}}^{(2;1)}]^{\alpha\beta\gamma\, abc}=(2\pi)^{d}\delta_{p_{1}+p_{2}+p_{3}}\,\frac{1}{\alpha}igf^{abc}\left(g^{\alpha\gamma}p_{2}^{\beta}-g^{\beta\gamma}p_{1}^{\alpha}\right)\,.\label{gauge_FD_9}
\end{equation}
Four mixed functional derivatives give the fluctuation-fluctuation-background-background
vertex:
\begin{equation}
[S_{gf\; p_{1},p_{2},p_{3},p_{4}}^{(2;2)}]^{\alpha\beta\gamma\delta\, abcd}=(2\pi)^{d}\delta_{p_{1}+p_{2}+p_{3}+p_{4}}\frac{1}{\alpha}g^{2}\left(f^{ace}f^{bde}g^{\alpha\gamma}g^{\beta\delta}+f^{ade}f^{bce}g^{\alpha\delta}g^{\beta\gamma}\right)\,.\label{gauge_FD_10}
\end{equation}
The vertices (\ref{gauge_FD_9}) and (\ref{gauge_FD_10}) will be
used in section 3.2.2. The ghost action (\ref{gauge_6}), when written
out explicitly, reads:
\begin{equation}
S_{gh}[a,\bar{c},c;\bar{A}]=\int d^{d}x\left(\partial_{\mu}\bar{c}^{a}+f^{abc}\bar{A}_{\mu}^{b}\bar{c}^{c}\right)\left(\partial^{\mu}c^{a}+f^{ade}\bar{A}^{d\mu}c^{e}+gf^{ade}a^{d\mu}c^{e}\right)\,.\label{gauge_FD_11}
\end{equation}
Note that (\ref{gauge_FD_11}) generates the three-vertices ghost-ghost-fluctuation
and ghost-ghost-background but only the four-vertex ghost-ghost-background-background.
The two three-vertices differ by a factor of two since the background
field enters both covariant derivatives while the fluctuation field
does not. We have: 
\begin{equation}
[S_{gh\; p_{1},p_{2},p_{3}}^{(1,1,1;0)}]^{\alpha\, abc}=-(2\pi)^{d}\delta_{p_{1}+p_{2}+p_{3}}igf^{abc}p_{2}^{\alpha}\label{gauge_FD_12}
\end{equation}
and
\begin{equation}
[S_{gh\; p_{1},p_{2},p_{3}}^{(0,1,1;1)}]^{\gamma\, abc}=(2\pi)^{d}\delta_{p_{1}+p_{2}+p_{3}}if^{abc}(p_{2}-p_{1})^{\gamma}\,.\label{gauge_FD_13}
\end{equation}
Finally the four-vertex is:\-
\begin{equation}
[S_{gh\; p_{1},p_{2},p_{3},p_{4}}^{(0,1,1;2)}]^{\gamma\delta\, abcd}=(2\pi)^{d}\delta_{p_{1}+p_{2}+p_{3}+p_{4}}g^{\gamma\delta}\left(f^{eac}f^{ebd}+f^{ead}f^{ebc}\right)\,.\label{gauge_FD_14}
\end{equation}
The gauge cutoff operator action in the type I case is
\begin{equation}
L[a;\bar{A}]=\frac{1}{2}\int d^{d}x\,\bar{D}_{\mu}a_{\nu}\,\bar{D}^{\mu}a^{\nu}\,;\label{gauge_FD_15}
\end{equation}
its vertices are:
\begin{eqnarray}
[L_{p_{1},p_{2},p_{3}}^{(2;1)}]^{\alpha\beta\gamma\, abc} & = & (2\pi)^{d}\delta_{p_{1}+p_{2}+p_{3}}if^{abc}g^{\alpha\beta}(p_{2}-p_{1})^{\gamma}\nonumber \\
{}[L_{p_{1},p_{2},p_{3},p_{4}}^{(2;2)}]^{\alpha\beta\gamma\delta\, abcd} & = & (2\pi)^{d}\delta_{p_{1}+p_{2}+p_{3}+p_{4}}g^{\alpha\beta}g^{\gamma\delta}\left(f^{eac}f^{ebd}+f^{ead}f^{ebc}\right)\,.\label{gauge_CA_3}
\end{eqnarray}
The ghost cutoff operator action is:
\begin{equation}
L[\bar{c},c;\bar{A}]=\int d^{d}x\,\bar{D}_{\mu}\bar{c}\,\bar{D}^{\mu}c\,;\label{gauge_CA_4}
\end{equation}
since $L[\bar{c},c;\bar{A}]=S_{gh}[0,\bar{c},c;\bar{A}]$ we have:
\begin{eqnarray}
[L_{p_{1},p_{2},p_{3}}^{(1,1;1)}]^{\alpha\, abc} & = & [S_{gh\; p_{1},p_{2},p_{3}}^{(0,1,1;1)}]^{\alpha\, abc}\nonumber \\
{}[L_{p_{1},p_{2},p_{3},p_{4}}^{(1,1;2)}]^{\gamma\delta\, abcd} & = & [S_{gh\; p_{1},p_{2},p_{3},p_{4}}^{(0,1,1;2)}]^{\gamma\delta\, abcd}\,.\label{gauge_CA_5}
\end{eqnarray}
With these we derived all variations and vertices that we use in sections
3.2.2 and 3.2.3.

\end{document}